\documentclass[%
reprint,
superscriptaddress,
showpacs,
nofootinbib,
amsmath,amssymb,
aps,prd
]{revtex4-1}

\usepackage{xcolor}
\usepackage{graphicx}
\usepackage{dcolumn}
\usepackage{bm}

\newcommand{\appropto}{\mathrel{\vcenter{
  \offinterlineskip\halign{\hfil$##$\cr
    \propto\cr\noalign{\kern2pt}\sim\cr\noalign{\kern-2pt}}}}}

\begin{document}

\title{Non-local parameter degeneracy in the intrinsic space\\ of gravitational-wave signals from extreme-mass-ratio inspirals}

\author{Alvin J. K. Chua}
\email{alvincjk@nus.edu.sg}
\affiliation{Department of Physics, National University of Singapore, Singapore 117551}
\affiliation{Department of Mathematics, National University of Singapore, Singapore 119076}
\affiliation{Theoretical Astrophysics Group, California Institute of Technology, Pasadena, CA 91125, U.S.A.}

\author{Curt J. Cutler}
\email{curt.j.cutler@jpl.nasa.gov}
\affiliation{Jet Propulsion Laboratory, California Institute of Technology, Pasadena, CA 91109, U.S.A.}

\date{\today}

\begin{abstract}
Extreme-mass-ratio inspirals will be prized sources for the upcoming space-based gravitational-wave observatory LISA. The hunt for these is beset by many open theoretical and computational problems in both source modeling and data analysis. We draw attention here to one of the most poorly understood: the phenomenon of non-local correlations in the space of extreme-mass-ratio-inspiral signals. Such correlations are ubiquitous in the continuum of possible signals (degeneracy), and severely hinder the search for actual signals in LISA data. However, they are unlikely to manifest in a realistic set of putative signals (confusion). We develop an inventory of new analysis tools in order to conduct an extensive qualitative study of degeneracy---its nature, causes, and implications. Previously proposed search strategies for extreme-mass-ratio inspirals are reviewed in the light of our results, and additional guidelines are suggested for the scientific analysis of such sources.

\end{abstract}

\maketitle

\section{Introduction}
\label{sec:introduction}

\subsection{Background}
\label{subsec:background}

In gravitational-wave (GW) astronomy, source modeling and data analysis come together on the manifold of possible signals that is described by a given waveform model and analysis setting. Our intuition and knowledge of such \emph{signal spaces} guides the development of techniques and strategies in GW scientific analysis: for example, local correlations in signal space might be used to perform approximate inference via the Fisher information, while the global structure of these correlations has utility in the fitting of numerical waveform data for the construction of faster approximate models. Many such modeling and analysis methods have been developed in the forge of contemporary ground-based observing \cite{LIGOScientific:2020ibl}; most will be highly relevant, if not directly transferable, to sources for near-future millihertz observatories such as the planned ESA--NASA mission LISA \cite{2017arXiv170200786A}. However, the nature of signal space---and thus the optimal approach to adopt in scientific analysis---remains poorly understood for one particular class of millihertz source.

Extreme-mass-ratio inspirals (EMRIs) are the late-stage orbits of astrophysical binaries with a small mass ratio $\epsilon\lesssim10^{-4}$. They arise as the capture of stellar-mass compact objects (white dwarfs, neutron stars or black holes) by massive black holes in galactic nuclei \cite{Amaro-Seoane:2012lgq}, and will be an important class of source for the LISA detector \cite{Babak:2017tow,Berry:2019wgg}. The GW signals from EMRIs that involve central black holes of $\sim10^5$--$10^7M_\odot$ can persist throughout the planned four-year lifetime of the LISA mission, typically with $\sim10^5$ observable cycles. At the same time, most EMRIs will likely exhibit extreme periapsis precession and Lense--Thirring precession, since the motion occurs deep in the strong field of the rotating central mass; those that are formed through traditional capture channels may also enter the LISA sensitivity band with high eccentricity \cite{Amaro-Seoane:2012jcd}. These effects endow EMRI signals with rich harmonic content over their many cycles.

With their combination of signal longevity and strong-field complexity, EMRIs (and their intermediate-mass-ratio cousins with $10^{-4}\lesssim\epsilon\lesssim10^{-2}$) have no analog in any other channel of GW astronomy. Our best EMRI waveform models will be constructed through black-hole-perturbation theory and multi-scale approaches \cite{Barack:2018yvs,Pound:2021qin} to leverage the extreme mass ratio and slow evolution, but the long duration of expected signals places exacting constraints on the accuracy and efficiency of calculations. Overcoming the theoretical and computational challenges in EMRI forward modeling remains an open and active area of research, with several promising recent developments \cite{Pound:2019lzj,Chua:2020stf,Warburton:2021kwk,Wardell:2021fyy}. In the inverse direction, the extraction and characterization of EMRI signals in LISA data has to contend with the information volume of the signal space, which is 20--30 orders of magnitude larger than in the case of comparable-mass binaries \cite{Gair:2004iv,Moore:2019pke}. Further understanding of the signal space has been limited by the lack of suitable modeling and analysis tools; little is known about its global correlation structure, or the representativeness of the simple proxy models \cite{Barack:2003fp,Babak:2006uv,Chua:2017ujo} used to date.

\subsection{Definition of key concepts}
\label{subsec:keyconcepts}

The structure of \emph{correlations} in the EMRI signal space is relevant to LISA scientific analysis in two distinct ways: \emph{confusion} and \emph{degeneracy}. All three of these terms are commonly and loosely used in the GW and LISA literature to refer to a variety of related concepts. Here we shall define them more concretely for the purposes of this work (and, it is hoped, for future use by the community). Given a waveform model $h:\Theta\to\mathcal{D}$ with some fixed sampling rate and duration, the canonical noise-weighted cross-correlation $\langle\cdot|\cdot\rangle$ \cite{Cutler:1994ys} between two signals $h_{1,2}:=h(\boldsymbol{\theta}_{1,2})$ in the signal space $\mathcal{S}:=h[\Theta]$ defines an inner product on the ambient data space $\mathcal{D}$ of fixed-length time/frequency series (in which $\mathcal{S}$ can be treated as an embedded submanifold). Of frequent interest is the normalized cross-correlation, or \emph{overlap} $\Omega(\cdot,\cdot)$; this can be interpreted as (the cosine of) the angle between $h_{1,2}$ in $\mathcal{D}$, and serves as a natural measure of signal similarity in that space. We will informally use the term ``correlation'' to mean the overlap rather than the cross-correlation, and further qualify the correlation between $h_{1,2}$ as \emph{local} if it is accompanied by a ``small'' distance between $h_{1,2}$ in $\mathcal{S}$ (equivalently, between $\boldsymbol{\theta}_{1,2}$ in $\Theta$). See Sec.\,\ref{subsec:measures} for the explicit definitions of $\langle\cdot|\cdot\rangle$, $\Omega$, and locality.

Confusion refers to the presence of non-negligible correlations among a finite set of putative signals in the LISA data stream. It is a defining feature of the full LISA catalog, which comprises many resolvable and unresolvable signals across multiple source types, and it provides the primary motivation for the LISA \emph{global fit}---a governing strategy to account for the multitude of correlated signals by searching for and characterizing them at the same time. The main contributor to global confusion is the Galactic population of compact binaries, as there are millions of such sources whose signals will form a foreground of astrophysical noise in the LISA band \cite{Nelemans:2001hp}. In the case of EMRIs, extreme estimates for the number of sub-threshold signals will also reduce the effective sensitivity of LISA, potentially quite significantly \cite{Barack:2004wc,Bonetti:2020jku}. Confusion between specific classes of source (e.g., a single EMRI and the Galactic-binary foreground) has been studied in some depth as well. In \cite{Racine:2007gv}, analytic arguments from large-deviations theory are used to show that when analyzing resolvable signals of a given source type, a combination of unresolvable signals from a different source type can be well approximated as Gaussian noise.

Our focus in this work is more on \emph{self-confusion}, where the signals under consideration are due to members from a single class of source, i.e., all sources (signals) are well described by points in the domain (image) of a single waveform model. The archetypal example of self-confusion is again provided by the Galactic binaries. Each source is described by a narrowband signal, so any signal-space correlations are effectively local; the problem is then one of resolving individual signals, with leading treatments employing the transdimensional sampling of a multi-source likelihood \cite{Littenberg:2020bxy}. Self-confusion is less of an immediate problem for EMRIs, because of their highly uncertain event rates \cite{Babak:2017tow}. It has yet to be established whether self-confusion might occur for a realistic number ($\lesssim10^4$) of detectable signals, but common intuition is that it will not, due to the sheer volume of the signal space. This is largely borne out by the results of an idealized calculation that we present in Sec.\,\ref{sec:confusion}.

We use the term ``degeneracy'' to mean the presence of non-negligible and \emph{non-local} correlations in the continuous signal space of a given waveform model. The non-local condition excludes its more informal usage in GW data analysis to describe extensive but essentially local regions of high correlation. Degeneracy manifests as disjoint secondary maxima in the overlap surface $\Omega(h_\mathrm{inj},\cdot)$ over the model parameter space, where the overlap is between a reference signal injection $h_\mathrm{inj}$ and the signal template at each point. We will also exclude from our definition the characteristic $\mathrm{sinc}$-like ``ringing'' of the overlap surface near the injection, as this is generally low-amplitude and still relatively local. Degeneracy does not affect the narrowband Galactic binaries, or massive-black-hole mergers with their rapid evolution and merger-dominated signal-to-noise ratio (SNR). It is a unique issue for EMRIs, and has significant implications even in the case of a lone EMRI signal in the data.

EMRI degeneracy occurs when different and non-local combinations of the source-\emph{intrinsic} parameters result in similar values for a subset of the three initial fundamental frequencies, as well as their starting time derivatives up to some order (see Sec.\,\ref{subsec:cause}). This causes the phasing of the dominant harmonic mode in both the injection and the degenerate template to be aligned for much of the inspiral duration. Such a criterion is highly improbable for a realistic set of putative signals, but we show in Sec.\,\ref{sec:degeneracy} that it is satisfied in a surprisingly large number of disjoint regions across the space of all possible signals. An alignment of dominant and sub-dominant modes can also cause degeneracy \cite{Cornish:2008zd,Babak:2009ua}---but to a lesser extent, since the relative distribution of mode amplitudes must also be similar for two signals to have a high overlap. This latter case might become more relevant when \emph{extrinsic} parameters are considered (again, see Sec.\,\ref{subsec:cause}).

Early hints of degeneracy in the EMRI signal space arose during the Mock LISA Data Challenges \cite{MockLISADataChallengeTaskForce:2009wir}, where participants either reported the presence of a few secondary peaks in the posterior surface, or misidentified one of them as the primary peak (even with fairly localized priors). At SNRs above the common estimate of 20 for the EMRI detection threshold \cite{Chua:2017ujo}, these posterior \emph{secondaries} are exponentially suppressed relative to the primary peak. However, they can pose practical difficulties for ``uninformed'' search and inference algorithms if the overlap against the injection at their locations is high, and/or if they are unexpectedly distant from the primary peak, and/or if they are inordinately numerous. Such information about the nature of secondaries has remained unavailable thus far; the global (log-)likelihood surface is extremely challenging to \emph{map out} even for a single injection, much less a globally representative set of injections. This is in turn due to the computational limitations of existing waveform models, as well as standard sampling and clustering methods being ill suited to the task of high-dimensional mapping and visualization.

\subsection{Synopsis of results}
\label{subsec:synopsis}

In this work, we investigate both confusion and degeneracy for EMRI signals, within the setting of LISA data analysis. The confusion study in Sec.\,\ref{sec:confusion} is the more straightforward of the two, as it is a conventional calculation that makes use of existing tools, and its results are perhaps somewhat unsurprising. We examine the pairwise overlaps among a set of $N$ detectable two-year signals, whose source parameters are distributed according to a simple astrophysical model (modified by selection based on some SNR threshold). For astrophysically realistic estimates of $N$, we conclude that self-confusion is unlikely to pose a problem for the extraction and characterization of EMRI signals. No signal in a set of $N\approx200$ should resemble any of the others beyond a $\sim1\%$ overlap level, while the root-mean-square correlation among signals is approximately constant at $\sim0.1\%$ for $N\lesssim200$.

The bulk of this work is devoted to an extensive (but by no means comprehensive) study of EMRI degeneracy. Such a study is hindered by several factors: i) the computational cost of generating analysis-length waveforms, even with simple proxy models; ii) the focus of modern sampling algorithms on optimization and density estimation rather than mapping; iii) the general difficulty of clustering and visualizing high-dimensional data; and iv) the intricate structure in the overlap surface that borders on noise in any region away from the injection parameters (which turns out to make the problem of localizing, counting or even defining ``secondaries'' somewhat ill posed). To overcome or at least circumvent these difficulties, we introduce various new tools (see Sec.\,\ref{sec:tools}) such as a stripped-down version of a fast semi-relativistic model \cite{Chua:2017ujo}, an approximation to the inner product $\langle\cdot|\cdot\rangle$, an exploratory sampling density for the recovery of high-overlap points, and a bespoke algorithm for the clustering of such points into approximately disjoint secondaries.

With these tools, we are able to unambiguously demonstrate the existence of degeneracy in the EMRI signal space for signals as long as two years, and can also shed some light on the severity and prevalence of secondaries. Our study is detailed in Sec.\,\ref{sec:degeneracy}; we map out the overlap surface $\Omega(h_\mathrm{inj},\cdot)$ for a single representative injection at varying magnification levels, from a starting region that is tightly centered on the posterior bulk, to a final region whose Euclidean volume is $\gtrsim10^{12}$ times larger. We find that secondaries exhibit varying degrees of connectivity to one another, and take on an assortment of shapes and scales with no immediately discernible global patterns. However, both their number density and their overlap against the injection appear to fall off with distance from the injection. Furthermore, case studies of specific secondaries also indicate that the posterior surface in those regions is unlikely to resemble the posterior corresponding to an actual injection at the same location.

In Secs \ref{sec:implications} and \ref{sec:suggestions}, we discuss the implications of our qualitative findings for EMRI data analysis, broadly review previously proposed search strategies, and provide our own suggestions to guide future work. We identify pathological scenarios that might arise from the unfortunate interaction of degeneracy with detector noise or the presence of multiple actual signals, but argue that they are highly improbable. Thus the main issues we foresee are practical in nature: the computational difficulty of stochastic search, and the verification of candidate signals. Past research on EMRI data analysis has mostly attempted to address the former, with some degree of success; here we propose several complementary strategies to tackle both issues, informed by the results of our confusion and degeneracy studies. These take the form of post-hoc vetoes for candidate signals, as well as a modified ``veto likelihood'' that is designed for search.

\subsection{Table of contents}

\tableofcontents

\section{Inventory of tools}
\label{sec:tools}

\subsection{Waveform and response models}
\label{subsec:models}

Fully relativistic EMRI waveform models that are both efficient and extensive enough for data-analysis studies are still under active development. The current state of the art for efficiency-oriented waveforms is a fast model describing eccentric inspirals in Schwarzschild spacetime \cite{Chua:2020stf}, whose frequency evolution is accurate up to leading \emph{adiabatic} order in the mass ratio $\epsilon$. In terms of extensiveness, recent work has introduced several models for various classes of inspirals in Kerr spacetime: eccentric and equatorial with adiabatic evolution \cite{Fujita:2020zxe}; fully generic (i.e., eccentric and inclined) with a post-Newtonian (PN) approximation to adiabatic evolution \cite{Isoyama:2021jjd}; and fully generic with adiabatic evolution \cite{Hughes:2021exa}. Such techniques will soon be combined to construct fast and fully generic adiabatic Kerr models---these in turn are precursors to the accurate, efficient and extensive \emph{post-adiabatic} models \cite{Miller:2020bft} that will enable us to achieve the EMRI-related science goals of LISA.

None of the existing models from above are suitable for the present work, which requires large-scale Monte Carlo simulations along with a sufficiently representative depiction of the full Kerr signal space. We instead employ a semi-relativistic model for fully generic Kerr inspirals, which combines adiabatic-fitted frequency evolution \cite{Gair:2005ih} with Newtonian instantaneous amplitudes (in the Peters--Mathews formalism \cite{PhysRev.131.435}). This model is known as the augmented analytic ``kludge'' (AAK) \cite{Chua:2017ujo}; it is based largely on an influential earlier construction for LISA data-analysis studies \cite{Barack:2003fp}, but correctly accounts for relativistic frequencies \cite{Chua:2015mua}. We refer the reader to \cite{Chua:2017ujo} for a full description of this model, along with a short review of its kludge predecessors \cite{Barack:2003fp,Babak:2006uv}.

As our study on EMRI confusion involves a straightforward analysis with an unambiguous answer, it is worth doing with slightly more ``realistic'' tools. There we use the AAK waveform integrated with a fast model for the time-delay-interferometry (TDI) \cite{Tinto:2004wu} response $h_\mathrm{A,E,T}$ of LISA to the signal, as implemented in \cite{Babak:2009ua,Chua:2019wgs}. In the degeneracy part of this work, we opt instead for the much simpler long-wavelength LISA response $h_\mathrm{I,I\!I}$ \cite{Cutler:1997ta} (essentially, approximating LISA as a point detector relative to the gravitational wavelength). One reason is that the AAK--TDI model was only developed well after the start of our investigation---but in any case, the choice of response model should not significantly alter the structure of source-intrinsic degeneracy, which is our focus here. This is due to the weak coupling between intrinsic and extrinsic parameters, as discussed in Sec.\,\ref{subsec:mapping}.

The full AAK model (i.e., including LISA response) is parametrized by a set of 14 source parameters, which can be further partitioned into the intrinsic parameters
\begin{equation}\label{eq:intrinsic}
    \boldsymbol{\theta}_\mathrm{int}:=(\mu,M,a/M,p_0/M,e_0,\iota,\Phi_0,\gamma_0,\alpha_0)
\end{equation}
and the extrinsic parameters
\begin{equation}\label{eq:extrinsic}
    \boldsymbol{\theta}_\mathrm{ext}:=(\theta_K,\phi_K,\theta_S,\phi_S,D),
\end{equation}
where all quantities are dimensionless. Explicitly:
\begin{itemize}
    \item $(\mu,M)$ are the detector-frame component masses, in Solar masses and with $\epsilon=\mu/M\ll1$;
    \item $a$ is the Kerr spin lengthscale for the central mass;
    \item $(p_0,e_0,\iota)$ are the quasi-Keplerian semi-latus rectum, eccentricity and inclination for the osculating geodesic to the inspiral at reference time $t_0$;
    \item $(\Phi_0,\gamma_0,\alpha_0)$ are phase angles describing the position of the small mass at reference time $t_0$;
    \item $(\theta_K,\phi_K)$ are polar and azimuthal angles describing the spin orientation in ecliptic coordinates;
    \item $(\theta_S,\phi_S)$ are polar and azimuthal angles describing the sky location in ecliptic coordinates;
    \item $D$ is the luminosity distance in Gpc.
\end{itemize}
Henceforth we will abuse the symbol $\boldsymbol{\theta}$ to denote any ordered combination of these parameters, as long as the specific parameters in question are clear from context.

There are several points to note about this choice of parametrization. First, the distinction between intrinsic and extrinsic EMRI parameters can be somewhat arbitrary, and also depends on whether one is working in the modeling or analysis context. It is a largely semantic distinction, though, and here we take more of a modeling viewpoint (as opposed to \cite{Chua:2017ujo})---but note that the sets $\{p_0,e_0,\iota\}$ and $\{\Phi_0,\gamma_0,\alpha_0\}$ each technically have one observer-dependent degree of freedom, corresponding to temporal translation and spatial rotation respectively. Second, the inclination angle used is $\iota:=\tan^{-1}(\sqrt{Q}/L_z)$ rather than $I:=\pi/2-\mathrm{sgn}(L_z)\theta_\mathrm{min}$ (where $(Q,L_z,\theta_\mathrm{min})$ are the Carter constant, the projection of angular momentum onto the spin axis, and the turning point of polar motion). Usage of the latter is becoming standard in the modeling community, but $\iota\approx I$ across much of the Kerr geodesic space \cite{Drasco:2005kz}. The rate of change for $\iota$ due to GW radiation is also generally small \cite{Hughes:1999bq}, and is thus approximated as zero in the AAK model. Finally, future post-adiabatic models are likely to have a qualitatively similar parametrization, apart from: i) the time evolution of $(M,a)$ \cite{Pound:2021qin}, which would simply change $(M,a)\to(M_0,a_0)$; and ii) secular or resonant effects due to the spin of the small mass \cite{Mathews:2021rod,Drummond:2022xej,Drummond:2022efc,Piovano:2020ooe,Piovano:2020zin,Zelenka:2019nyp}, although more detailed studies are required to determine how measurable this quantity will be with LISA.

The AAK model is efficiency-oriented, with a computational wall time of $\sim10\,\mathrm{s}$ for a two-year signal sampled at $0.1\,\mathrm{Hz}$. While this is adequate for the confusion study, our investigation of EMRI degeneracy calls for multiple Monte Carlo simulations, each with up to billions of template evaluations (if this seems excessive, recall that our aim here is high-resolution mapping---not search or inference). For the degeneracy study, we thus rely on an amplitude-and-phase representation of the AAK model, further stripped down to just four strong harmonic modes. Although the semi-relativistic and quadrupolar AAK waveform falls well short of realistic harmonic content, this still reduces the number of modes by about an order of magnitude. The amplitude and phase trajectories also vary smoothly over the radiation-reaction timescale $M/\epsilon$, and are downsampled by a factor of $\gtrsim10^3$ for direct use in an approximate inner product (see Sec.\,\ref{subsec:measures}). With these simplifications, the wall time for a single template evaluation (plus typical operations on said template) is slashed to $\lesssim10\,\mathrm{ms}$.

An explicit description of harmonic modes in the AAK model will depend on the specific choice of harmonic basis. The model is constructed from Keplerian orbits with artificially induced precession; this is reflected in the phase parametrization $(\Phi,\gamma,\alpha)$, where $\Phi$ is the quasi-Keplerian mean anomaly and $(\gamma,\alpha)$ are two precession-related angles. Data-analysis studies that involve the AAK (and its predecessor \cite{Barack:2003fp}) typically use the frequencies $(\dot{\Phi},\dot{\gamma},\dot{\alpha})$ as the harmonic basis, where $\dot{\gamma}+\dot{\alpha}$ and $\dot{\alpha}$ are the rates of periapsis and Lense--Thirring precession respectively. The phases $(\Phi,\gamma,\alpha)$ have a simple relation to the \emph{fundamental phases} $(\varphi_r,\varphi_\theta,\varphi_\phi)$, i.e., the generalized coordinates associated with the action-angle variables for Kerr geodesic motion \cite{Schmidt:2002qk}:
\begin{equation}\label{eq:phasemap}
    (\varphi_r,\varphi_\theta,\varphi_\phi)=(\Phi,\Phi+\gamma,\Phi+\gamma+\alpha).
\end{equation}
Another harmonic basis is thus provided by the Kerr \emph{fundamental frequencies}
$(\omega_r,\omega_\theta,\omega_\phi)$---the derivatives of $(\varphi_r,\varphi_\theta,\varphi_\phi)$ with respect to coordinate time. We choose the latter basis for this work. The phase and angular frequency of a mode $(m,k,n)$ are given respectively by
\begin{equation}\label{eq:modephase}
    \varphi_{mkn}:=m\varphi_\phi+k\varphi_\theta+n\varphi_r,
\end{equation}
\begin{equation}\label{eq:modefreq}
    \omega_{mkn}:=m\omega_\phi+k\omega_\theta+n\omega_r,
\end{equation}
where $\omega$ is used here and henceforth to denote dimensionful frequencies with units of Hz.

\begin{figure}[!tbp]
\centering
\includegraphics[width=\columnwidth]{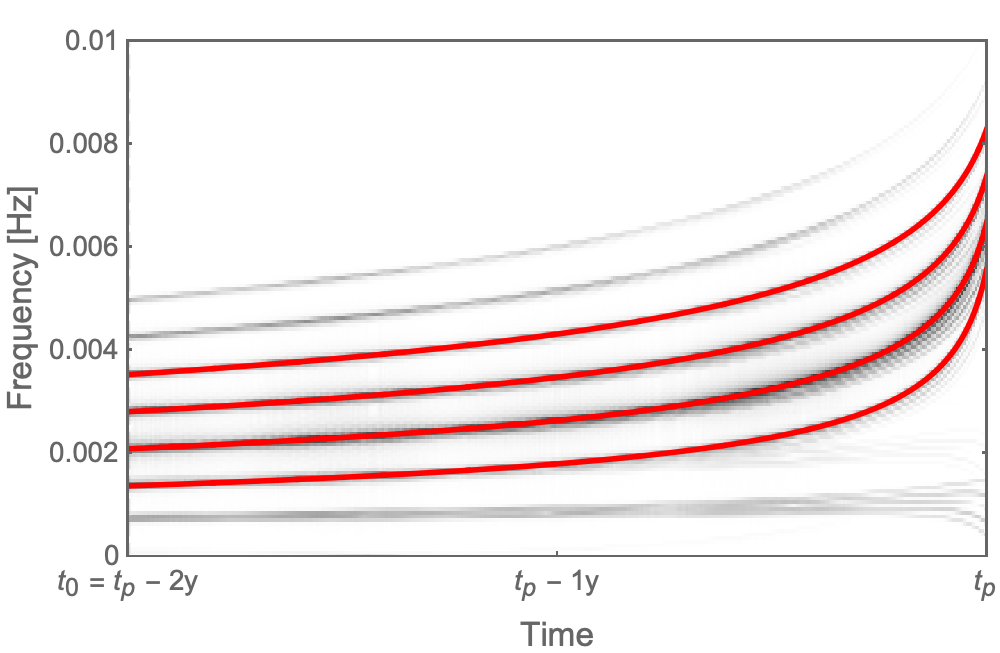}
\caption{Spectrogram of reference signal ($h_I$ channel only) with intrinsic parameters \eqref{eq:injpars}. Short-time Fourier amplitude (grayscale value) is in log scale. Overlaid in red are frequency trajectories for the four strong harmonic modes \eqref{eq:strongmodes}.}
\label{fig:spectrogram}
\end{figure}

\begin{figure}[!tbp]
\centering
\includegraphics[width=0.99\columnwidth]{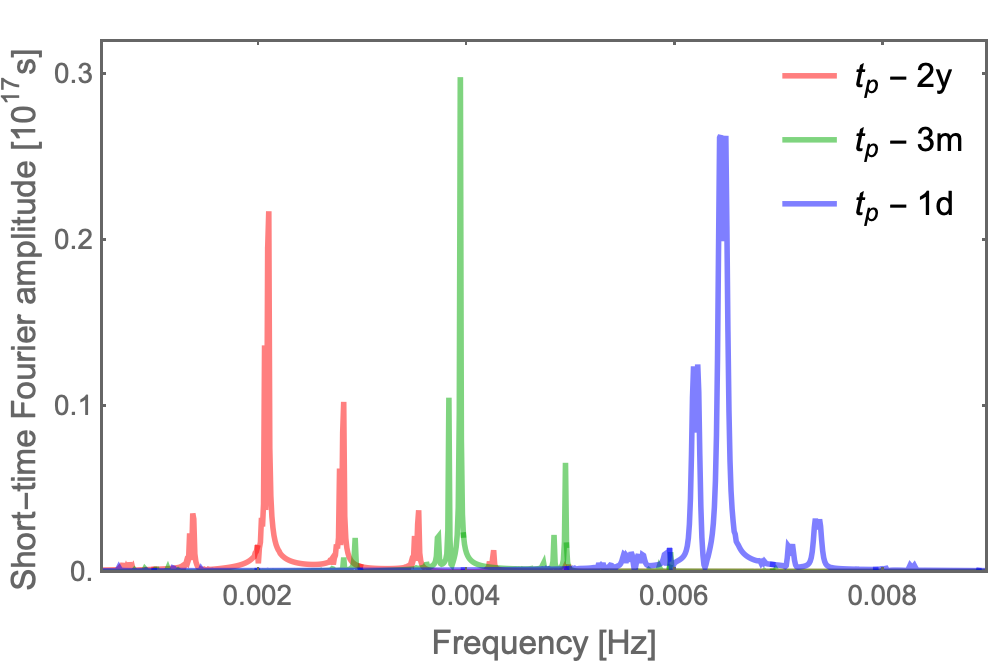}
\caption{Linear-scale cross sections of the spectrogram in Fig.\,\ref{fig:spectrogram}, at two years / three months / one day before plunge. The four modes \eqref{eq:strongmodes} are mostly representative of the signal, but strong sidebands become more resolvable in the final months.}
\label{fig:fourieramps}
\end{figure}

For illustrative purposes, we introduce here the reference signal injection $h_\mathrm{inj}$ that is used in the degeneracy study of Sec.\,\ref{sec:degeneracy}; its intrinsic parameters are
\begin{equation}\label{eq:injpars}
\boldsymbol{\theta}_\mathrm{inj}=\left(10,10^6,0.5,9.5,0.2,\pi/6,0,0,0\right),
\end{equation}
with randomly chosen extrinsic parameters. The masses and spin are assigned ``central'' values in the LISA-relevant range, while the initial semi-latus rectum is fixed by requiring that the small mass plunges (reaches the Kerr separatrix) at time $t_p=t_0+2\,\mathrm{y}$. We focus on low-to-moderate starting eccentricity due to the trimming of modes, and examine the slightly inclined prograde case. Over its two-year duration, much of the power in the signal is contributed by the four modes
\begin{equation}\label{eq:strongmodes}
    m=2,\quad k=0,\quad -1\leq n\leq2,
\end{equation}
and in particular by the $(2,0,0)$ mode. This can be visualized through the time--frequency plots in Figs \ref{fig:spectrogram} and \ref{fig:fourieramps}, for the $h_\mathrm{I}$ channel of the long-wavelength response $h_\mathrm{I,I\!I}$.

The exact AAK implementation used in this work is that from the now-discontinued \texttt{EMRI Kludge Suite} (v0.5.2) \cite{KS}, which is largely faithful to the original presentation in \cite{Chua:2017ujo}. During the course of this work, an updated AAK model with 5PN evolution \cite{Fujita:2020zxe} and GPU support was included in the \texttt{Fast EMRI Waveforms} software package that will house the next-generation of EMRI models \cite{Katz:2021yft,FEW}. The AAK--5PN waveform features several improvements such as exact fundamental frequencies (rather than their PN expansions) and an evolving inclination, and is also significantly accelerated over the CPU version in \cite{KS}. Future follow-up studies without the approximations made in this work will likely employ that model, or a relativistic 5PN model \cite{Isoyama:2021jjd} that is currently being implemented within \cite{FEW}. However, our present approach is still an order of magnitude faster than the GPU generation and manipulation of templates at full sampling resolution---and more importantly, we do not expect our main results to change qualitatively for other EMRI models (see extended discussion in Sec.\,\ref{subsec:representativeness}).

\subsection{Similarity measures}
\label{subsec:measures}

In GW data analysis, the detection of a signal $h$ in noisy time-series strain data $x=h+n$ relies on the linear filtering of $x$ against a signal template that matches $h$. If the detector noise $n$ is approximated as a zero-mean and stationary process, this matched-filtering procedure can be expressed as a noise-weighted cross-correlation
\begin{equation}\label{eq:fullinner}
    \langle x|h\rangle:=4\,\mathrm{Re}\sum_\chi\sum_{f>0}^{f_N}\delta\!f\,\frac{\tilde{x}_\chi(f)^*\tilde{h}_\chi(f)}{S_{n,\chi}(f)},
\end{equation}
where the outer sum is over all independent data channels $\chi$ (for this work, $\chi=\mathrm{A,E,T}$ or $\chi=\mathrm{I,I\!I}$); $f_N$ is the Nyquist frequency; $\delta\!f$ is the frequency resolution; overtildes denote discrete Fourier transforms (multiplied by the time resolution $\delta t$); and $S_{n,\chi}$ is the one-sided power spectral density of the channel noise $n_\chi$ (provided here by analytic models that correspond to the science requirements for the LISA mission \cite{SciRD,Petiteau:2008zz,Robson:2018ifk}). The noise assumptions and the form of Eq.\,\eqref{eq:fullinner} give rise to the identities
\begin{equation}\label{eq:noisemean}
    \mathrm{E}[\langle n|a\rangle]=0,
\end{equation}
\begin{equation}\label{eq:noisevar}
    \mathrm{E}[\langle n|a\rangle\langle n|b\rangle]=\langle a|b\rangle,
\end{equation}
valid for all time series $a,b$ with the same length as $n$.

Eq.\,\eqref{eq:fullinner} satisfies the conditions for an inner product on the data space $\mathcal{D}$ of fixed-length time/frequency series, which unlocks a useful geometric picture of other common concepts in GW data analysis. For one, the overlap between two signals $h_{1,2}$ is simply their normalized inner product, or the cosine of the angle between them in $\mathcal{D}$:
\begin{equation}\label{eq:overlap}
    \Omega(h_1,h_2):=\frac{\langle h_1|h_2\rangle}{\sqrt{\langle h_1|h_1\rangle\langle h_2|h_2\rangle}}.
\end{equation}
The optimal SNR of a signal template $h$ is its norm:
\begin{equation}\label{eq:optsnr}
    \rho_\mathrm{opt}(h):=\frac{\mathrm{E}[\langle x|h\rangle]}{\sqrt{E[\langle n|h\rangle\langle n|h\rangle]}}=\sqrt{\langle h|h\rangle},
\end{equation}
and its detection SNR is the scalar projection of $x$ on $h$:
\begin{equation}\label{eq:detsnr}
    \rho_\mathrm{det}(h):=\frac{\langle x|h\rangle}{\sqrt{\langle h|h\rangle}}.
\end{equation}
Maximum-likelihood estimation with the standard GW likelihood $L(\boldsymbol{\theta}|x)$ (see Sec.\,\ref{subsec:sampling}) boils down to distance minimization between $x$ and the signal space $\mathcal{S}$ in $\mathcal{D}$:
\begin{equation}\label{eq:mle}
    \boldsymbol{\theta}_\mathrm{ML}=\underset{\boldsymbol{\theta}}{\mathrm{argmin}}\,\langle x-h(\boldsymbol{\theta})|x-h(\boldsymbol{\theta})\rangle.
\end{equation}
Finally, the Fisher information matrix $\mathcal{I}(\boldsymbol{\theta}_\mathrm{inj})$ \cite{Vallisneri:2007ev} for $L(\boldsymbol{\theta}|x)\propto p(x|\boldsymbol{\theta}_\mathrm{inj})$ coincides precisely (component-wise) with the pullback by $h(\boldsymbol{\theta})$ of the flat metric on $\mathcal{D}$:
\begin{equation}\label{eq:fisher}
    [\mathcal{I}(\boldsymbol{\theta}_\mathrm{inj})]_{ij}=\langle\partial_ih(\boldsymbol{\theta})|\partial_jh(\boldsymbol{\theta})\rangle|_{\boldsymbol{\theta}_\mathrm{inj}}.
\end{equation}

This geometric picture extends to the concept of local correlations in signal space, and thus to that of degeneracy---the presence of non-negligible and non-local correlations. From the informal definition in Sec.\,\ref{subsec:keyconcepts}, the correlation between two signals is local if there is a ``small'' distance between their corresponding points in parameter space $\Theta$ (with respect to the pullback metric at one of the points). To make this slightly more concrete, we will say that two parameter points $\boldsymbol{\theta}_{1,2}$ are local with respect to each other if their metric distance in $\Theta$ approximately equals the Euclidean distance in $\mathcal{D}$ between their associated signals $h_{1,2}$, i.e.,
\begin{equation}\label{eq:locality}
    \sqrt{(\boldsymbol{\theta}_1-\boldsymbol{\theta}_2)^T\mathcal{I}(\boldsymbol{\theta}_1-\boldsymbol{\theta}_2)}\approx\sqrt{\langle h_1-h_2|h_1-h_2\rangle},
\end{equation}
where $\mathcal{I}$ is also approximately invariant: $\mathcal{I}(\boldsymbol{\theta}_1)\approx\mathcal{I}(\boldsymbol{\theta}_2)$. In other words, $h_{1,2}$ are local if they coexist in a region where the signal manifold is nearly flat. From a data-analysis perspective, the definition \eqref{eq:locality} holds in a region around the maximum-likelihood parameters where the likelihood is near-Gaussian, and to a lesser extent in the immediate vicinity of this region, with decaying likelihood oscillations in parameter directions that affect frequency. Degeneracy on the other hand occurs when the specific embedding $h(\boldsymbol{\theta})$ admits (regions of) signals that have high overlap against some $h_\mathrm{inj}\in\mathcal{S}$, but whose parameters are non-local to the neighborhood of $\boldsymbol{\theta}_\mathrm{inj}\in\Theta$. 

The inner product $\langle\cdot|\cdot\rangle$ is defined for time series at full sampling resolution, and can be computationally unwieldy even without accounting for the cost of template generation. For the degeneracy study in Sec.\,\ref{sec:degeneracy}, we introduce an approximation to $\langle\cdot|\cdot\rangle$ that acts directly on amplitude and phase trajectories from the AAK model. Beyond its original partial decomposition into harmonic modes of only radial motion, the AAK waveform with LISA response $h_\mathrm{I,I\!I}$ is more fully decomposed as
\begin{equation}\label{eq:aakdecomp}
    h_\mathrm{I}(t)+ih_\mathrm{I\!I}(t)=\sum_jA_j(t)\exp{(i\varphi_j(t))},
\end{equation}
where $1\leq(j:=n+2)\leq4$ is a re-indexing of the four strong modes \eqref{eq:strongmodes}, and $A_j$ is a complex amplitude for mode $j$ \cite{KS}. The mode phasing $\varphi_j$ is obtained from the time integration of frequency trajectories $\omega_j(t)$, whose availability also allows us to fold the noise-weighting of $\langle\cdot|\cdot\rangle$ directly into the time-domain signal \cite{Barack:2003fp}. Using overbars to denote noise-weighted quantities, $\bar{h}_\mathrm{I,I\!I}$ is given by the analog of Eq.\,\eqref{eq:aakdecomp} with $A_j\to\bar{A}_j$, where
\begin{equation}
    \bar{A}_j(t):=\frac{A_j(t)}{\sqrt{S_n(f_j(t))}};
\end{equation}
i.e., the mode amplitudes at time $t$ are essentially reduced by the noise estimate $S_n:=S_{n,\mathrm{I}}=S_{n,\mathrm{I\!I}}$ at the corresponding instantaneous mode frequencies $f_j:=\omega_j/(2\pi)$.

\begin{figure}[!tbp]
\centering
\includegraphics[width=\columnwidth]{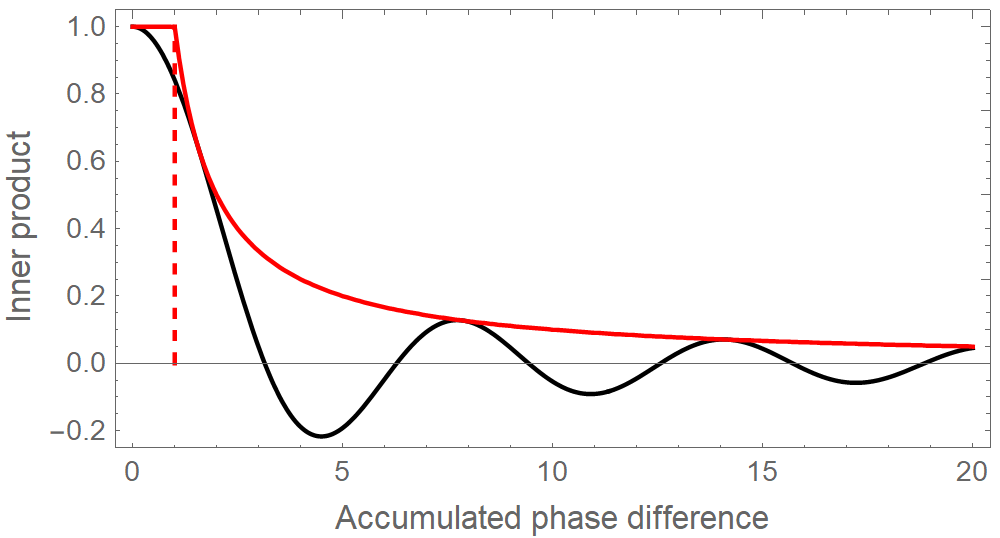}
\caption{Full (black) and approximate (red) inner product between $\cos{(\omega t)}$ and $\cos{(\omega't)}$ over duration $T=1$, as function of accumulated phase difference $(\omega-\omega')T$. Vertical red dashed line indicates specified phase tolerance of 1.}
\label{fig:innprod}
\end{figure}

In the above representation, the mode amplitude and phase trajectories are smooth enough to be downsampled significantly. For our degeneracy study, we consider a signal injection specified by the source parameters \eqref{eq:injpars}, with an analysis duration of $T=t_p-t_0=2\,\mathrm{y}$ and a fixed trajectory timestep of $\Delta t=T/10^3$. All injection--template comparisons are performed on the common analysis interval $\{t:t_0\leq t\leq t_0+T\}$, using the same trajectory timestamps. Note that $\varphi_j$ is technically ill defined at a set of times in this interval for templates that plunge before $t_0+T$, but we also have $A_j=0$ at those times. An approximation to the noise-weighted inner product $\langle\cdot|\cdot\rangle$ between two signals $h,h'$ is then defined as
\begin{equation}\label{eq:appinner}
(h|h'):=\mathrm{Re}\sum_{jj'}\sum_{t\in\tau_{jj'}}\Delta t\,\bar{A}_j(t)^*\bar{A}'_{j'}(t),
\end{equation}
\begin{equation}\label{eq:phasetol}
\tau_{jj'}:=\{t:|\varphi_j(t)-\varphi'_{j'}(t)|<1\},
\end{equation}
with $\tau_{jj'}$ being the set of times at which mode $j$ is ``in phase'' with mode $j'$. The phase tolerance value of 1 in Eq.\,\eqref{eq:phasetol} is chosen such that $(\cdot|\cdot)$ gives an upper envelope for $\langle\cdot|\cdot\rangle$ between two sinusoidal signals, as it varies with the difference in their frequencies (see Fig.\,\ref{fig:innprod}). More generally, we have
\begin{equation}\label{eq:envelope}
    (h|h')\gtrsim2\sum_{\chi=\mathrm{I,I\!I}}\sum_{t=t_0}^{t_0+T}\delta t\,\bar{h}_\chi(t)\bar{h}'_\chi(t)\approx\langle h|h'\rangle.
\end{equation}

Since Eq.\,\eqref{eq:appinner} is undefined for general time series in the data space $\mathcal{D}$, it is not strictly an inner product, but rather approximates the restriction of $\langle\cdot|\cdot\rangle$ to the signal space $\mathcal{S}$. For convenience, however, we will call $(\cdot|\cdot)$ the approximate ``inner product'', and $\langle\cdot|\cdot\rangle$ the full inner product. The approximate inner product is only used in the degeneracy study of Sec.\,\ref{sec:degeneracy}, and even there only in bulk calculations for which the full inner product would be unfeasible. Qualitative results obtained with $(\cdot|\cdot)$ are verified with $\langle\cdot|\cdot\rangle$ if practical; e.g., after a set of high-overlap points from sampling has been distilled to a smaller set of representative points by clustering, we report their overlaps with respect to $\langle\cdot|\cdot\rangle$. Thus we will in general refer to both $(\cdot|\cdot)$ and $\langle\cdot|\cdot\rangle$ simply as the inner product, and will only distinguish between them (and in derived quantities such as the overlap) if important qualitative differences arise due to the approximation.

\subsection{Sampling densities}
\label{subsec:sampling}

Of central importance in GW inference is the standard Whittle likelihood function $L(\boldsymbol{\theta}|x)\propto p(x|\boldsymbol{\theta}_\mathrm{inj})$ (with the distinction between the random argument $\boldsymbol{\theta}$ and the fixed quantity $\boldsymbol{\theta}_\mathrm{inj}$ being explicitly highlighted). As in Eq.\,\eqref{eq:fullinner}, we take the data as $x=h_\mathrm{inj}+n$ with zero-mean and stationary noise, but here $n$ is additionally assumed to be a Gaussian process \cite{Finn:1992wt}. The natural logarithm of $L$ (informally, the ``log-likelihood'') is given by
\begin{equation}\label{eq:fullloglike}
    \ln L_\mathrm{full}(\boldsymbol{\theta}|x):=-\frac{1}{2}\langle x-h(\boldsymbol{\theta})|x-h(\boldsymbol{\theta})\rangle.
\end{equation}
Functional form notwithstanding, Eq.\,\eqref{eq:fullloglike}
describes a non-Gaussian density function on the parameter space $\Theta$; however, from the definition of locality \eqref{eq:locality}, a Gaussian approximation to $L$ is reasonably valid in the local neighborhood of $\boldsymbol{\theta}_\mathrm{ML}$ from Eq.\,\eqref{eq:mle}. Since Eq.\,\eqref{eq:fullloglike} is defined in terms of the full inner product $\langle\cdot|\cdot\rangle$, we will refer to it in the context of this work as the full (log-)likelihood.

The full likelihood is not used in our studies beyond the initial validation of an approximate likelihood, which is simply given by Eq.\,\eqref{eq:fullloglike} with $\langle\cdot|\cdot\rangle\to(\cdot|\cdot)$:
\begin{equation}\label{eq:apploglike}
    \ln L_\mathrm{app}(\boldsymbol{\theta}|x):=-\frac{1}{2}(x-h(\boldsymbol{\theta})|x-h(\boldsymbol{\theta})).
\end{equation}
With the reduced cost of both the template-generation and inner-product operations, $L_\mathrm{app}$ is $\gtrsim10^3$ times faster to evaluate than $L_\mathrm{full}$. Efficiency considerations aside, Eq.\,\eqref{eq:apploglike} also facilitates the degeneracy study in Sec.\,\ref{sec:degeneracy} by smoothing out local oscillations in the posterior surface, such that the maxima identified by our algorithms are generally due to non-local correlations. By design, the profile of $L_\mathrm{app}$ over $\Theta$ corresponds roughly to the upper envelope of $L_\mathrm{full}$; this relation is not exact as depicted for the inner products in Fig.\,\ref{fig:innprod}, since the signals here have evolving frequency, and the log-likelihood contains normalization terms in addition to the inner product.

For conceptual ease, let us consider the specific noise realization $n=0$ without much loss of generality, such that $x=h_\mathrm{inj}$ lies in the signal space $\mathcal{S}$ with SNR $\rho_\mathrm{inj}:=\rho_\mathrm{opt}(h_\mathrm{inj})=\rho_\mathrm{det}(h_\mathrm{inj})$. Eq.\,\eqref{eq:apploglike} may be written as
\begin{equation}\label{eq:apploglike2}
    \ln L_\mathrm{app}(\boldsymbol{\theta}|x)=\rho_\mathrm{inj}\,\rho_{\boldsymbol{\theta}}\,\Omega(x,h(\boldsymbol{\theta}))-\frac{1}{2}\left(\rho_\mathrm{inj}^2+\rho_{\boldsymbol{\theta}}^2\right),
\end{equation}
where $\rho_{\boldsymbol{\theta}}:=\rho_\mathrm{opt}(h(\boldsymbol{\theta}))$. In the tail regions of the likelihood, where $\mathrm{E}[\Omega(x,h)]\approx0$, we see that $L_\mathrm{app}\appropto\exp{(-1/2\,\rho_{\boldsymbol{\theta}}^2)}$ is dominated by large-scale gradients arising from the slow variation of $\rho_{\boldsymbol{\theta}}$ over $\Theta$ (relative to $(x|h(\boldsymbol{\theta})$). As these gradients can hinder a global analysis of intrinsic parameter degeneracy, it is useful to introduce a \emph{matched-SNR} version of $L_\mathrm{app}$, with templates that have been renormalized to match the injection SNR:
\begin{equation}\label{eq:matchloglike}
    \ln L_\mathrm{match}(\boldsymbol{\theta}|x):=\rho_\mathrm{inj}^2(\Omega(x,h(\boldsymbol{\theta}))-1).
\end{equation}
This matched-SNR likelihood is proportional to the overlap surface $\Omega(h_\mathrm{inj},\cdot)$ over $\Theta$; in the vicinity of $\boldsymbol{\theta}_\mathrm{inj}$, we also have $L_\mathrm{match}\approx L_\mathrm{app}$. Our degeneracy study leverages the absence of large-scale SNR gradients in Eq.\,\eqref{eq:matchloglike} to search more effectively for secondary peaks in $L_\mathrm{match}$ (and the overlap surface), which then have the same locations as secondaries in $L_\mathrm{app}$ (and the posterior surface). This is because the intra-secondary variation of $\rho_{\boldsymbol{\theta}}$ is generally small, as opposed to its inter-secondary variation.

As it turns out, straightforward sampling of the matched-SNR likelihood is unsuitable for finding, resolving and visualizing secondaries in $L_\mathrm{match}$; even if these secondaries were cleanly localized and separated (they are not), they would be exponentially suppressed at realistic values of $\rho_\mathrm{inj}\gtrsim10$ and contiguously congealed at artificially lowered values---or high temperatures, in the language of annealing Markov-chain Monte Carlo (MCMC) methods \cite{Kirkpatrick671,2005PThPS.157..317W}. Our solution is to introduce an \emph{exploratory} ``likelihood'' that is used only for global exploration, and to obtain a large set of high-overlap points for follow-up clustering analysis. This is given by $L_\mathrm{match}$ with both injection and template normalized to unit SNR, plus artificially suppressed tails:
\begin{equation}\label{eq:explloglike}
    \ln L_\mathrm{expl}(\boldsymbol{\theta}|x):=\frac{1}{\rho_\mathrm{inj}^2}\ln L_\mathrm{match}(\boldsymbol{\theta}|x)+\mathrm{ramp}(\boldsymbol{\theta}|x),
\end{equation}
\begin{equation}\label{eq:ramp}
    \mathrm{ramp}(\boldsymbol{\theta}|x):=100\min{\{\Omega(x,h(\boldsymbol{\theta}))-0.5,0\}}.
\end{equation}
Eq.\,\eqref{eq:explloglike} is designed to reduce the volume in the tail regions of the matched-(unit-)SNR likelihood, so as to better locate and flesh out individual secondaries. The exact specifications of the ramp function \eqref{eq:ramp} are empirically chosen, and not necessarily optimal. While we will continue referring to Eq.\,\eqref{eq:explloglike} as a (Bayesian) likelihood, it can no longer be linked to a natural probabilistic statement based on noise and signal assumptions. This last point also holds true for a \emph{veto} ``likelihood'' that we propose as a mitigation of EMRI degeneracy, but whose presentation we delay until Sec.\,\ref{sec:suggestions} (as it is a consequence, rather than component, of the degeneracy study).

\subsection{Clustering algorithm}
\label{subsec:clustering}

With the exploration aspect of signal-space mapping covered by the stochastic sampling of the exploratory likelihood in Sec.\,\ref{subsec:sampling}, the remaining analysis boils down to visualizing and making sense of the resultant high-dimensional data. Central to this is the task of clustering. Some of the commonly used sampling algorithms in GW data analysis \cite{2013PASP..125..306F,2015MNRAS.453.4384H,2020MNRAS.493.3132S} employ in-built clustering to aid convergence on multi-modal distributions; this form of clustering is more of an intermediate means to an end, with the byproduct set of identified ``clusters'' being of limited utility for realistic examples. A specialized clustering algorithm is required here, but the main off-the-shelf options are unsuitable as well. \emph{Centroid-based} methods such as $k$-means clustering \cite{kaufman2009finding} and its many variants use geometric proximity to define cluster membership. They typically require the number of clusters to be pre-specified, and are thus immediately inadequate for our purposes. \emph{Density-based} methods \cite{kriegel2011density} perform clustering based on the density of points in the data set. In this context, they would rely on the sampler being able to populate secondaries in the correct proportion---which is in general non-trivial to guarantee even for ``clean'' multi-modal distributions. Such methods also tend to rely on arbitrarily tuned criteria for cluster membership, which can significantly affect the cluster count itself.

We introduce here a bespoke clustering algorithm that can be classified as a \emph{connectivity-based} method \cite{kaufman2009finding}. For a general signal space, the overlap between two signals provides a measure of connectivity between their associated parameter points. It can thus be used directly in clustering applications; for example, to group posterior samples in multi-source transdimensional-MCMC searches for Galactic-binary signals in LISA data \cite{Littenberg:2020bxy}, by assigning sample points to the same cluster if there is a high overlap between their associated templates. This works well for resolving sources in the absence of parameter degeneracy, because it agrees with the natural measure of connectivity defined by the metric distance \eqref{eq:locality} in $\Theta$ (which is required for sensible results). When non-local correlations are present, this sort of clustering might instead be used as a veto for spurious source candidates that arise due to secondaries (see Sec.\,\ref{subsec:vetoes}).

In the context of our degeneracy study, however, an alternative notion of connectivity is required. As discussed in Sec.\,\ref{subsec:sampling}, the key quantity in the understanding of posterior secondaries is the overlap function $\Omega(h_\mathrm{inj},\cdot)$. We are not restricted to the overlap information in the raw data set that is obtained by sampling $L_\mathrm{expl}$, since our access to a generative model for the overlap can be used to provide additional information during clustering. Let us first define the \emph{connection} $\ell$ between two points $\boldsymbol{\theta}_{1,2}\in\Theta$: a vector of $l$ equally spaced evaluations of $\Omega(h_\mathrm{inj},\cdot)$ along the connecting line between $\boldsymbol{\theta}_{1,2}$ (inclusive). We set $l=7$ in our algorithm, which is empirically determined to be the minimum value required for satisfactory convergence of results. The connection $\ell$ is then used to construct a symmetric \emph{pre-metric} $d$ on parameter space, satisfying only $d(\boldsymbol{\theta}_1,\boldsymbol{\theta}_2)\geq0$, $d(\boldsymbol{\theta}_1,\boldsymbol{\theta}_1)=0$ and $d(\boldsymbol{\theta}_1,\boldsymbol{\theta}_2)=d(\boldsymbol{\theta}_2,\boldsymbol{\theta}_1)$ \cite{arkhangel2012general}. This pre-metric is defined as
\begin{equation}\label{eq:premetric}
    d(\boldsymbol{\theta}_1,\boldsymbol{\theta}_2):=1-\frac{\min_i{\ell_i(\boldsymbol{\theta}_1,\boldsymbol{\theta}_2)}}{\min\{\ell_1(\boldsymbol{\theta}_1,\boldsymbol{\theta}_2),\ell_l(\boldsymbol{\theta}_1,\boldsymbol{\theta}_2)\}},
\end{equation}
such that two points with ``pre-distance'' zero have no intermediate point with a lower overlap value than either.

With the existence of a pre-metric, we may define various degrees of ``connectedness'' for use in the clustering algorithm. We will say that two points are \emph{strictly connected} if they have a pre-distance $d=0$, that they are \emph{connected} if $d<0.5$, and that they are \emph{not connected} if $d\geq0.5$. The additional category of being connected, rather than only strictly connected or not strictly connected, is required in our algorithm to handle intra-secondary structure. In other words, we are defining secondaries as disjoint clusters of points that are uniquely connected to some representative point (generally a cluster maximum), instead of simply calling all local maxima secondaries (because these can and often do coexist in contiguous regions of high overlap). Since we will only be clustering high-overlap points with $\Omega(h_\mathrm{inj},\cdot)\gtrsim0.5$, any two connected points will have no intermediate point with $\Omega(h_\mathrm{inj},\cdot)\lesssim0.25$, which admits an interpretation of connectedness in terms of absolute overlap. The pre-distance between two points also depends more strongly on their overlap values when $d\geq0.5$, such that disconnected points with higher overlaps tend to be ``further apart''. This allows connectedness to be used in locating cluster maxima, in addition to the more obvious function of determining cluster membership.

Using the above definitions of connectedness as a notion of connectivity leads to a clustering problem with a significant degree of noise. In graph-theory terms, this is because the graph for a set of parameter points (vertices) and their pairwise connectedness (edges) has high vertex connectivity \cite{diestel2000graph}, and cannot be partitioned into disjoint subgraphs without removing a large number of vertices. Another issue is that standard connectivity-based algorithms are generally computationally expensive, with their complexity scaling as $\mathcal{O}(N^2)$ or worse for a data set of size $N$---for example, if one were to compute the full similarity matrix of pairwise pre-distances for use in spectral-clustering methods \cite{von2007tutorial}. To combat these difficulties, we take a greedy approach \cite{cormen2001introduction} to the identification of candidate \emph{nodes} (representative points for clusters), with global and local steps to ensure that these nodes pick out any clear maxima without the need for full-blown optimization. Another design objective is for the algorithm to run in $\mathcal{O}(N)$ rather than $\mathcal{O}(N^2)$ time. A single iteration of our clustering algorithm is described below in pseudocode:

\begin{enumerate}
    \item At the $i$-th iteration, there exists an $(i-1)\times N$ matrix containing the pre-distance of all points in the data set to each of the existing $i-1$ nodes. Each point has a minimal pre-distance to the set of existing nodes. (For the set of nodes themselves and the points that are strictly connected to them, this minimal pre-distance will be zero.) Choose the point with the largest minimal pre-distance as the preliminary $i$-th node. This is the point that is ``least connected'' to the existing nodes, and the process of finding it is akin to a global-search step.
    \item Compute the pre-distance of all points to the preliminary $i$-th node.
    \item Define the strict $i$-th cluster as the set of points that are strictly connected to the preliminary node, and not connected to any of the existing $i-1$ nodes. Choose the point in the strict $i$-th cluster with the highest overlap value as the actual $i$-th node. This process is akin to a local-maximization step.
    \item If the preliminary $i$-th node is not the same point as the actual $i$-th node, ``re-center'' the cluster by computing the pre-distance of all points to the actual $i$-th node. In principle, the actual node might still not be the highest-overlap point in its cluster after re-centering; however, additional re-centering steps are found to offer marginal gain for their cost.
    \item Append the list of pre-distances for the $i$-th node to the pre-distance matrix, as a new row.
    \item Compute the cluster coverage, which is the fraction of points that are connected to at least one of the existing $i$ nodes. If the coverage equals unity, end the algorithm. Defining the stopping criterion in terms of being connected rather than strictly connected causes the cluster count to depend on the arbitrary connectedness threshold $d=0.5$, but this just corresponds to setting a limit on what is defined as a clear cluster---which is never fully avoidable in any non-textbook clustering task.
\end{enumerate}

\begin{figure}[!tbp]
\centering
\includegraphics[width=\columnwidth]{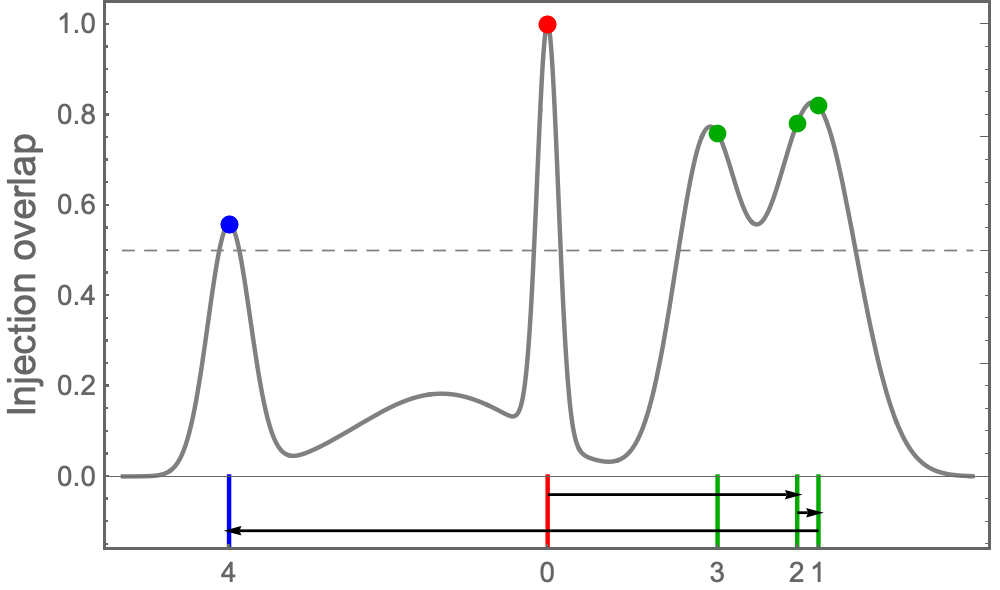}
\caption{Application of the clustering algorithm to an artificial one-dimensional overlap surface, and a data set of $N=5$ high-overlap points (with zero-based index). At the first iteration, Point 0 (the injection) is the identified node, and the cluster coverage is $0.2$. At the second iteration, Point 2 ($\Omega=0.78$) is the preliminary node with largest minimal pre-distance $d=0.95$; note that this is neither the point with the next highest overlap, nor the point with the largest metric distance. Point 1 ($\Omega=0.82$) is the actual node after re-centering, and the cluster coverage is $0.8$ (Points 2 and 3 are both connected to Point 1). At the third iteration, Point 4 is the identified node with largest minimal pre-distance $d=0.88$ (to Point 0), and the algorithm terminates with a cluster coverage of $1$. Color grouping of points indicates the final clusters for each node.}
\label{fig:clustering}
\end{figure}

The only special iteration is the initial one, where both the preliminary and actual nodes in Steps 1 and 3 are simply chosen to be the point in the data set with the highest overlap value $\Omega(h_\mathrm{inj},\cdot)$. Computing new rows of the pre-distance matrix in Steps 2 and 4 dominates the algorithm cost, which thus scales as $\mathcal{O}(N)$ times the number of identified nodes. With the final pre-distance matrix in hand, it is straightforward to construct a set of clusters post hoc. The final (strict) cluster for each node is defined as the set of points that are (strictly) connected to the node, and not (strictly) connected to any of the other nodes. Note that the strict cluster here differs slightly from that defined in Step 3, where the shortlist of candidates for the actual node is more stringent by design. For illustrative purposes, we provide in Fig.\,\ref{fig:clustering} a heuristic example that depicts the application of the algorithm to an artificial one-dimensional overlap surface.

Our clustering algorithm uses information about $\Omega(h_\mathrm{inj},\cdot)$ beyond the original data set, and thus its performance does not rely strongly on the distribution of the data. The number and composition of clusters are both robust to the choices of connection length $l\gtrsim7$ and initial node (provided it is close to a local maximum), with variations in the results that are far smaller than errors due to imperfect sampling. Non-trivial structure in the high-dimensional overlap surface can be unearthed even from relatively small data sets with $N\lesssim10^4$, as demonstrated particularly clearly in Sec.\,\ref{subsec:secondaryB}. Although the algorithm was developed for our specific purposes, it essentially only requires the surface under examination to be evaluable separately from the data, and thus might find utility in more general clustering applications. In the degeneracy study of Sec.\,\ref{sec:degeneracy}, we employ it both for large-scale mapping to identify disjoint secondary regions of high overlap, and for localized mapping in these regions to resolve and visualize the finer structure.

\section{Confusion study}
\label{sec:confusion}

Before turning our attention to correlations in the continuous space of possible signals, we seek to address a simpler (two-part) question: For an astrophysically realistic set of EMRI signals in LISA data, i) how correlated are the most correlated pair of signals; and ii) how does the overall degree of correlation among the set vary with the size of the set? The answer to this bears on both EMRI search and inference; large overlaps between actual signals will hinder their resolution during search, and will necessitate a combined inference of their source parameters. This in turn affects the design of data-analysis strategies, as well as the prospects of EMRI science. The conclusion of our study is unequivocal, at least for a middle-ground estimate of 200 detectable signals---self-confusion is unlikely to be an issue for EMRIs.

\subsection{Astrophysical model}
\label{subsec:astro}

Models for the astrophysical population of EMRIs that LISA will observe are weakly constrained by existing knowledge, and thus span a broad range of predictions. Some of the main uncertainties lie in the massive-black-hole mass function, in the fraction of such black holes hosted in dense stellar cusps, as well as in the intrinsic rate of EMRI formation per massive black hole \cite{Babak:2017tow}. In this work, we consider a population of detectable sources from a single representative model, since the distribution of these sources is dominated by detector-specific selection effects anyway. Furthermore, our intuition a priori is that the results of the confusion study will be virtually independent of model choice, as they are simply determined by the intrinsic nature of signal space---specifically, the large information volume expected to be retained by any astrophysically restricted subspace.

We now define a population model that corresponds approximately to M1 from \cite{Babak:2017tow}, which is often used as a representative model in LISA Science Group work-package studies (e.g., \cite{Seoane:2021kkk}). Our model specifies independent distributions for each of the AAK parameters \eqref{eq:intrinsic} and \eqref{eq:extrinsic}, but with the source-frame masses $(\mu',M')$ and redshift $z$ in place of the detector-frame masses $(\mu,M)$ and luminosity distance $D$. The map $(\mu',M',z)\to(\mu,M,D)$ is given by the usual relations
\begin{equation}\label{eq:redmass}
    \mu(\mu',z)=(1+z)\mu',\quad M(M',z)=(1+z)M',
\end{equation}
as well as \cite{peebles1993principles}
\begin{equation}\label{eq:lumdist}
    D(z)=(1+z)\int_0^zdz'\,\frac{c}{H(z')},
\end{equation}
\begin{equation}\label{eq:hubble}
    H(z)=H_0\sqrt{(1-\Omega_\Lambda)(1+z)^3+\Omega_\Lambda},
\end{equation}
with $c/H_0=4280\,\mathrm{Mpc}$ and $\Omega_\Lambda=0.7$.

\begin{table}[]
\begin{tabular}{|l|l|l|}
\hline
$\boldsymbol{\theta}_i$ \hspace{1cm} & $\Delta\boldsymbol{\theta}_i$ \hspace{2cm} & Distribution \hspace{0.5cm} \\ \hline\hline
$\mu'$ & $\{10\}$ & Fixed \\ \hline
$M'$ & $[3,30]\times10^5$ & Eq.\,\eqref{eq:massfunction} \\ \hline
$a/M$ & $\{0.99\}$ & Fixed \\ \hline
$p_0/M$ & $\{11\}$ & Fixed \\ \hline
$e_0$ & $(0,0.2]$ & Uniform \\ \hline
$\cos{\iota}$ & $(-1,0)\cup(0,1)$ & Uniform \\ \hline
$\Phi_0$ & $[-\pi,\pi]$ & Uniform \\ \hline
$\gamma_0$ & $[-\pi,\pi]$ & Uniform \\ \hline
$\alpha_0$ & $[-\pi,\pi]$ & Uniform \\ \hline
$\cos{\theta_K}$ & $(-1,1)$ & Uniform \\ \hline
$\phi_K$ & $[-\pi,\pi]$ & Uniform \\ \hline
$\cos{\theta_S}$ & $(-1,1)$ & Uniform \\ \hline
$\phi_S$ & $[-\pi,\pi]$ & Uniform \\ \hline
$z$ & $(0,4.5]$ & Eq.\,\eqref{eq:redshift} \\ \hline
\end{tabular}
\caption{Source-parameter extents and distributions for the simple astrophysical-population model in Sec.\,\ref{subsec:astro}.}
\label{tab:astroprior}
\end{table}

The probability density of the joint distribution for $\boldsymbol{\theta}$ is the product of the individual parameter densities $p(\boldsymbol{\theta}_i)$, which are supported on the corresponding sets $\Delta\boldsymbol{\theta}_i\subset\mathbb{R}$ (to consolidate notation, $\Delta\boldsymbol{\theta}_i$ can have measure zero). These sets are listed in Tab.\,\ref{tab:astroprior} for each source parameter; note the reparametrization of all polar angles to their cosine values. With the exception of $M'$ and $z$, all parameters are either fixed to or distributed uniformly on  $\Delta\boldsymbol{\theta}_i$. Although the initial semi-latus rectum is chosen as $p_0=11M$ for all sources (to accommodate retrograde inspirals), each is assumed to be observed for $T=2\,\mathrm{y}$, i.e., their evolution is extended backward in time from the reference time $t_0=0$ if they plunge before $t_0+T$. Near-spherical inspirals with $e_0\approx0$ are excluded from the analysis for technical reasons, as well as near-equatorial ($\cos{\iota}\approx\pm1$) or near-polar ($\cos{\iota}\approx0$) ones.

Following M1 from \cite{Babak:2017tow}, the density for $M'$ is almost independent of redshift, and thus taken to be \cite{Barausse:2012fy}
\begin{equation}\label{eq:massfunction}
    p(M')\propto(M')^{-1.3}\,\boldsymbol{1}_{\Delta M'}(M'),
\end{equation}
where $\boldsymbol{1}_{\Delta\boldsymbol{\theta}_i}$ denotes the indicator function of $\Delta\boldsymbol{\theta}_i$. Finally, for the redshift itself, we have \cite{Cutler:2005qq}
\begin{equation}\label{eq:redshift}
    p(z)\propto\frac{D(z)^2\dot{\sigma}(z)}{(1+z)^3H(z)}\,\boldsymbol{1}_{\Delta z}(z),
\end{equation}
where $\dot{\sigma}(z)$ is the EMRI event rate per unit proper time, per unit co-moving volume. The variation of $\dot{\sigma}$ is neglected in our model since it is highly uncertain to begin with, and is also expected to be nearly constant at low redshifts of $z\lesssim2$ (where most detectable events occur).

\subsection{Monte Carlo analysis}
\label{subsec:montecarlo}

\begin{figure}[!tbp]
\centering
\includegraphics[width=\columnwidth]{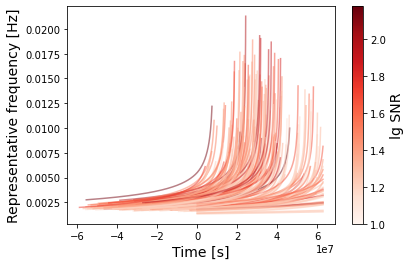}
\caption{Time evolution of representative frequency $\omega_\phi/\pi$ over two years, for 200 detectable sources drawn from the population model in Sec.\,\ref{subsec:astro} with an SNR threshold of 20.}
\label{fig:signals}
\end{figure}

Let us consider a set of $N$ detectable EMRIs, obtained in practice by applying a minimum-SNR cutoff to a sufficiently large set of $\geq N$ sources that is distributed according to the population model in Sec.\,\ref{subsec:astro}. Each simulated source is described by a random draw from the joint distribution in Tab.\,\ref{tab:astroprior}; its associated optimal SNR is computed using the AAK--TDI waveform model \cite{Chua:2019wgs} from Sec.\,\ref{subsec:models}, and the full inner product \eqref{eq:fullinner} with a noise model corresponding to the LISA science requirements. We adopt the standard SNR-threshold value of 20 \cite{Chua:2017ujo}, which results in a detection efficiency of 0.06. A set of $N=200$ detectable sources is simulated for this study, as a middle ground between estimates of $\sim1$ and $\sim10^4$ such sources from a range of astrophysical models \cite{Babak:2017tow,Chua:2017ujo}. Their signals are visualized as a population in Fig.\,\ref{fig:signals}, where only the time evolution of the representative (dominant) frequency $\omega_\phi/\pi$ is plotted for each signal.

\begin{figure}[!tbp]
\centering
\includegraphics[width=\columnwidth]{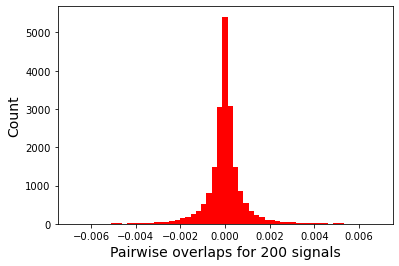}
\caption{Histogram of the 19900 pairwise-overlap values for the 200 detectable signals depicted in Fig.\,\ref{fig:signals}.}
\label{fig:pairovers}
\end{figure}

For this (and any) set of $N$ independent and identically distributed EMRIs, we may define a vector-valued statistic $v_N$ of pairwise overlaps, i.e., $v_N$ is a vector comprising $N_p:=N(N-1)/2)$ components, each taking a value within the interval $[-1,1]$. Note the important conceptual distinction between the components of $v_N$, and the sequence of overlaps for $N_p$ pairs of i.i.d.\! sources; the latter is itself a set of i.i.d.\! random variables, but does not relate to the question at hand (except in the trivial case $N=2$). The sampling distribution of the statistic $v_N$ is intractable---analytically and even numerically for modest $N$. Nevertheless, a single realization of $v_N$ still provides some insight into the matter of EMRI self-confusion. Fig.\,\ref{fig:pairovers} shows a histogram of the $N_p=19900$ pairwise-overlap values for our set of 200 sources. (We reiterate that this histogram cannot be construed as a sampling distribution, since the components of $v_N$ are not i.i.d.) As might be expected, the set of overlaps is tightly and symmetrically centered about zero, with a ``standard deviation'' of 0.001 and a maximum absolute value of 0.012. This addresses the first part of our question: in a set of $\approx200$ detectable sources, we expect no two of their signals to have a $\gtrsim1\%$ resemblance.

The above root-mean-square overlap value of $0.001$ may be explained by the following order-of-magnitude estimate. First, idealize each of the EMRI signals visualized in Fig.\,\ref{fig:signals} as comprising only its depicted dominant harmonic mode $(2,0,0)$, with time-evolving frequency $f:=\omega_\phi/\pi$. Pairs of signals whose frequency trajectories do not cross have an approximately vanishing overlap, while the overlap value for those that do cross is largely determined by the length of time that both signals are concurrently oscillating at similar frequencies, i.e., some interval centered on the instant of crossing. This duration is given by $\tau\approx|\Delta\dot{f}|^{-1/2}$, where $\Delta\dot{f}$ is the difference in the frequency derivatives of the two signals at the crossing time. At leading order in a PN expansion, $\dot{f}$ for a binary-inspiral signal is $(3/8)(f/T_m)$, where $T_m$ ($\lesssim T=2\,\mathrm{y}$) is the time to merger from frequency $f$. For a typical (crossing) pair of signals in Fig.\,\ref{fig:signals}, $\Delta\dot{f}$ then falls within an order of magnitude of $(3/8)(f_c/T)$ (where $f_c$ is the crossing frequency), so let us simply say that $|\Delta\dot{f}|\approx f_c/T$. Finally, the overlap is approximately $\tau/T\approx(f_cT)^{-1/2}$. Since $f_c$ falls between $10^{-3}\,\mathrm{Hz}$ and $10^{-2}\,\mathrm{Hz}$ for the vast majority of signal pairs in Fig.\,\ref{fig:signals}, we recover an estimated overlap of $\sim10^{-3}$ as expected.

For the second part of our question, it is useful to form the matrix $M_N$ of pairwise overlaps (including self-overlaps) for a set of $N$ sources. This matrix has unit diagonal elements, and off-diagonal elements corresponding to the components of $v_N$. Thus it is effectively a correlation matrix for the set of sources (viewed as random variables), and various standard matrix norms can provide different summary notions of the overall correlations among the set; e.g., the Frobenius norm $||\cdot||_F$ (the vector norm of the vectorized matrix), or the spectral norm $||\cdot||_S$ (the largest eigenvalue of the matrix). We use such norms to examine the behavior of $M_N$ as sources are incrementally added to a starting population, which we take to be a subset of 30 detectable sources from our full set of 200. More precisely, we consider nested subsets of our full set such that $M_N$ is a submatrix of $M_{N+1}$ for $N$ ranging from 30 to 199.

\begin{figure}[!tbp]
\centering
\includegraphics[width=\columnwidth]{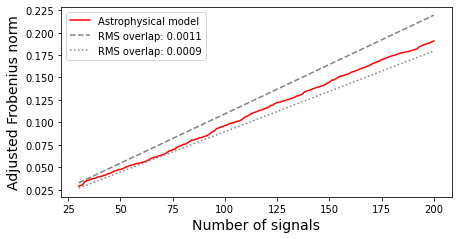}
\includegraphics[width=\columnwidth]{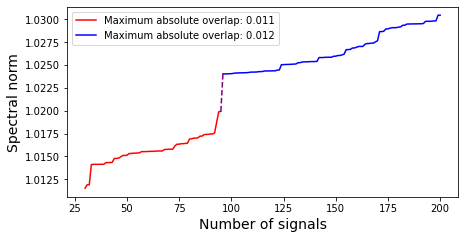}
\caption{Matrix norms for the pairwise overlaps among $N$ of the 200 detectable signals in Fig.\,\ref{fig:signals}, for $30\leq N\leq200$. Top: Adjusted Frobenius norm, indicating that the root-mean-square overlap is approximately constant at $\approx0.001$. Bottom: Spectral norm, illustrating how the overall correlation of the set is driven by individual signals. The largest jump coincides with the addition of a signal that raises the maximum absolute overlap among the set (dashed purple line).}
\label{fig:norms}
\end{figure}

An adjusted Frobenius norm $(||M_N||_F^2-N)^{1/2}$ (essentially just $|v_N|$) is plotted in the top panel of Fig.\,\ref{fig:norms}, as a function of $N$ from 30 to 200. We bound the observed sequence of norms by two other sequences for hypothetical $M_N$, where the root mean squares of the off-diagonal terms are held constant at $\{0.9,1.1\}\times10^{-3}$. A linear trend is evident, which indicates that the root mean square of pairwise overlaps remains approximately constant at $\approx0.001$ as the number of sources is raised from 30 to 200. As $N\to\infty$, linear scaling will not hold as the set of signals saturates the (astrophysically restricted) signal space. This can be made more intuitive by examining the spectral norm of $M_N$ (bottom panel of Fig.\,\ref{fig:norms}), which would be unity for a completely uncorrelated set of sources. A large jump in $||M_N||_S$ occurs when a signal with high overlap against any of the previous ones is added to the set. This occurrence naturally becomes more frequent as $N\to\infty$, and is what drives the eventual super-linear scaling of overall correlation against $N$. The value of $N$ at which super-linear scaling becomes an issue (say, leading to a root-mean-square overlap of $\gtrsim0.1$) remains undetermined due to computational constraints, although we conjecture that such a scenario will not arise for astrophysically relevant values of $N\lesssim10^4$.

\subsection{Discussion}
\label{subsec:confusiondiscussion}

As context, we now briefly describe how the present work relates to some previous studies on EMRI confusion noise. The spectral density of the effective noise from unresolvable EMRI signals in LISA data was first estimated in \cite{Barack:2004wc}, in order to quantify its impact on LISA's ability to detect sources of other types---principally, Galactic binaries and massive-black-hole mergers. Since EMRI event rates as functions of the source parameters were (and remain) poorly understood, the analysis relied on assumptions based on the theoretical models available at the time; the resultant noise curve was then multiplied by a range of overall rate factors (for different classes of compact objects from white dwarfs to black holes), to give approximate bounds on the potential impact of an EMRI confusion background. The main conclusion of \cite{Barack:2004wc} is that unresolvable EMRIs will likely be a significant contributor to LISA's noise budget, and for the highest event rates, might actually dominate the total noise at frequencies where detector noise is minimal.

More recently, an updated analysis of EMRI confusion noise for LISA was performed with several key differences \cite{Bonetti:2020jku}: i) the current LISA baseline noise curve is used, with a trough that is located at higher frequencies; ii) the assumed scaling of event rates with the central mass is based on newer numerical simulations; and iii) the EMRI confusion noise curve is computed for a range of astrophysical population models in the literature. While the updated noise curves differ in shape from that found in \cite{Barack:2004wc}, the study arrives at the same basic conclusion: EMRI confusion noise will be significant for LISA, and might be dominant at its most sensitive frequencies if EMRI event rates are on the high end of estimates. In both \cite{Barack:2004wc} and \cite{Bonetti:2020jku}, the stochastic process $n_\mathrm{EMRI}$ that describes the sum of unresolvable signals from an astrophysical EMRI background is treated as approximately Gaussian. This assumption of near-Gaussianity is largely justified by the study in \cite{Racine:2007gv}; there, the Edgeworth expansion and large-deviations theory are used to show that the distribution of the cross-correlation between $n_\mathrm{EMRI}$ and a Galactic-binary or massive-black-hole-merger signal template is indeed very close to Gaussian.

The above papers all deal with sums of $N$ unresolvable EMRI signals, and their impact on searches for resolvable signals. In our confusion study, we are instead concerned with the distinguishability of a set of $N$ resolvable signals, which can be described summarily by the root mean square of their pairwise overlaps. These two concepts are of course related, but quite distinct. To draw a connection between them, consider the mean square of the unnormalized cross-correlations (rather than overlaps) between some normalized template, and $N$ unresolvable signals with some astrophysical distribution of sub-threshold SNRs. This quantity times $N$ is then approximately the variance in the detection SNR of that template due to the associated EMRI confusion noise, and thus will be significant if it is $\gtrsim1$. We do not include such an analysis in this work, although sums of $N$ resolvable signals are explicitly studied in Sec.\,\ref{subsec:multiple}. There, however, the focus is on how often such sums have a higher cross-correlation with some completely separate template than with any individual template in the sum.

\section{Degeneracy study}
\label{sec:degeneracy}

In this study, we examine a single representative signal injection with intrinsic source parameters given by Eq.\,\eqref{eq:injpars}, and delve deeply into the tail structure of its associated posterior distribution. The bulk properties of secondaries in the posterior will no doubt vary for different injections across the parameter space. Instead of attempting to characterize the entire statistical manifold of posterior distributions (which will provide conclusions that are model-specific to a greater extent), we focus here on using our sole posterior to build up general strategies for EMRI posterior mapping. This is largely uncharted territory; for example, even visualizing the local posterior around secondaries turns out to be quite counterintuitive (see Secs \ref{subsec:secondaryA} and \ref{subsec:secondaryB}). Also, our results indicate that our original aim of localizing and counting secondaries (in order to estimate their coverage of parameter space) is not particularly well posed. Nevertheless, we expect that the qualitative statements we are able to make from this study will in general be model-independent, and thus representative of the EMRI signal space. Various arguments to that end are put forth in Sec.\,\ref{subsec:corollaries}.

\subsection{Mapping analysis}
\label{subsec:mapping}

We restrict our analysis to the six intrinsic parameters
\begin{equation}\label{eq:analysispars}
    \boldsymbol{\theta}=(\lg{\mu},\lg{M},a/M,p_0/M,e_0,\cos{\iota})
\end{equation}
and the associated six-dimensional subspaces of $\Theta$ and $\mathcal{S}$, for fixed values of the remaining parameters. (Note the reparametrization $(\mu,M,\iota)\to(\lg{\mu},\lg{M},\cos{\iota})$.) This is motivated partly by computational constraints, and partly by our initial focus on the degeneracy in frequency evolution---which is most strongly determined by these six parameters. We have verified from a sampling of the full-dimensional (approximate) posterior that local covariances between the sets of intrinsic and extrinsic EMRI parameters are low, at least in the case of the simple long-wavelength response used here. This is also observed in a more recent but unrelated EMRI study \cite{Speri:2022upm}. In other words, the conditional posterior for the intrinsic parameters (given fixed extrinsic parameters) is locally not too dissimilar from the marginal posterior. Non-locally, additional degrees of degeneracy might arise due to the interplay of intrinsic and extrinsic parameters; this possibility is discussed briefly in Sec.\,\ref{subsec:cause}.

The three phase angles $(\Phi_0,\gamma_0,\alpha_0)$ are also excluded from consideration in our mapping analysis. With only the four strong harmonic modes from Eq.\,\eqref{eq:strongmodes}, the stripped-down AAK model loses all sidebands due to Lense--Thirring precession, and thus its ability to constrain $\alpha_0$. For simplicity, we set the other two phase angles to zero in the injection and all templates, such that the phasing in all signals lines up at time $t_0=0$ (the initial time in the degeneracy study). This alignment essentially forces any crossing of mode phasing between two signals to occur at the earliest time, which generally maximizes their overlap over $(\Phi_0,\gamma_0)$---even if the signals are non-degenerate, since the frequencies for the four modes \eqref{eq:strongmodes} take on their minimal values at that time. An inclusion of $(\Phi_0,\gamma_0)$ in the analysis would likely reveal a small degree of additional degeneracy in those parameter directions, but this would not significantly alter the intrinsic degeneracy observed here (as initial phasing does not affect frequency evolution in this simple model).

We explore the signal space over a variety of hyperrectangular regions in (the six-dimensional subspace of) $\Theta$, centered on $\boldsymbol{\theta}_\mathrm{inj}$ from Eq.\,\eqref{eq:injpars}. To be explicit here, the injection parameters in the form of Eq.\,\eqref{eq:analysispars} are
\begin{equation}\label{eq:analysisinjpars}
    \boldsymbol{\theta}_\mathrm{inj}=(1,6,0.5,9.5,0.2,0.866),
\end{equation}
with the signal renormalized to an optimal SNR of 20. Results are presented for three regions in particular: a starting region around the primary posterior peak, another with 10 times its extent in each parameter direction, and a third with $\sim100$ times its extent. In the latter two regions, we obtain a set of representative locations for secondary peaks in the overlap surface; from the line of reasoning given in Sec.\,\ref{subsec:sampling}, these correspond almost exactly to secondaries in the actual posterior for the full likelihood (with a flat or diffuse prior). Two high-overlap secondaries are further singled out as case studies, where we directly map out the (approximate) posterior in their vicinity, instead of the overlap surface. A summary plot that depicts the relative scales and locations for all of these analyses is provided at the end of Sec.\,\ref{subsec:mapping}.

Throughout the degeneracy study, we employ a combination of different implementations for the two main classes of stochastic sampling algorithms used in GW data analysis: nested sampling (as implemented in \texttt{PolyChord} \cite{2015MNRAS.453.4384H} and \texttt{Dynesty} \cite{2020MNRAS.493.3132S}), and parallel-tempering MCMC (as implemented in \texttt{PTMCMCSampler} \cite{justin_ellis_2017_1037579}). The primary motivation for this redundancy is to provide a cross-algorithm and cross-implementation validation of our sampling results, where feasible. For the posterior sampling (using $L_\mathrm{app}$) in Secs \ref{subsec:1x}, \ref{subsec:secondaryA} and \ref{subsec:secondaryB}, we run the samplers to the point of convergence---typically determined by built-in stopping criteria, but also easily verified through cross-sampler checks. Convergence to the target distribution is less relevant for the exploratory searches in Secs \ref{subsec:10x} and \ref{subsec:100x}. There we simply seek to obtain $\sim10^5$--$10^6$ independent samples from $L_\mathrm{expl}$, a large fraction of which correspond to signals with $\Omega(h_\mathrm{inj},\cdot)>0.5$ (since that is the threshold used in Eq.\,\eqref{eq:ramp}). For the most extensive searches in Sec.\,\ref{subsec:100x}, this can require $\gtrsim10^9$ ``likelihood'' evaluations.

\subsubsection{Starting region: The posterior bulk}
\label{subsec:1x}

Our starting region $\mathcal{R}_0$ is chosen to encompass the Gaussian-analogous 2-$\sigma$ contours of the primary posterior peak, with some room. More precisely, we consider a ``prior'' with probability density proportional to the indicator function on the Cartesian product of intervals
\begin{equation}\label{eq:r0}
    \mathcal{R}_0:=\prod\limits_i\,[\boldsymbol{\theta}_{\mathrm{inj},i}-\delta\boldsymbol{\theta}_i/2,\boldsymbol{\theta}_{\mathrm{inj},i}+\delta\boldsymbol{\theta}_i/2],
\end{equation}
where the half-extents $\delta\boldsymbol{\theta}/2$ are 2--3 times larger than the sample standard deviations. (This is then only a prior in practice but not in principle, as it is defined post-hoc.) At an SNR of 20, suitable values for the prior extents are
\begin{equation}\label{eq:priorextents}
    \delta\boldsymbol{\theta}=(0.5,1,3,15,0.3,5)\times10^{-3}.
\end{equation}
The Euclidean volume enclosed within $\mathcal{R}_0$ is a miniscule fraction of the six-dimensional parameter subspace: $\lesssim10^{-18}$, since we have $\delta\boldsymbol{\theta}_i/\Delta\boldsymbol{\theta}_i\lesssim10^{-3}$ for global extents $\Delta\boldsymbol{\theta}_i$ corresponding to the range of LISA-relevant signals.

\begin{figure*}[!tbp]
\centering
\includegraphics[width=0.9\paperwidth,trim={2.5cm 0 0 0},clip]{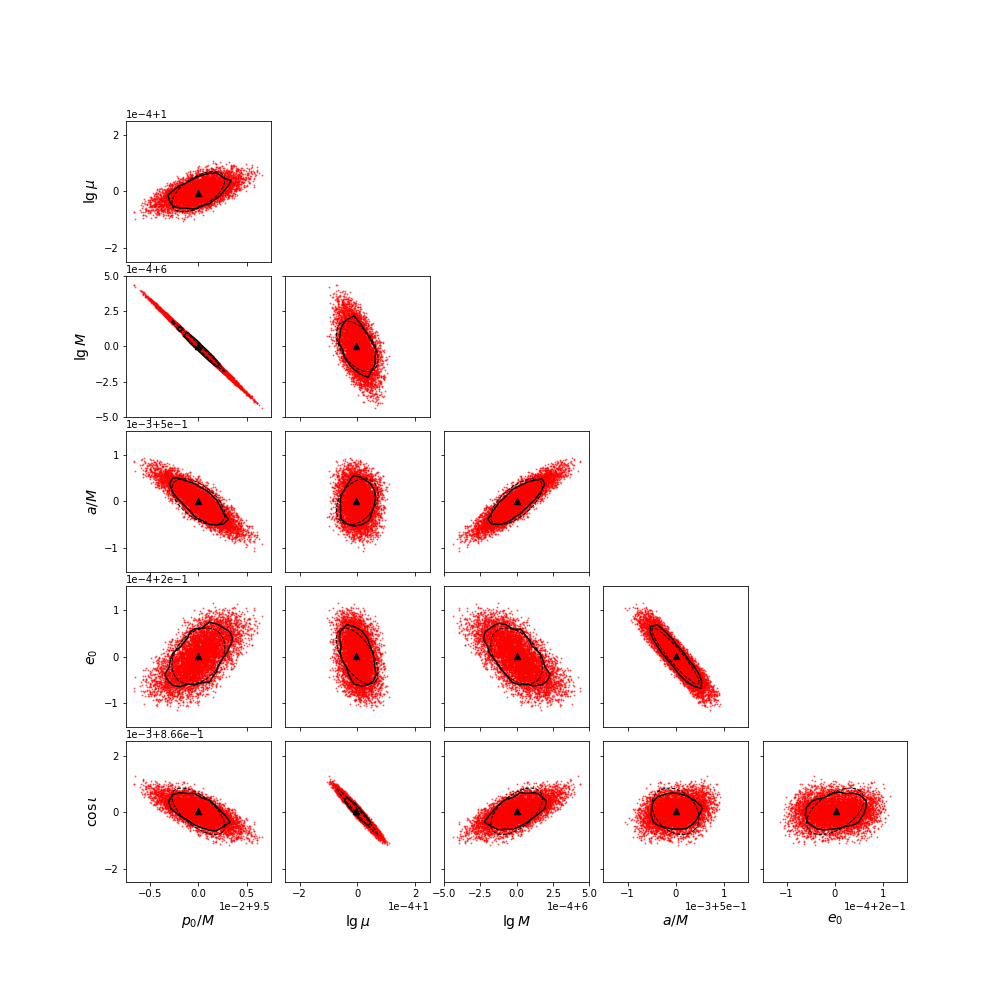}
\caption{Various visual indicators of posterior distribution over $\mathcal{R}_0$ (around the posterior bulk), computed using the approximate likelihood and a flat hyperrectangular prior. Black triangle: MAP estimate (this agrees with the injection parameters at the prior centroid). Solid black curves: Level sets \eqref{eq:marglevel} of the marginal posterior densities $p_{ij}$ (1-$\sigma$ level for a bivariate Gaussian distribution). Dashed black ellipses: 1-$\sigma$ ellipses corresponding to projections of the sample mean and covariance. Red points: Projections of posterior samples in the super-level set \eqref{eq:fulllevel} of the joint posterior density $p$ (1-$\sigma$ level for a six-dimensional Gaussian distribution). All four visual indicators will coincide exactly for a multivariate Gaussian distribution.}
\label{fig:true}
\end{figure*}

The sampling of $L_\mathrm{app}$ over the region $\mathcal{R}_0$ is extremely straightforward, as the locality condition in Eq.\,\eqref{eq:locality} ensures that $L_\mathrm{app}$ is near-Gaussian. A direct visual examination of the conditional densities $L_\mathrm{app}(\boldsymbol{\theta}_i|\boldsymbol{\theta}_{j\neq i}=\boldsymbol{\theta}_{\mathrm{inj},j})$ reveals that the non-Gaussianity takes the form of a flatter peak and heavier tails, which is largely due to the usage of $(\cdot|\cdot)$ rather than $\langle\cdot|\cdot\rangle$; we refer the reader again to Fig.\,\ref{fig:innprod} for intuition about why this is the case. The traditional way of visualizing a sampled $(d>2)$-dimensional posterior distribution is simply to examine plots of its $d(d-1)/2$ bivariate marginal distributions (and the $d$ univariate marginals). In Fig.\,\ref{fig:true}, we instead characterize the sampled posterior using four visual indicators: i) the maximum a posteriori (MAP) estimate; ii) level sets of the marginal posterior densities $p_{ij}$, defined analogously to the 1-$\sigma$ level set for a bivariate Gaussian density:
\begin{equation}\label{eq:marglevel}
    \left\{(\boldsymbol{\theta}_i,\boldsymbol{\theta}_j):p_{ij}(\boldsymbol{\theta}_i,\boldsymbol{\theta}_j)=e^{-1/2}\max_{(\boldsymbol{\theta}_i,\boldsymbol{\theta}_j)}{p_{ij}(\boldsymbol{\theta}_i,\boldsymbol{\theta}_j)}\right\};
\end{equation}
iii) the sample mean and covariance; and iv) a level set of the joint posterior density $p$, defined analogously to the 1-$\sigma$ level set for a six-dimensional Gaussian density. This last indicator is itself five-dimensional, so it is more conveniently represented by the projections of posterior samples within the corresponding super-level set
\begin{equation}\label{eq:fulllevel}
    \left\{\boldsymbol{\theta}:p(\boldsymbol{\theta})>e^{-1/2}\max_{\boldsymbol{\theta}}{p(\boldsymbol{\theta})}\right\}.
\end{equation}

This whole exercise may seem somewhat trivial, as all four visual indicators will coincide exactly when $p$ describes a multivariate Gaussian---the MAP estimate with the sample mean, the marginal level sets with the 1-$\sigma$ sample covariance ellipses, and the convex hulls of the projected super-level set with said ellipses. The indicators do however provide a useful way of visualizing high-dimensional non-Gaussian posteriors, especially when the deviation from Gaussianity is severe (see Secs \ref{subsec:secondaryA} and \ref{subsec:secondaryB}). For the posterior over $\mathcal{R}_0$, the incongruence of the final indicator clearly highlights the non-Gaussianity that would not be immediately apparent from examining traditional contour plots for the marginal distributions (essentially, the second indicator). Nevertheless, the posterior bulk appears to be relatively well-behaved, i.e., unimodal as expected. We verify this by applying our clustering algorithm to the posterior samples in the super-level set (the red points in Fig.\,\ref{fig:true}). All samples are not just connected to the MAP point, but strictly connected with a maximal pre-distance of zero.

\subsubsection{Starting region $\times10^6$}
\label{subsec:10x}

We now consider the Cartesian product of intervals
\begin{equation}\label{eq:r1}
    \mathcal{R}_1:=\prod\limits_i\,[\boldsymbol{\theta}_{\mathrm{inj},i}-5\,\delta\boldsymbol{\theta}_i,\boldsymbol{\theta}_{\mathrm{inj},i}+5\,\delta\boldsymbol{\theta}_i],
\end{equation}
which is a six-dimensional hyperrectangle with $10^6$ times the Euclidean volume of $\mathcal{R}_0$. It is useful to examine the sampling of posteriors over this region for the various likelihood functions in Sec.\,\ref{subsec:sampling}. Even in this intermediate region, large-scale SNR gradients are already manifest in $L_\mathrm{app}$; this shifts the posterior bulk toward certain edges of the prior (and away from the injection parameters, even if the sampler manages to ``find'' that point along the way). When sampling from $L_\mathrm{match}$ over $\mathcal{R}_1$, the posterior bulk does remain tightly centered on the injection parameters, and the samplers we use are able to locate it without any fine-tuning. However, no information about any secondaries in the region can be gleaned from the sampled posterior, even for $L_\mathrm{match}$ at higher annealing temperatures (as discussed in Sec.\,\ref{subsec:sampling}).

By sampling from the exploratory likelihood $L_\mathrm{expl}$, we obtain a large set of points in $\mathcal{R}_1$ whose associated signals have a high overlap against the injection. We consider only the subset of points with approximate overlaps of $>0.5$ for clustering. The approximate overlap typically overestimates full overlaps that are $>0.1$ by a factor of $\lesssim3$, such that these points actually have full overlaps of $\gtrsim0.25$ (which is still reasonably considered ``non-negligible''). Since we know the cluster content in $\mathcal{R}_0$, any points that fall in the previously analyzed region are also removed; this is done in Sec.\,\ref{subsec:100x} as well, where $\mathcal{R}_1$ is excised instead. We find it convenient to perform such an excision after sampling, rather than through a modification of the prior. The resultant set of $N\gtrsim10^5$ points is generally still too large for our clustering algorithm to handle in an acceptable time frame---although the computational cost scales as $\mathcal{O}(N)$, each iteration involves template generation and thus is more expensive than a typical step in a standard algorithm. We further distill the set to exactly $N=10^4$ points, by sorting it in order of overlap and then downsampling the sequence (such that the distribution of the data is loosely preserved).

\begin{figure*}[!tbp]
\centering
\includegraphics[width=0.9\paperwidth,trim={2.5cm 0 0 0},clip]{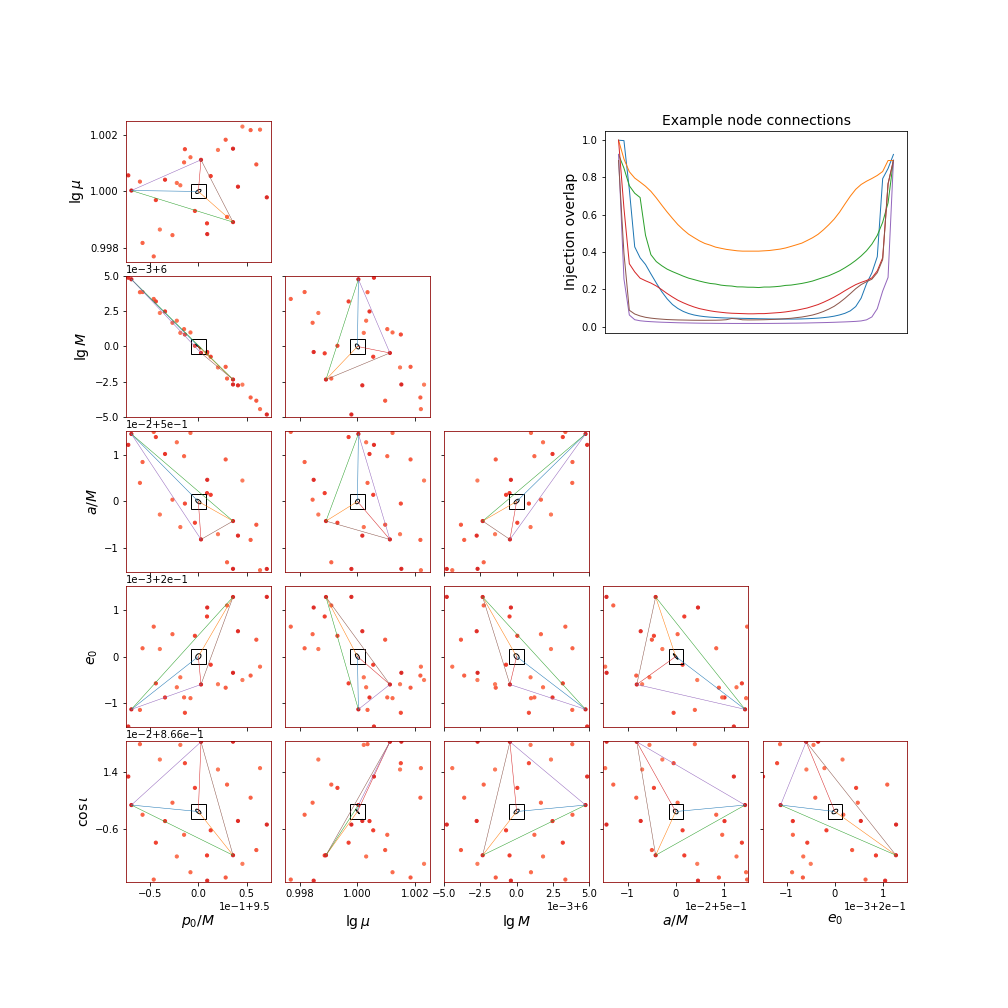}
\caption{Locations of 30 secondary nodes (red points) identified in the intermediate region $\mathcal{R}_1\setminus\mathcal{R}_0$. Relative overlap values at these nodes are indicated by color saturation. The black square and ellipse in each panel corresponds to the plot range and dashed ellipse in each panel of Fig.\,\ref{fig:true}. Inset: The six pairwise connections among the injection parameters and the three highest-overlap nodes in $\mathcal{R}_1$, i.e., the overlap value against the injection along the corresponding colored lines in each panel.}
\label{fig:10x}
\end{figure*}

In the region $\mathcal{R}_1$ (minus $\mathcal{R}_0$), the clustering algorithm identifies 30 secondary nodes, with full overlap values ranging from 0.45 to 0.72. The locations of these nodes, projected onto each of the 15 parameter-pair planes, are plotted in Fig.\,\ref{fig:10x}. It is not particularly useful to present the actual clusters associated with the nodes, due to visual congestion. To give a flavor of the overlap surface in this region (and some assurance that the nodes are indeed adequate representations of local overlap maxima), we instead consider the six pairwise connections $\ell$ among the injection parameters and the three nodes in $\mathcal{R}_1$ with the highest overlap values. These connections are shown in the inset of Fig.\,\ref{fig:10x}, with an increased resolution of $l=50$ (see Sec.\,\ref{subsec:clustering}) for visualization purposes.

\begin{figure}[!tbp]
\centering
\includegraphics[width=\columnwidth,trim={1.8cm 6.8cm 1.8cm 7.2cm},clip]{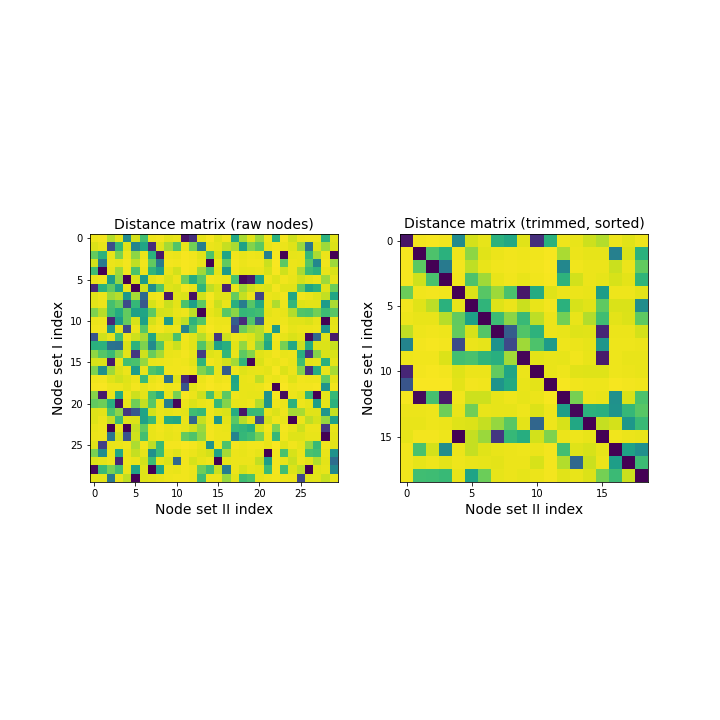}
\includegraphics[width=0.8\columnwidth]{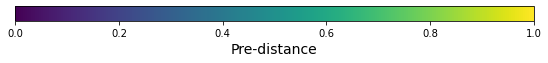}
\caption{Distance matrix between the two sets of secondary nodes identified in $\mathcal{R}_1$. Left: Raw sets with 30 nodes each. Right: Trimmed and sorted sets with 19 node pairs.}
\label{fig:distmat}
\end{figure}

To verify the robustness of our results against the stochastic error in sampling, we repeat the analysis on the output of a second sampling run, and compare the first set (I) of 30 nodes to the new set (II). Set II turns out to have the exact same count; surprisingly, however, the locations of its nodes do not match up well with those in set I. We seek to pair up nodes in the two sets, and to determine which pairs are more ``distinct''. This is done by computing the distance matrix of pairwise pre-distances between two sets of points---not to be confused with the pre-distance matrix from the clustering algorithm, which is essentially a (partial) distance matrix between a single set of points and itself. With the distance matrix between sets I and II in hand (Fig.\,\ref{fig:distmat}; left panel), it is straightforward to identify pairs of nodes that are in one-to-one correspondence with each other. More precisely, if node $a$ from set I has node $b$ as its closest node in set II, we demand that node $a$ is also the closest node in set I to node $b$. There are 19 such pairs (which we index in decreasing order of set-I overlap, such that the set-I point in pair 0 has the highest overlap value in set I). The distance matrix may then be trimmed accordingly and sorted by pair index, to visually indicate the inter-connectivity among the reduced set of node pairs (Fig.\,\ref{fig:distmat}; right panel).

\begin{figure}[!tbp]
\centering
\includegraphics[width=\columnwidth]{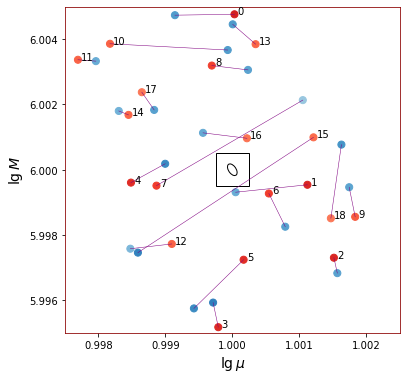}
\includegraphics[width=0.9\columnwidth]{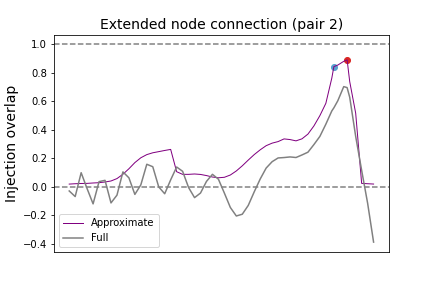}
\includegraphics[width=0.9\columnwidth]{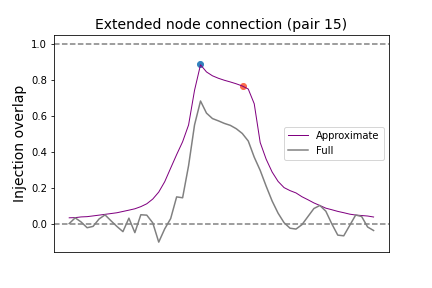}
\caption{Trimmed and sorted pairs from sets I (red) and II (blue) of secondary nodes identified in $\mathcal{R}_1$. Relative overlap values at these nodes are indicated by color saturation. Top: Projection of node-pair locations onto the $(\lg{\mu},\lg{M})$ plane (corresponding to Fig.\,\ref{fig:10x}, second row, second panel). Middle: Extended connection for pair 2, which has the third-highest set-I overlap value. Bottom: Extended connection for pair 15, which has the largest extent in the mass--mass plane.}
\label{fig:pairs}
\end{figure}

The locations of the 19 node pairs and intra-pair connections are plotted in the top panel of Fig.\,\ref{fig:pairs}, after projection onto the mass--mass plane. Another surprise at this stage is the variety of orientations and lengthscales on display; all of the pairs are strictly or nearly strictly connected (see diagonal elements of reduced distance matrix in Fig.\,\ref{fig:distmat}), indicating that the local maxima they represent also take on an assortment of shapes and sizes. While these pairs clearly correspond to secondaries in accordance with our search criteria, we do not find any that can be cleanly localized---i.e., where the posterior peak is contained exclusively in a topologically connected subset of parameter space. We will revisit the problem of trying to localize secondaries in Secs \ref{subsec:secondaryA} and \ref{subsec:secondaryB}. For now, we examine the connections for a couple of example node pairs, extended outward until the (approximate) overlap $\Omega(h_\mathrm{inj},\cdot)$ is $\approx0$. These extended connections are shown in the middle and bottom panels of Fig.\,\ref{fig:pairs} for both the approximate and full overlaps, and they provide some intuition (albeit cross-sectional) as to the structure of the overlap surface at and around secondaries.

\subsubsection{Starting region $\times10^{12}$ (and beyond)}
\label{subsec:100x}

A similar sampling and clustering analysis is performed for the Cartesian product of intervals
\begin{equation}\label{eq:r2}
    \mathcal{R}_2:=\prod\limits_i\,[\boldsymbol{\theta}_{\mathrm{inj},i}-50\,\delta\boldsymbol{\theta}_i,\boldsymbol{\theta}_{\mathrm{inj},i}+50\,\delta\boldsymbol{\theta}_i].
\end{equation}
Note that $\mathcal{R}_2$ is truncated by the upper bound $\cos{\iota}=1$, and thus the enclosed volume is $\lesssim10^{12}$ times that of $\mathcal{R}_0$. In this large region, the direct sampling of $L_\mathrm{match}$ now presents difficulties for the samplers; they are unable to locate the posterior bulk either at termination, or after $\sim10^9$ likelihood evaluations. This result likely indicates the rough scale at which posterior sampling can become challenging without prior localization of the source parameters (even if large-scale SNR gradients are accounted for)---although it may be possible to improve the performance of specific samplers through further tuning.

\begin{figure*}[!tbp]
\centering
\includegraphics[width=0.9\paperwidth,trim={2.5cm 0 0 0},clip]{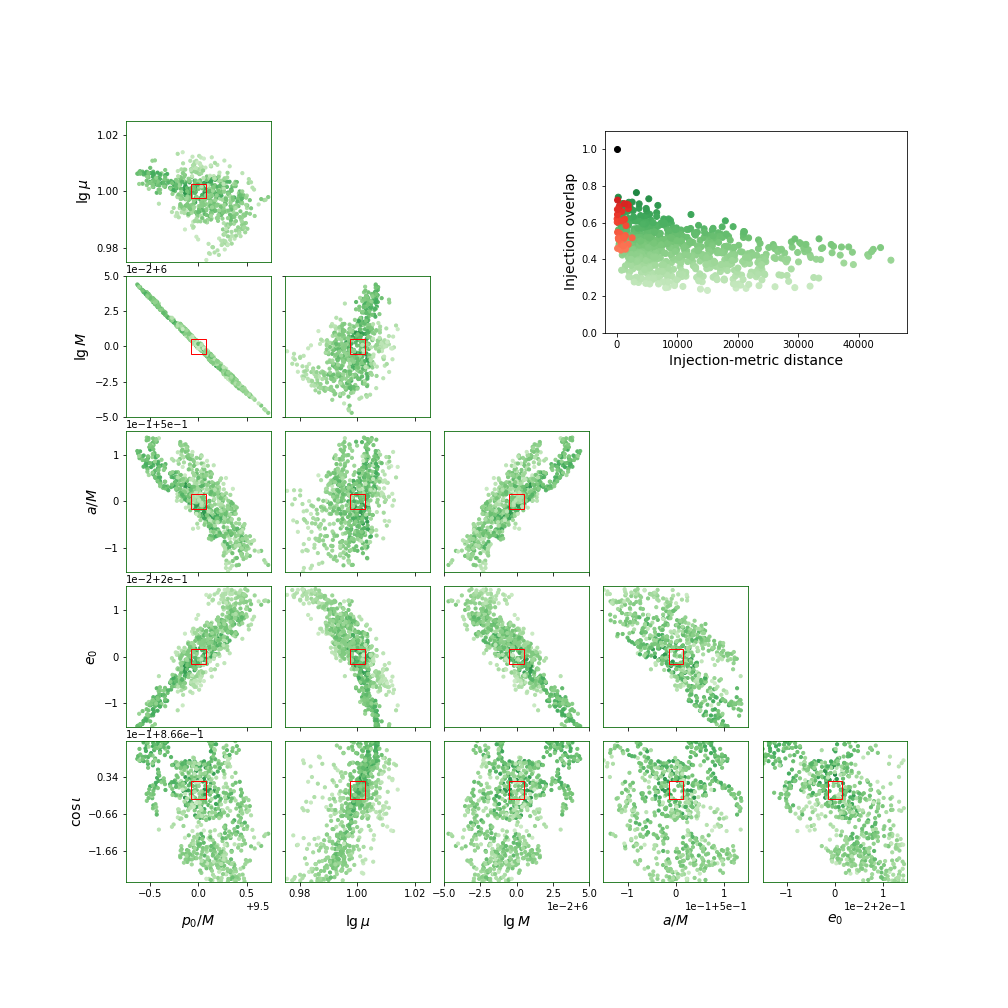}
\caption{Locations of 675 secondary nodes (green points) identified in the large region $\mathcal{R}_2\setminus\mathcal{R}_1$. Relative overlap values at these nodes are indicated by color saturation. The red rectangle in each panel corresponds to the plot range in each panel of Fig.\,\ref{fig:10x}. Inset: Full overlap values for these nodes, plotted against their distance from the injection parameters (black point) with respect to the pullback (Fisher) metric there. The set-I nodes shown in Fig.\,\ref{fig:10x} (red points) are also included in this plot.}
\label{fig:100x}
\end{figure*}

As in Sec.\,\ref{subsec:10x}, the previous analysis region (now $\mathcal{R}_1$) is excised after sampling to avoid double-counting, and the data set for clustering is trimmed to $10^4$ points with approximate overlaps of $>0.5$. A total of 675 secondary nodes are identified in the region $\mathcal{R}_2\setminus\mathcal{R}_1$, with full overlap values ranging from 0.23 to 0.76. Their projected locations are shown in Fig.\,\ref{fig:100x}. The broad correlation structure of secondaries in the large region---as visually indicated by the plane convex hull for each projection of the set---generally does not resemble that observed in Fig.\,\ref{fig:10x}, or the local structure in Fig.\,\ref{fig:true} (one exception being the strong correlation between $\lg{M}$ and $p_0/M$). Another notable feature in Fig.\,\ref{fig:100x} is that the distribution of nodes appears somewhat ``patchy'', although this may be attributable to imperfect sampling rather than the actual absence of secondaries in some of the empty regions.

With a large set of secondary nodes from the combined analysis in $\mathcal{R}_{1,2}$, it is natural to examine the relationship between overlap (against the injection) and distance (from the injection parameters) for any discernible trends. Instead of using the Euclidean distance in parameter space $\Theta$ or the (computationally intractable) geodesic distance in signal space $\mathcal{S}$, we consider the distance with respect to the pullback metric at the injection parameters. This is given component-wise by Eq.\,\eqref{eq:fisher}, where the Fisher information $\mathcal{I}(\boldsymbol{\theta}_\mathrm{inj})$ is in turn approximated by the sample covariance of the posterior bulk in Sec.\,\ref{subsec:1x}. A plot of overlap against the injection-metric distance is shown in the inset of Fig.\,\ref{fig:100x}, for the 30 set-I nodes in $\mathcal{R}_1\setminus\mathcal{R}_0$ and the 675 nodes in $\mathcal{R}_2\setminus\mathcal{R}_1$. There is a slight negative correlation between overlap and distance, as well as between number density and distance; however, these trends are not particularly pronounced (and even less so if the Euclidean distance is used).

The results from the analysis of $\mathcal{R}_2$ are admittedly less reliable than those for $\mathcal{R}_1$, since we do not (and cannot) increase sampling resolution by anywhere close to the factor of $10^6$ that is required to compensate for the larger volume. Even if we were able to, the size of the data set for clustering would also have to be raised accordingly for high-resolution sampling to be meaningful. As a heuristic assessment of sampling error and its increased impact in extended regions, we briefly discuss our analysis of a final region: the Cartesian product of intervals
\begin{equation}\label{eq:r3}
    \mathcal{R}_3:=\prod\limits_i\,[\boldsymbol{\theta}_{\mathrm{inj},i}-100\,\delta\boldsymbol{\theta}_i,\boldsymbol{\theta}_{\mathrm{inj},i}+100\,\delta\boldsymbol{\theta}_i],
\end{equation}
whose results we do not present explicitly here (although we do use one of the identified secondary nodes as a case study in Sec.\,\ref{subsec:secondaryA}). Two identically initialized sampling runs are conducted for $L_\mathrm{expl}$ over this region, each with $\gtrsim10^6$ independent samples drawn after $\gtrsim10^9$ likelihood evaluations. It is immediately clear from their density plots that the two sets of posterior samples are concentrated in very different sub-regions, and thus that clustering will give inconsistent results as well. This also points to a systematic undercounting of clusters, which worsens as the analysis region is expanded. Nevertheless, the convex hulls of the raw posterior samples from the two runs are similar in both shape and volume ($\sim10^{-6}$ relative to $\mathcal{R}_3$), which gives a more reliable indication of where secondaries are absent, as well as a very conservative upper bound on their coverage of parameter space.

\subsubsection{Case study A}
\label{subsec:secondaryA}

We now examine in detail the actual posterior surface over the vicinity of an example secondary node in $\mathcal{R}_3$---specifically, the node with the highest overlap value found outside $\mathcal{R}_2$. Relative to $\boldsymbol{\theta}_\mathrm{inj}$ from Eq.\,\eqref{eq:analysisinjpars}, this node (labeled A) is given by
\begin{equation}\label{eq:asecpars}
    \boldsymbol{\theta}_\mathrm{A}-\boldsymbol{\theta}_\mathrm{inj}=(-30,13,228,-53,13,-482)\times10^{-3}.
\end{equation}
Its associated signal has an approximate overlap of 0.81 against the injection, but a full overlap value of only 0.39. This is nevertheless a significant example of degeneracy, especially considering its large injection-metric distance from the injection parameters ($\gtrsim2\times10^5$), and it is not inconceivable for an uninformed search algorithm to flag the secondary represented by node A as a separate candidate source (with a detection SNR of $\approx0.4\,\rho_\mathrm{inj}$).

\begin{figure*}[!tbp]
\centering
\includegraphics[width=0.9\paperwidth,trim={2.5cm 0 0 0},clip]{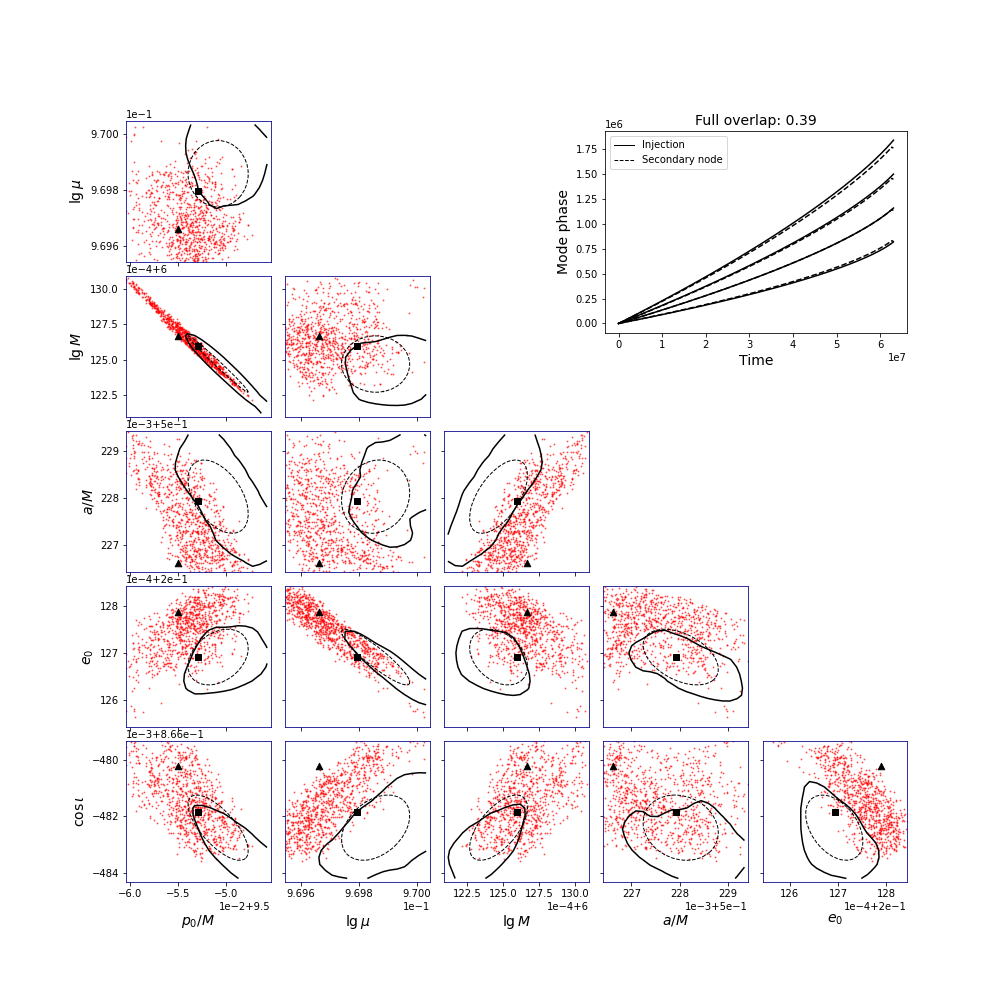}
\caption{Various visual indicators of posterior distribution over $\mathcal{R}_\mathrm{A}$ (a small region around secondary node A), computed using the approximate likelihood and a flat hyperrectangular prior. The prior extents are the same as in Fig.\,\ref{fig:true}. Black square: Node A (the prior centroid). Black triangle: MAP estimate. Solid black curves: Level sets \eqref{eq:marglevel} of the marginal posterior densities $p_{ij}$ (1-$\sigma$ level for a bivariate Gaussian distribution). Dashed black ellipses: 1-$\sigma$ ellipses corresponding to projections of the sample mean and covariance. Red points: Projections of posterior samples in the super-level set \eqref{eq:fulllevel} of the joint posterior density $p$ (1-$\sigma$ level for a six-dimensional Gaussian distribution). All four visual indicators will coincide exactly for a multivariate Gaussian distribution. Inset: Comparison of phase trajectories $\varphi_j(t)$ between the signal injection and the signal template at node A. Only the dominant $(2,0,0)$ modes (third from top) are in phase for a substantial fraction of the analysis duration.}
\label{fig:furthest_1x}
\end{figure*}

\begin{figure*}[!tbp]
\centering
\includegraphics[width=0.9\paperwidth,trim={2.5cm 0 0 0},clip]{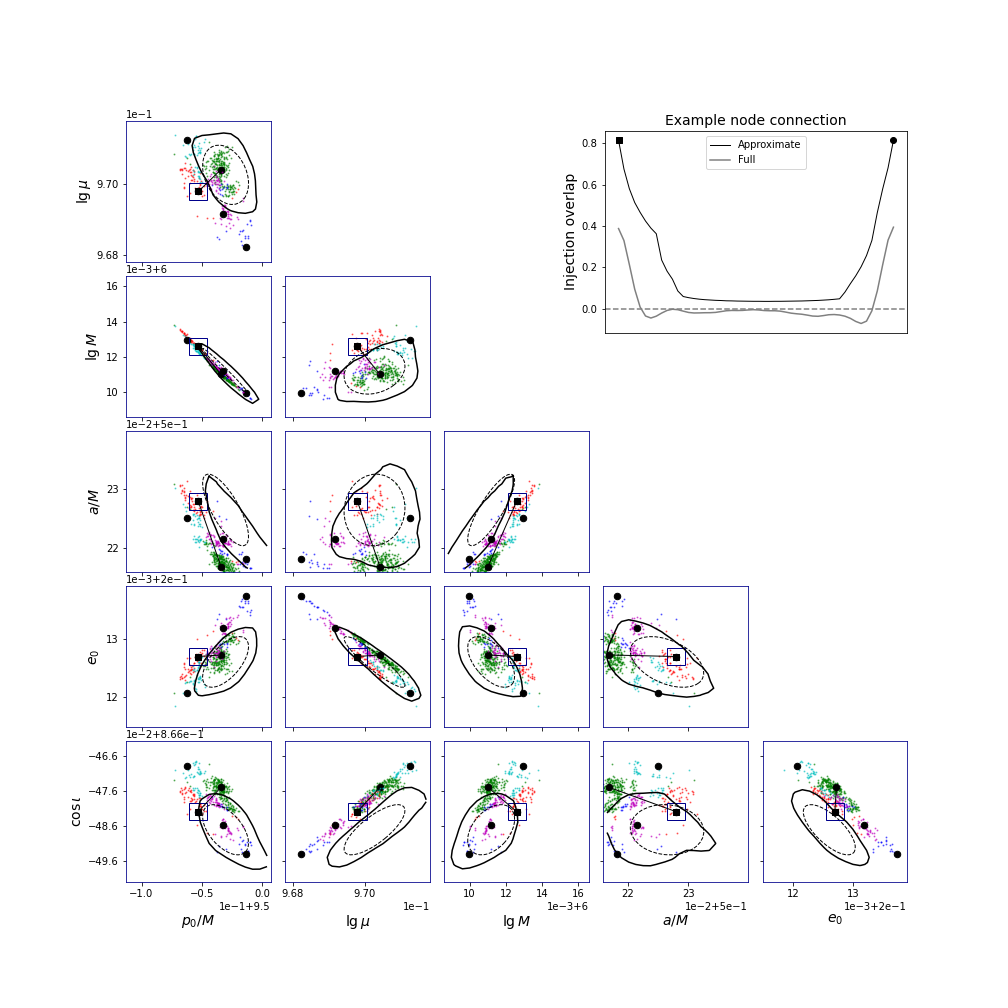}
\caption{Various visual indicators of posterior distribution over $\mathcal{R}_\mathrm{A}'$ (a larger region around secondary node A), computed using the approximate likelihood and a flat hyperrectangular prior. The blue square in each panel corresponds to the plot range in each panel of Fig.\,\ref{fig:furthest_1x}. Black square: Node A (the prior centroid). Black points: Representative nodes for four other secondaries identified in $\mathcal{R}_\mathrm{A}'$. Solid black curves: Level sets \eqref{eq:marglevel} of the marginal posterior densities $p_{ij}$. Dashed black ellipses: 1-$\sigma$ ellipses corresponding to projections of the sample mean and covariance. Colored points: Clusters of posterior samples in the super-level set \eqref{eq:fulllevel} of the joint posterior density $p$. These are the subsets of samples that are connected to each node, and not connected to any of the other nodes. Inset: Connection between node A and the highest-overlap node in $\mathcal{R}_\mathrm{A}'$, i.e., the overlap value against the injection along the black line in each panel.}
\label{fig:furthest_8x}
\end{figure*}

The approximate likelihood $L_\mathrm{app}$ is sampled over a flat-prior region $\mathcal{R}_\mathrm{A}$ centered on $\boldsymbol{\theta}_\mathrm{A}$, with the same extents as $\mathcal{R}_0$. A traditional inspection of the marginal posterior densities, either of their Gaussian-analogous 1-$\sigma$ contours (the solid black curves in Fig.\,\ref{fig:furthest_1x}) or their density plots (not shown), presents nothing particularly out of the ordinary; the posterior might be construed at this stage as a moderately deformed Gaussian distribution, centered on a point that is close to node A. We find however that any attempt to localize this (or any other) secondary by simply enlarging and re-centering the prior region---``chasing'' the secondary, as we not-so-affectionately term it---turns out to be futile. Hints as to why this is the case are provided in Fig.\,\ref{fig:furthest_1x} by the other visual indicators introduced in Sec.\,\ref{subsec:1x}. While the sample mean and covariance are at least somewhat consistent with the marginal 1-$\sigma$ contours, the projections of the super-level set (points within the Gaussian-analogous 1-$\sigma$ contour for the joint posterior) are not. The MAP estimate also falls in a location that defies reasonable prediction.

Our findings seem to indicate that secondary A cannot be localized; i.e., there does not exist a topologically connected subset of parameter space $U\ni\boldsymbol{\theta}_\mathrm{A}$ with some neighborhood $V\supset U$ such that the posterior probability is approximately zero over $V\setminus U$. We do not make this argument rigorous, but offer an anecdotally supported conjecture that it holds generally for all secondaries, as well as a partial explanation for the phenomenon (see Sec.\,\ref{subsec:nongaussian}). Another important qualitative distinction between the primary posterior peak and a typical secondary might be the presence of fine structure. The posterior surface over the region $\mathcal{R}_\mathrm{A}$ is not quite unimodal, and appears to comprise multiple local maxima that are strongly but not strictly connected to one another. This is inferred from the fact that the maximal pre-distance between node A and all posterior samples in the super-level set is $\approx0.1$. Node A itself is at least a stationary point of the posterior, from visual examination of the conditional densities $L_\mathrm{app}(\boldsymbol{\theta}_i|\boldsymbol{\theta}_{j\neq i}=\boldsymbol{\theta}_{\mathrm{A},j})$, but we are unable to rule out the saddle-point case due to severe instability in the numerical derivatives.

As we ``zoom out'' on secondary A by a factor of eight (i.e., sample from $L_\mathrm{app}$ over a larger flat-prior region $\mathcal{R}_\mathrm{A}'$ with extents $8\,\delta\boldsymbol{\theta}$), we encompass additional local maxima that are now disconnected to one another. The existence and proximity of these other secondaries is likely the main impediment to the localization of secondary A; indeed, their presence cannot even be discerned from the traditional approach of considering only the marginal posteriors (see solid black curves in Fig.\,\ref{fig:furthest_8x}). To resolve and visualize the multiple secondaries in this region, we again apply our clustering algorithm to all samples in the super-level set of the posterior over $\mathcal{R}_\mathrm{A}'$. Five nodes (including node A) are identified, and their associated clusters are shown grouped by color in Fig.\,\ref{fig:furthest_8x}. These clusters do not exhibit a high degree of order in their spatial structure, and are also underpopulated due to increased inter-connectivity, i.e., many points in the super-level set are connected to more than one node, and thus do not appear in any cluster by definition. Nevertheless, the clustering analysis provides clear evidence of additional maxima in the region (see inset of Fig.\,\ref{fig:furthest_8x}), and demonstrates how secondaries can congeal into larger ones especially when viewed after projection (marginalization).

\subsubsection{Case study B}
\label{subsec:secondaryB}

\begin{figure*}[!tbp]
\centering
\includegraphics[width=0.9\paperwidth,trim={2.5cm 0 0 0},clip]{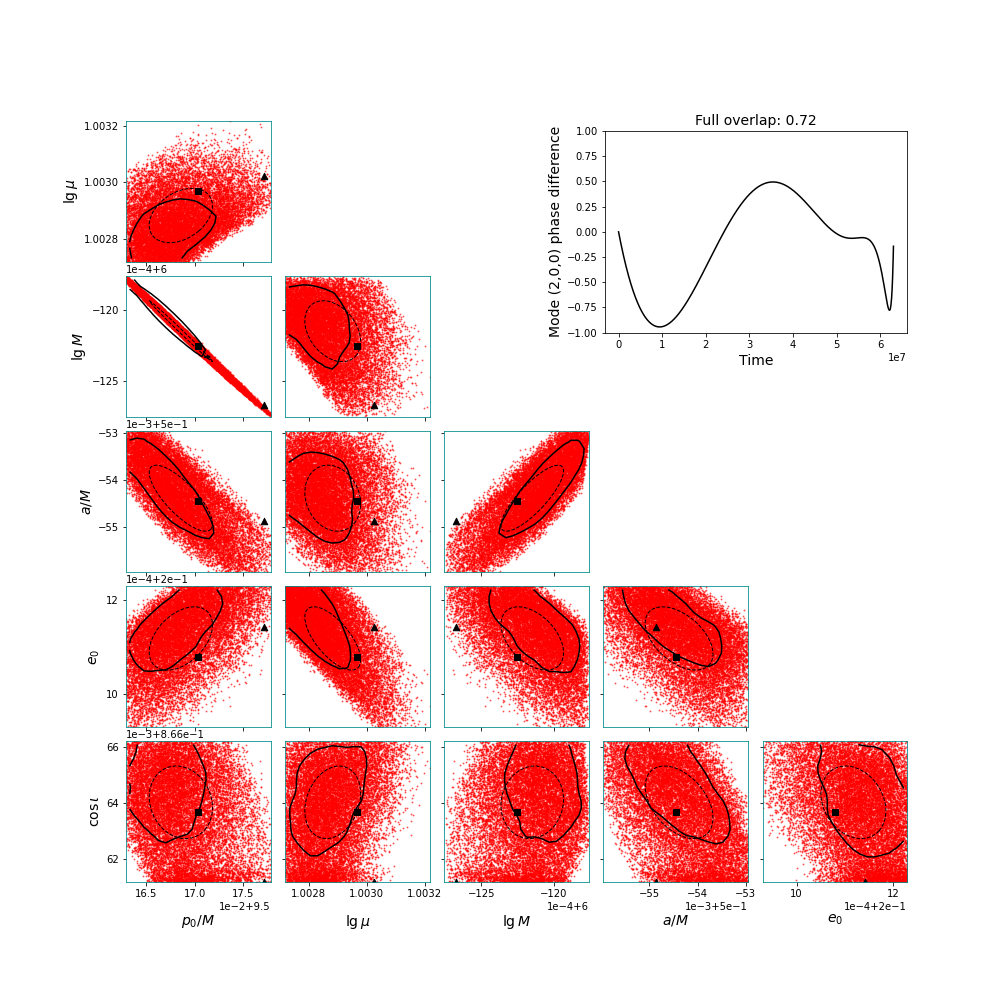}
\caption{Various visual indicators of posterior distribution over $\mathcal{R}_\mathrm{B}$ (a small region around secondary node B), computed using the approximate likelihood and a flat hyperrectangular prior. The prior extents are the same as in Figs \ref{fig:true} and \ref{fig:furthest_1x}. Black square: Node B (the prior centroid). Black triangle: MAP estimate. Solid black curves: Level sets \eqref{eq:marglevel} of the marginal posterior densities $p_{ij}$ (1-$\sigma$ level for a bivariate Gaussian distribution). Dashed black ellipses: 1-$\sigma$ ellipses corresponding to projections of the sample mean and covariance. Red points: Projections of posterior samples in the super-level set \eqref{eq:fulllevel} of the joint posterior density $p$ (1-$\sigma$ level for a six-dimensional Gaussian distribution). All four visual indicators will coincide exactly for a multivariate Gaussian distribution. Inset: Phase difference between the dominant $(2,0,0)$ modes of the signal injection and the signal template at node B, showing that the two signals are approximately in phase over the full analysis duration.}
\label{fig:highest_1x}
\end{figure*}

\begin{figure*}[!tbp]
\centering
\includegraphics[width=0.9\paperwidth,trim={2.5cm 0 0 0},clip]{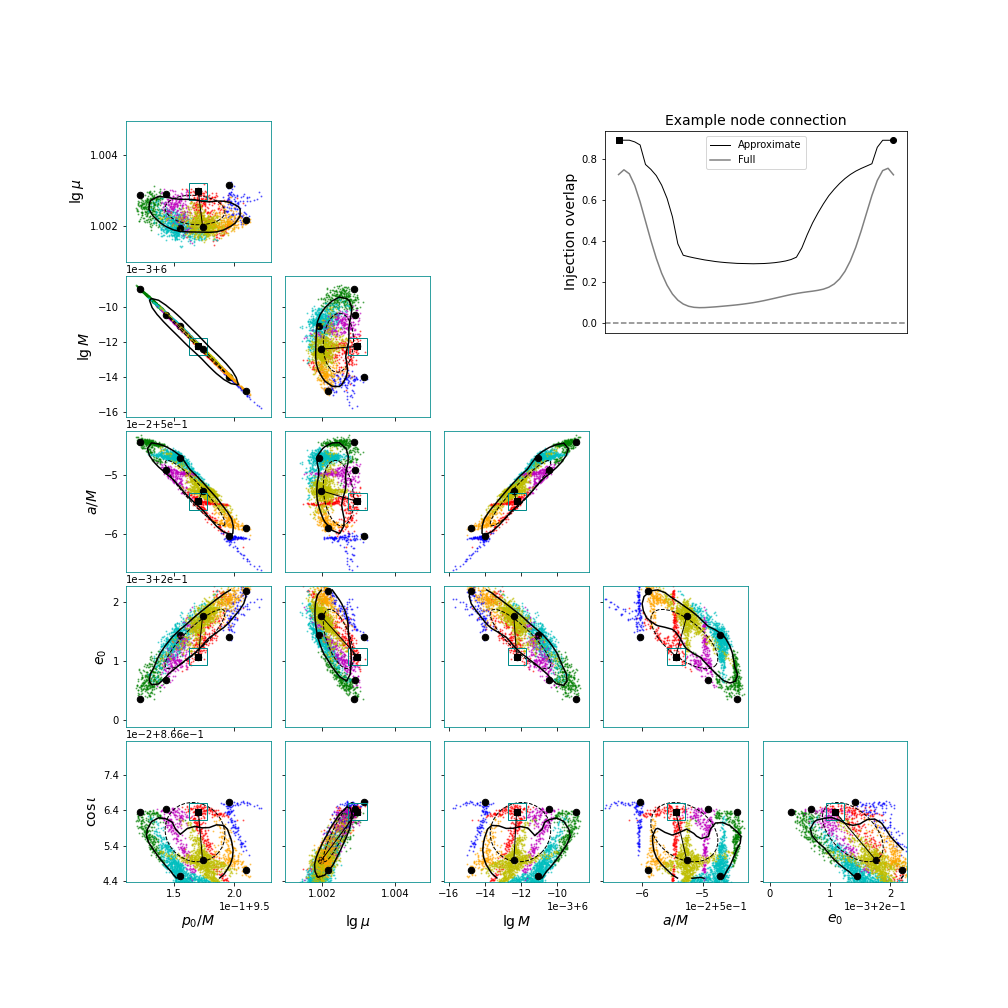}
\caption{Various visual indicators of posterior distribution over $\mathcal{R}_\mathrm{B}'$ (a larger region around secondary node B), computed using the approximate likelihood and a flat hyperrectangular prior. The green square in each panel corresponds to the plot range in each panel of Fig.\,\ref{fig:highest_1x}. Black square: Node B (the prior centroid). Black points: Representative nodes for six other secondaries identified in $\mathcal{R}_\mathrm{B}'$. Solid black curves: Level sets \eqref{eq:marglevel} of the marginal posterior densities $p_{ij}$. Dashed black ellipses: 1-$\sigma$ ellipses corresponding to projections of the sample mean and covariance. Colored points: Clusters of posterior samples in the super-level set \eqref{eq:fulllevel} of the joint posterior density $p$. These are the subsets of samples that are connected to each node, and not connected to any of the other nodes. Inset: Connection between node B and the highest-overlap node in $\mathcal{R}_\mathrm{B}'$, i.e., the overlap value against the injection along the black line in each panel.}
\label{fig:highest_8x}
\end{figure*}

For our second case study, we select one of the highest-overlap nodes found in the region $\mathcal{R}_2\setminus\mathcal{R}_1$. Relative to $\boldsymbol{\theta}_\mathrm{inj}$ from Eq.\,\eqref{eq:analysisinjpars}, this node (labeled B) is given by
\begin{equation}\label{eq:bsecpars}
    \boldsymbol{\theta}_\mathrm{B}-\boldsymbol{\theta}_\mathrm{inj}=(3,-12,-54,170,1,64)\times10^{-3}.
\end{equation}
It has a full (approximate) overlap value of 0.72 (0.89), and its injection-metric distance from the injection parameters is $\approx8000$. The analysis in Sec.\,\ref{subsec:secondaryA} is repeated around $\boldsymbol{\theta}_\mathrm{B}$---first in a small region $\mathcal{R}_\mathrm{B}$ with the same size as $\mathcal{R}_0$ and $\mathcal{R}_\mathrm{A}$ (Fig.\,\ref{fig:highest_1x}), then in a larger region $\mathcal{R}_\mathrm{B}'$ with the same size as $\mathcal{R}_\mathrm{A}'$ (Fig.\,\ref{fig:highest_8x}).

As its overlap value would indicate, the signal template at node B strongly resembles the signal injection. The two signals can no longer be distinguished in a plot analogous to the inset of Fig.\,\ref{fig:furthest_1x}, and their dominant $(2,0,0)$ modes remain ``in phase'' over the full analysis duration (see inset of Fig.\,\ref{fig:highest_1x}). This increased resemblance over node A might partly explain why the visual indicators in Fig.\,\ref{fig:highest_1x} are slightly less inconsistent with one another than in Fig.\,\ref{fig:furthest_1x}. However, secondary B is still distinctly non-Gaussian, and like secondary A it resists all attempts at localization. Fine structure is present here as well---for all samples in the super-level set of the posterior over $\mathcal{R}_\mathrm{B}$, the maximal pre-distance from node B is $\approx0.1$. For the posterior over $\mathcal{R}_\mathrm{B}'$, we see once again from Fig.\,\ref{fig:highest_8x} that the marginal densities show only a single contiguous secondary peak without any hints of internal structure. Clustering reveals seven nodes (including node B) in this region; more interestingly, it is now possible to identify clear patterns in the shapes, sizes and relative locations of the associated clusters. The general existence of ordered structure in the posterior surface
is expected a priori, and its verification in this case study also serves to demonstrate the efficacy of our clustering algorithm.

\subsubsection{Summary plot}
\label{subsec:summaryplot}

\begin{figure*}[!tbp]
\centering
\includegraphics[width=0.9\paperwidth,trim={2.5cm 0 0 0},clip]{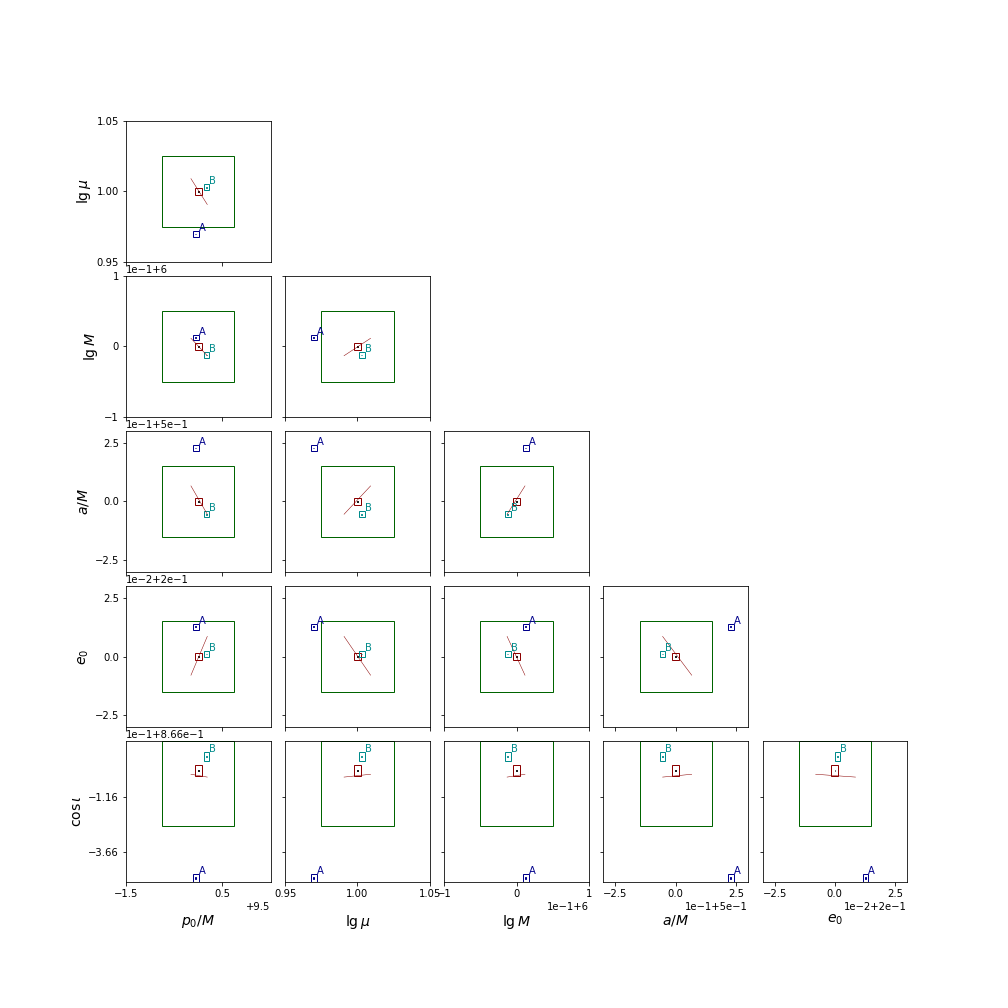}
\caption{Summary plot showing the various analysis sub-regions in Sec.\,\ref{subsec:mapping}: $\mathcal{R}_0$ (black points, corresponding to the plot range in each panel of Fig.\,\ref{fig:true}); $\mathcal{R}_1$ (red rectangles, Fig.\,\ref{fig:10x}); $\mathcal{R}_2$ (green rectangles, Fig.\,\ref{fig:100x}); $\mathcal{R}_3$ (plot range in each panel here); $\mathcal{R}_\mathrm{A}$ (blue points, Fig.\,\ref{fig:furthest_1x}); $\mathcal{R}_\mathrm{A}'$ (blue rectangles, Fig.\,\ref{fig:furthest_8x}); $\mathcal{R}_\mathrm{B}$ (cyan points, Fig.\,\ref{fig:highest_1x}); $\mathcal{R}_\mathrm{B}'$ (cyan rectangles, Fig.\,\ref{fig:highest_8x}). The red line in each panel indicates the projected extent of the extended connection for node-pair 15 in Fig.\,\ref{fig:pairs}.}
\label{fig:summary}
\end{figure*}

The relative scales of the various sub-analyses in Sec.\,\ref{subsec:mapping} are depicted by the summary plot of Fig.\,\ref{fig:summary}, where the starting region $\mathcal{R}_0$ (along with $\mathcal{R}_\mathrm{A,B}$) is unresolvable and represented as a point. We draw particular attention to the secondary associated with pair 15 from Sec.\,\ref{subsec:10x} (see bottom panel of Fig.\,\ref{fig:pairs}). This is the pair of nodes in $\mathcal{R}_1$ with the largest separation---overtly in the mass parameters, as shown in the top panel of Fig.\,\ref{fig:pairs}, but in many other parameter pairs as well. Its extended connection indicates at least a ridge-like region of high posterior probability relative to the immediate neighborhood, and possibly one with non-negligible width (which cannot be determined from the connection alone). As seen in Fig.\,\ref{fig:summary}, this secondary actually extends well beyond $\mathcal{R}_1$, and up to a significant fraction of $\mathcal{R}_2$.

Our findings in Sec.\,\ref{subsec:mapping} are clear evidence of highly non-trivial degeneracy in the EMRI signal space---this manifests as the presence of many posterior secondaries with different shapes, sizes, and degrees of connectivity to one another. While the tools we have brought to bear in the study can provide insight into specific occurrences of degeneracy, the cost and reliability of high-resolution sampling (and the follow-up clustering of data) remains the impediment to a fully global analysis. There is nevertheless ample room for improvement, and the present work provides the
foundation for extended studies in the future. Such studies will be especially relevant after science-adequate waveform models become available (see additional comments in Secs \ref{subsec:representativeness} and \ref{sec:conclusion}).

\subsection{Interpretations}
\label{subsec:corollaries}

\subsubsection{What actually causes degeneracy?}
\label{subsec:cause}

Broadly speaking, degeneracy arises when strong modes in the signal injection and template have similar initial frequencies and time derivatives of these frequencies, such that their phasing is aligned for much of the analysis duration. Treating the effect of initial phasing on degeneracy as negligible, we will find it useful to interpret any given secondary (node) $\boldsymbol{\theta}_\mathrm{sec}$ as an approximate root for a nonlinear system of equations in $\boldsymbol{\theta}$:
\begin{equation}\label{eq:system}
    \left(\frac{d}{dt}\right)^k\omega_j(t;\boldsymbol{\theta})|_{t_0}\approx\left(\frac{d}{dt}\right)^k\omega_{j'}(t;\boldsymbol{\theta}_\mathrm{inj})|_{t_0},
\end{equation}
where $(j,j')$ is a single fixed pair of general mode indices, and the order $k$ of the time derivative ranges from zero to some small positive integer. (To be more precise, the notation $x\approx y$ in Eq.\,\eqref{eq:system} represents $|x-y|<\epsilon$ for some given $\epsilon(j,j',k)\ll1$.) In principle, a secondary might satisfy a system specified by a range of pairs $(j,j')$, i.e., two or more modes are matched well between injection and template. This is far less common in the signal space, however, and the injection overlap is likely to be more strongly determined by a single pair in any case.

\begin{figure}[!tbp]
\centering
\includegraphics[width=\columnwidth]{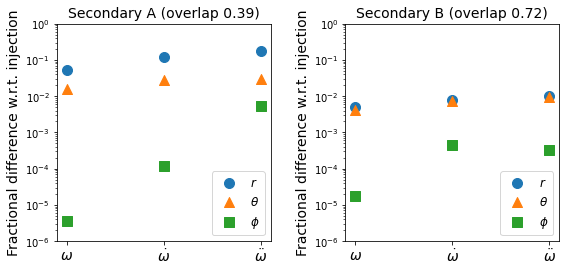}
\caption{Fractional differences in the fundamental frequencies and their low-order time derivatives at time $t_0$, for secondaries A (Eq.\,\eqref{eq:asecpars}) and B (Eq.\,\eqref{eq:bsecpars}) with respect to the injection.}
\label{fig:freqderivs}
\end{figure}

From the results of our mapping analysis, we may draw several anecdotal conclusions about which modes are typically aligned between injection and template, and to what degree. For all of the 705 secondary nodes identified in the large analysis region $\mathcal{R}_2$, it is the dominant mode in both injection and template that is matched, i.e., $j=j'=2$ in Eq.\,\eqref{eq:system}, with $j$ as defined just after Eq.\,\eqref{eq:aakdecomp}. In other words, only the azimuthal fundamental frequency $\omega_\phi$ and its time derivatives (up to at least second order) are similar in the injection and these secondaries. As an example, the fractional differences in $(d/dt)^k\omega_{r,\theta,\phi}$ (henceforth denoted by overdots) with respect to the injection values are shown in Fig.\,\ref{fig:freqderivs} for the case-study secondaries A and B. A matched azimuthal mode of motion is a sufficient condition for the existence of a strong secondary (at least in the regime of low eccentricity and modest initial separation, where $j=2$ is the dominant mode), and our results seem to indicate that it is a necessary one as well. This is not too surprising, since our considered injection has around 70--90\% of its overall power in the dominant mode and 10--20\% in the second strongest (see Fig.\,\ref{fig:fourieramps}); thus the expected overlap from a perfect alignment of the dominant modes is $\gtrsim0.7$, versus $\lesssim0.2$ for the second-strongest modes.

As highlighted in \cite{Cornish:2008zd,Babak:2009ua}, it is also possible for secondaries to arise from the matching of different modes between injection and template, i.e., $j\neq j'$ in Eq.\,\eqref{eq:system}. These are not encountered in our analysis---partly because we do not include the sideband modes from Lense--Thirring precession, and partly because templates with a $j\neq j'$ matching of modes tend to occur at greater separations from the injection in parameter space (as the initial mode frequencies are now different). It is also unclear whether such secondaries will be more common than the type observed here; while they admit additional possible combinations of matched modes, their prevalence really depends on the measure of the set of approximate solutions to Eq.\,\eqref{eq:system}. Nevertheless, the $j\neq j'$ case (excluding strong sidebands) is generally less of a factor for matched-SNR searches because it does not cause secondaries that are very pronounced. For our considered injection, the expected overlap from a perfect alignment of the dominant and second-strongest modes is $\lesssim0.4$.

When large-scale SNR gradients in the posterior surface are taken into account, the relationship between the strength of a secondary (relative to the posterior tails) and its overlap against the injection is no longer straightforward. For example, consider two secondaries that correspond to nodes $\boldsymbol{\theta}_\mathrm{sec},\boldsymbol{\theta}_\mathrm{sec}'$ with the same injection-overlap value, but different optimal SNRs $(\rho_\mathrm{sec},\rho_\mathrm{sec}')=(\rho_\mathrm{inj},2\,\rho_\mathrm{inj})$. The difference in log-likelihood between $\boldsymbol{\theta}_\mathrm{sec}$ and a nearby tail region (where $\rho_\mathrm{opt}\approx\rho_\mathrm{inj}$ and $\Omega(h_\mathrm{inj},\cdot)\approx0$) is half that of the difference between $\boldsymbol{\theta}_\mathrm{sec}'$ and its nearby tails (where $\rho_\mathrm{opt}\approx2\,\rho_\mathrm{inj}$). In other words, low-overlap secondaries (be they from the $j=j'$ or $j\neq j'$ case) are more pronounced relative to the large-scale gradients if they occur in regions of high SNR. This fact becomes particularly relevant when extrinsic source parameters are added to the mix, since they have a larger effect on SNR relative to their measurement precision, and thus might lead to the (uneven) boosting of low-overlap secondaries within a typical search region.

\subsubsection{Why are secondaries non-Gaussian?}
\label{subsec:nongaussian}

Speaking from our own experience, it is natural to hold two related notions about the nature of EMRI degeneracy: i) posterior secondaries are unimodal peaks that can be cleanly localized; and ii) even congealed or deformed secondaries can still be characterized as ``Gaussian'' around a local maximum, with covariances given by the pullback metric at that point (i.e., the Fisher information matrix for a different posterior with that signal as the injection). This intuition is largely inherited from the behavior of the posterior bulk around the injection parameters---not just for EMRIs, but for GW sources in general. Our mapping analysis now provides empirical evidence that both notions are invalid, and here we give a simple theoretical argument to support this conclusion.

Recall from Sec.\,\ref{sec:tools} that the standard GW log-likelihood \eqref{eq:fullloglike} with $n=0$ is proportional to the squared Euclidean distance between signal injection and template in the data space $\mathcal{D}$ (with equipped inner product $\langle\cdot|\cdot\rangle$). For $\boldsymbol{\theta}$ near the global likelihood maximum $\boldsymbol{\theta}_\mathrm{inj}$ such that $|\delta\boldsymbol{\theta}_\mathrm{inj}|\ll|\boldsymbol{\theta}_\mathrm{inj}|$ with $\delta\boldsymbol{\theta}_\mathrm{inj}:=\boldsymbol{\theta}_\mathrm{inj}-\boldsymbol{\theta}$, we have
\begin{align}\label{eq:truepeak}
    \ln{L}(\boldsymbol{\theta})\propto&\;\langle h_\mathrm{inj}-h(\boldsymbol{\theta})|h_\mathrm{inj}-h(\boldsymbol{\theta})\rangle\nonumber\\
    =&\;\delta\boldsymbol{\theta}_\mathrm{inj}^T\,\mathcal{I}_\mathrm{inj}\,\delta\boldsymbol{\theta}_\mathrm{inj}+\mathcal{O}(|\delta\boldsymbol{\theta}_\mathrm{inj}|^3),
\end{align}
where the Fisher information $\mathcal{I}_\mathrm{inj}:=\mathcal{I}(\boldsymbol{\theta}_\mathrm{inj})$ from Eq.\,\eqref{eq:fisher} is symmetric and positive-definite. With an uninformative prior, the posterior can thus be approximated as Gaussian over a local neighborhood of $\boldsymbol{\theta}_\mathrm{inj}\in\Theta$.

An oft-overlooked subtlety is that Eq.\,\eqref{eq:truepeak} alone does not explain why GW posteriors can strongly resemble Gaussian distributions up to 1- or 2-$\sigma$. Over a sufficiently small region around a maximum point, the likelihood surface can be well described by only the leading-order piece of Eq.\,\eqref{eq:truepeak}; this does not guarantee that the associated lengthscales are smaller than the region itself, which is a requirement for the surface to ``look'' Gaussian rather than flat. The other requirement is that the higher-order central moments must be similar to the multivariate-normal values for the leading-order likelihood. In particular, the third-order moments that determine skewness must nearly vanish. This is indeed satisfied in typical GW posteriors for the intrinsic source parameters, since those moments scale approximately in size with the components of the third-order tensor $\langle\partial_{\boldsymbol{\theta}} h|\partial_{\boldsymbol{\theta}}^2h\rangle$, which are $\approx0$ for parameters that affect signal phasing.

Now consider a secondary likelihood maximum $\boldsymbol{\theta}_\mathrm{sec}$, with corresponding signal $h_\mathrm{sec}$. For $\boldsymbol{\theta}$ near $\boldsymbol{\theta}_\mathrm{sec}$ such that $|\delta\boldsymbol{\theta}_\mathrm{sec}|\ll|\boldsymbol{\theta}_\mathrm{sec}|$ with $\delta\boldsymbol{\theta}_\mathrm{sec}:=\boldsymbol{\theta}_\mathrm{sec}-\boldsymbol{\theta}$, we may write
\begin{align}\label{eq:secexpand}
    \ln{L}(\boldsymbol{\theta})\propto&\;\langle h_\mathrm{inj}-h(\boldsymbol{\theta})|h_\mathrm{inj}-h(\boldsymbol{\theta})\rangle\nonumber\\
    =&\;\langle h_\mathrm{inj}-h_\mathrm{sec}|h_\mathrm{inj}-h_\mathrm{sec}\rangle\nonumber\\
    &\;+2\langle h_\mathrm{inj}-h_\mathrm{sec}|h_\mathrm{sec}-h(\boldsymbol{\theta})\rangle\nonumber\\
    &\;+\langle h_\mathrm{sec}-h(\boldsymbol{\theta})|h_\mathrm{sec}-h(\boldsymbol{\theta})\rangle\\
    =&\;\delta\boldsymbol{\theta}_\mathrm{sec}^T\,\mathcal{I}_\mathrm{sec}\,\delta\boldsymbol{\theta}_\mathrm{sec}\nonumber\\
    &\;+\delta\boldsymbol{\theta}_\mathrm{sec}^T\langle h_\mathrm{inj}-h_\mathrm{sec}|H_\mathrm{sec}\rangle\delta\boldsymbol{\theta}_\mathrm{sec}\nonumber\\
    &\;+\mathcal{O}(|\delta\boldsymbol{\theta}_\mathrm{sec}|^3)+\mathrm{const.},\label{eq:secpeak}
\end{align}
where $\mathcal{I}_\mathrm{sec}:=\mathcal{I}(\boldsymbol{\theta}_\mathrm{sec})$ and $H_\mathrm{sec}:=\partial_{\boldsymbol{\theta}}^2h(\boldsymbol{\theta})|_{\boldsymbol{\theta}_\mathrm{sec}}$. Note that $H$ here denotes the Hessian tensor of the waveform model, not to be confused with the Hessian matrix of the leading-order likelihood (which is $-\mathcal{I}$). Also, the linear-in-$\delta\boldsymbol{\theta}_\mathrm{sec}$ term in Eq.\,\eqref{eq:secexpand} vanishes because
\begin{equation}\label{eq:linearterm}
    \langle h_\mathrm{inj}-h_\mathrm{sec}|\partial_{\boldsymbol{\theta}}h(\boldsymbol{\theta})\rangle|_{\boldsymbol{\theta}_\mathrm{sec}}\propto\partial_{\boldsymbol{\theta}}L(\boldsymbol{\theta})|_{\boldsymbol{\theta}_\mathrm{sec}}=0.
\end{equation}

The third-order term in Eq.\,\eqref{eq:secexpand} includes the usual contribution from $\langle\partial_{\boldsymbol{\theta}} h|\partial_{\boldsymbol{\theta}}^2h\rangle$, which is again $\approx0$ at $\boldsymbol{\theta}_\mathrm{sec}$, but picks up an additional piece that scales in size with the components of $\langle h_\mathrm{inj}-h_\mathrm{sec}|\partial_{\boldsymbol{\theta}}^3h\rangle\approx\langle h_\mathrm{inj}|\partial_{\boldsymbol{\theta}}^3h\rangle$. This quantity generally does not vanish since $h_\mathrm{inj}$ is not perfectly proportional to $h_\mathrm{sec}$; although its impact on the local deviation from Gaussianity is specific to the behavior of $h$ at $\boldsymbol{\theta}_\mathrm{sec}$, it does indicate that secondaries with higher injection overlaps will tend to look more Gaussian (as might be expected). Higher-order likelihood terms aside, the local covariance structure for a secondary is altered even at leading order, as seen from the second term in Eq.\,\eqref{eq:secpeak}. This quadratic form is not necessarily positive-definite, but let us say that the sum of the first two terms in Eq.\,\eqref{eq:secpeak} still is, and that all higher-order terms are negligible. The inverse covariance matrix is then not $\mathcal{I}_\mathrm{sec}$ as expected but $\mathcal{I}_\mathrm{sec}+\langle h_\mathrm{inj}-h_\mathrm{sec}|H_\mathrm{sec}\rangle$, which encodes the local embedding curvature of the signal manifold $\mathcal{S}$ through its dependence on $H_\mathrm{sec}$.

\subsubsection{How representative are these results?}
\label{subsec:representativeness}

As discussed at the start of Sec.\,\ref{sec:degeneracy}, a more extensive study of degeneracy is not really warranted at this stage, since any results obtained with existing tools will change quantitatively for the next generation of science-adequate EMRI waveform models---or even qualitatively, if the nature of signal space is severely altered by distinctive higher-order effects such as transient self-force resonances \cite{Flanagan:2010cd,Ruangsri:2013hra,Berry:2016bit,Speri:2021psr}. Thus it is important to question the robustness of our results to the expected differences in waveform models, and whether our conclusions are generally representative of the EMRI signal space. In particular, might systematic errors in PN (theoretical) or adiabatic-fitted (computational) evolution schemes introduce artificial degrees of degeneracy that would otherwise be non-existent in more accurate models? Here we give a heuristic argument for why this is unlikely.

Regardless of the underlying evolution, a sufficient condition for a strong secondary to arise (at low eccentricity and modest initial separation) is simply Eq.\,\eqref{eq:system} with $j=j'=2$ and $k\leq2$. We write this more concisely as
\begin{equation}\label{eq:azisystem}
    (\omega_\phi,\dot{\omega}_\phi,\ddot{\omega}_\phi)(t_0;\boldsymbol{\theta})\approx(\omega_\phi,\dot{\omega}_\phi,\ddot{\omega}_\phi)(t_0;\boldsymbol{\theta}_\mathrm{inj}).
\end{equation}
The time evolution of the azimuthal fundamental frequency $\omega_\phi$ (along with its radial and polar counterparts) over the inspiral is fully described by a trajectory of osculating geodesics $G(t):=(p(t)/M,e(t),\iota(t))$, whose governing equations are specific to the EMRI model in question. (Recall that the evolution of $\iota$ is only neglected in the AAK model.) In other words, we may decouple its explicit dependence on time: $\omega_\phi(t)=\omega_\phi(G(t))$. Both the instantaneous frequency $\omega_\phi(G)$ and the trajectory $G(t)$ depend explicitly on the intrinsic source parameters $\boldsymbol{\theta}$, but the former does so in a model-independent way (it is simply a characteristic of Kerr geodesic motion \cite{Schmidt:2002qk}). With such a description, we have
\begin{equation}\label{eq:omegadot}
    \dot{\omega}_\phi(t)=\partial\omega_\phi(G)\cdot\dot{G}(t),
\end{equation}
\begin{equation}\label{eq:omegaddot}
    \ddot{\omega}_\phi(t)=\dot{G}(t)^T\cdot\partial^2\omega_\phi(G)\cdot\dot{G}(t)+\partial\omega_\phi(G)\cdot\ddot{G}(t),
\end{equation}
where all dependence on $\boldsymbol{\theta}$ is hidden for compactness.

We may examine the ``approximate solution set'' for the underdetermined nonlinear system \eqref{eq:azisystem} by casting it as a optimization problem. Some shorthand notation is useful here: let $\omega_0(\boldsymbol{\theta}):=\omega_\phi(t_0,\boldsymbol{\theta})$ and $\delta\omega_0(\boldsymbol{\theta}):=|\omega_0(\boldsymbol{\theta})-\omega_0(\boldsymbol{\theta}_\mathrm{inj})|/\omega_0(\boldsymbol{\theta}_\mathrm{inj})$, with analogous notation for the time derivatives of $\omega_\phi$. (The set $\{\delta\omega_0,\delta\dot{\omega}_0,\delta\ddot{\omega}_0\}$ corresponds to the ordinate values of the green squares in Fig.\,\ref{fig:freqderivs}.) In essence, a secondary node is then simply a point that minimizes the objective function
\begin{equation}\label{eq:objective}
    f(\boldsymbol{\theta}):=|(\delta\omega_0,\delta\dot{\omega}_0,\delta\ddot{\omega}_0)(\boldsymbol{\theta})|,
\end{equation}
for $f$ below some unspecified threshold value.

\begin{figure}[!tbp]
\centering
\includegraphics[width=\columnwidth,trim={0 0 0 1cm},clip]{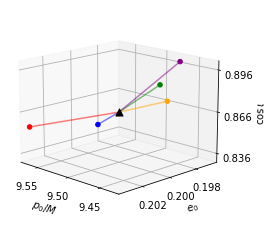}
\raggedright
\includegraphics[width=0.85\columnwidth,trim={0 1cm 0 1cm},clip]{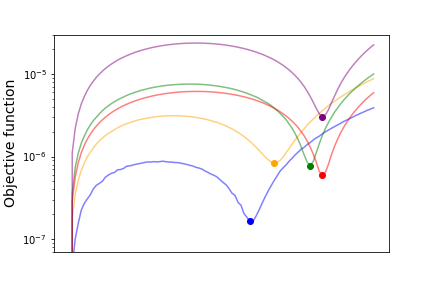}
\caption{Minimization of objective function \eqref{eq:objective} with constant $\dot{G}(\boldsymbol{\theta})$ and $\ddot{G}(\boldsymbol{\theta})$, over a five-dimensional cross section of the intermediate region $\mathcal{R}_1$ in Sec.\,\ref{subsec:10x} (setting $\lg{\mu}=1$). Around 50 randomly drawn starting points are used, with each leading to a different local minimum; only five of these are presented here for clarity. Top: Projection of minimum locations (colored points) onto the $(p_0/M,e_0,\cos{\iota})$ space, together with the lines connecting them to the injection parameters (black triangle). Bottom: Value of objective function along the (extended) connecting lines in the top panel.}
\label{fig:roots}
\end{figure}

The objective surface over parameter space will of course depend on the variation of $\dot{G}(t_0;\boldsymbol{\theta})$ and $\ddot{G}(t_0;\boldsymbol{\theta})$ through Eqs \eqref{eq:omegadot} and \eqref{eq:omegaddot}, but the trajectories themselves are qualitatively similar across different EMRI models---modulo the inclusion of transient resonances, which cannot be treated with Eq.\,\eqref{eq:azisystem} anyway. To demonstrate that degeneracy does not arise solely from using any specific trajectory model, let us consider the case where $\dot{G}(t_0;\boldsymbol{\theta})$ and $\ddot{G}(t_0;\boldsymbol{\theta})$ are artificially assigned fixed values over parameter space (say, their values at $\boldsymbol{\theta}_\mathrm{inj}$). The objective surface is then driven purely by the variation of $\omega_\phi$ and its partial derivatives with respect to $G$, but it is straightforward to verify that there still exist multiple distant minima in $f$ (see Fig.\,\ref{fig:roots}). This strongly indicates that degeneracy occurs under generic conditions, and is not merely an artifact of the adiabatic-fitted AAK trajectories used in our study.

While our conclusions on the nature of EMRI degeneracy should generalize to future waveform models, the single signal injection that we consider is not fully representative of injections in other regions of parameter space. Degeneracy is expected to be most severe at low eccentricity, since the prevalence of strong posterior secondaries should be reduced when higher harmonic modes have a larger contribution to the overall signal power, i.e., the injection becomes more distinctive with added degrees of freedom to fit. Another consideration is the effect of the analysis duration $T$. In regions of parameter space where the inspiral lifetime is significantly longer than $T$, an extension of the latter will also mitigate degeneracy (this is not applicable here since all sources plunge after around $T=2\,\mathrm{y}$). Finally, transient resonant ``jumps'' in the trajectories $G(t)$ will be a generically occurring feature of models with post-adiabatic evolution \cite{Flanagan:2010cd}, and thus of the true EMRI signal space; the presence of one such jump in the signal injection effectively shifts the evolution of its post-jump part onto a different set of phase trajectories. This again increases the complexity of the injection, and should reduce degeneracy as well.

\section{Implications for data analysis}
\label{sec:implications}

\subsection{Fundamental implications}
\label{subsec:general}

\subsubsection{Interaction with detector noise}
\label{subsec:noise}

With the qualitative nature of degeneracy established in Sec.\,\ref{subsec:mapping}, we now turn to various implications that our results pose for EMRI data analysis (both in a general sense, and in the context of the standing wisdom on search, inference, and modeling approaches). The first question we address is: How likely is it that detector noise will combine with the signal template at a secondary node to give a higher detection SNR than the template corresponding to the injection? Under the standard noise assumptions laid out in Sec.\,\ref{subsec:measures}, the answer is straightforward---a false determination of the best-fit template due to noise is extremely improbable.

As in Eq.\,\eqref{eq:fullinner}, let the data be $x=h_\mathrm{inj}+n$, and further assume that $n$ is a zero-mean and stationary Gaussian process. From Eqs \eqref{eq:noisemean}, \eqref{eq:noisevar} and \eqref{eq:detsnr}, the detection SNR $\rho_1$ of the injection template $h_1:=h_\mathrm{inj}$ is normally distributed with mean $\rho_\mathrm{inj}$ and unit variance. Consider some secondary template $h_2$ with an injection overlap of $\Omega$; its detection SNR $\rho_2$ is also normally distributed with mean $\Omega\,\rho_\mathrm{inj}$ and unit variance, while the correlation coefficient (normalized covariance) of $\rho_1$ and $\rho_2$ is $\Omega$. The probability distribution of $\rho_2-\rho_1$ is then $\mathcal{N}(\rho_\mathrm{inj}(\Omega-1),2(1-\Omega))$, and the probability that $\rho_2\geq\rho_1$ is given by
\begin{equation}\label{eq:noisesec}
\mathrm{P}(\rho_2\geq\rho_1)=\frac{1}{2}\left(1-\mathrm{erf}\left(\frac{1}{2}\,\rho_\mathrm{inj}\sqrt{1-\Omega}\right)\right).
\end{equation}
Thus for threshold injections with $\rho_\mathrm{inj}=20$, noise will lead to false determination of the injection parameters $\gtrsim1\%$ of the time only if the offending secondary has an injection overlap of $\gtrsim0.97$. For the highest-overlap secondary identified in the analysis regions of Sec.\,\ref{subsec:mapping} ($\Omega=0.76$), the probability of this occurring is $\sim10^{-12}$.

\subsubsection{Interaction with multiple signals}
\label{subsec:multiple}

The degeneracy study of Sec.\,\ref{sec:degeneracy} focuses on non-local signal templates that strongly resemble a single signal injection. We now use analytic arguments to extend our results to the scenario where templates have non-negligible and non-local correlations with the sum of two injections (which themselves are non-local to each other). Let us restrict to the case where the injections are uncorrelated, since i) two injections with non-negligible overlap have a sum that lies approximately in the span of either, so they reduce effectively to the case of a single injection; and ii) injections with near-zero overlap are far more typical anyway, as shown by the confusion study of Sec.\,\ref{sec:confusion}. Our analysis remains well motivated in the uncorrelated-injection case, since it is quite plausible for some template to match the dominant mode in one injection with its own strongest mode, while matching the dominant mode in the other with its second strongest mode.

To pose a more precise question, let us consider two uncorrelated injections $h_{1,2}$ with optimal SNRs $\rho_{1,2}$, where $\rho_1\geq\rho_2$ without loss of generality. Then we ask: What is the probability that some non-local template $h(\boldsymbol{\theta})$ (i.e., a template that is not local to either $h_1$ or $h_2$) has a high enough overlap with $h_1+h_2$ such that its (expected) detection SNR is $\rho_{\boldsymbol{\theta}}\geq\rho_1$? If this scenario turned out to be even moderately likely, then clearly one would have to design search strategies with the possibility in mind. However, the order-of-magnitude argument we sketch below strongly suggests that the probability of such an occurrence is negligible.

Denoting the individual injection--template overlaps by $\Omega_1(\boldsymbol{\theta}):=\Omega(h_1,h(\boldsymbol{\theta}))$ and $\Omega_2(\boldsymbol{\theta}):=\Omega(h_2,h(\boldsymbol{\theta}))$, the detection SNR of $h(\boldsymbol{\theta})$ given the sum of injections $h_1+h_2$ may be written as $\rho_{\boldsymbol{\theta}}=\Omega_1(\boldsymbol{\theta})\rho_1+\Omega_2(\boldsymbol{\theta})\rho_2$. Our calculation relies on some assumptions about the functions $\Omega_{1,2}(\boldsymbol{\theta})$ over parameter space, which for random $\boldsymbol{\theta}$ may be approximated as two independent and identically distributed random variables $\Omega_{1,2}$. Based on the overall population of identified secondaries in Sec.\,\ref{subsec:mapping}, we will assume that $\Omega_{1,2}\leq\Omega_\mathrm{max}\approx0.8$ (beyond the local neighborhoods of $\boldsymbol{\theta}_{1,2})$. For a lower bound on $\Omega_{1,2}$, note first that our formal definition of detection SNR in Eq.\,\eqref{eq:detsnr} admits negative values; implicitly, we are treating any template as being fully specified by its parameters with no freedom for maximization over phase (e.g., the translation of a template in time is viewed as a distinct template corresponding to a different point in parameter space). We then consider only $\Omega_{1,2}\geq\Omega_\mathrm{min}\approx0$ in this calculation, which is conservative as $\rho_{\boldsymbol{\theta}}$ is constrained to be positive, and thus the probability $\mathrm{P}(\rho_{\boldsymbol{\theta}}\geq\rho_1)$ is overestimated. This streamlines our argument, and connects it more naturally to results obtained with the approximate inner product (which is also positive by definition).

Another important assumption is to specify the probability distribution of the random variables $\Omega$. Our mapping analysis does not directly provide an empirical determination of the probability density function $p(\Omega)$, but it enables us to posit a rough functional form for $p$. For $\Omega_\mathrm{min}\leq\Omega\leq\Omega_\mathrm{max}$, we write
\begin{equation}\label{eq:overlappdf}
    p(\Omega):=-\frac{1}{V_0}\frac{dV}{d\Omega},
\end{equation}
where $V(\Omega)$ is the volume of the set of points in $\Theta$ with injection-overlap values that are $\geq\Omega$, and $V_0:=V(\Omega_\mathrm{min})$. Since $\Omega_\mathrm{min}\approx0$, $V_0$ is of a similar order of magnitude to the volume of the full (six-dimensional) parameter space $\Theta$. The function $V$ is defined with respect to some natural measure on $\Theta$; we will choose a measure such that the volume associated with a typical posterior secondary is $\sim1$, give or take a few orders of magnitude (to account for the diversity in the extents of secondaries with different injection overlaps). This allows us to make the simplification that $V(\Omega)$ is given by the number of secondaries with injection-overlap values $\geq\Omega$, so that $dV/d\Omega$ can be estimated from our mapping analysis. With such a measure, the volume of the posterior-bulk region examined in Sec.\,\ref{subsec:1x} is also $\sim1$, and thus $V_0\sim10^{18}$ from the discussion after Eq.\,\eqref{eq:priorextents}.

\begin{figure}[!tbp]
\centering
\includegraphics[width=\columnwidth]{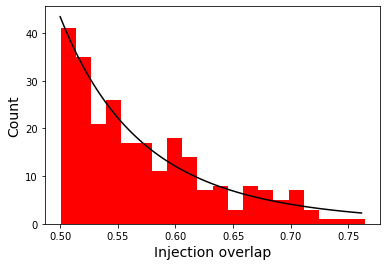}
\caption{Histogram of injection-overlap values for 251 signals (with overlaps $>0.5$) from inset of Fig.\,\ref{fig:100x}. The probability density for the histogram distribution is well fit by the power law \eqref{eq:powerlaw} with $\alpha\approx7$, and is robust to the choice of bin size.}
\label{fig:powerlaw}
\end{figure}

The above simplification is clearly invalid for $\Omega\gtrsim0.75$ (where there are no secondaries), while selection effects in both sampling and clustering also lead to vastly lower counts for secondaries with overlaps $\lesssim0.5$ (by design). Nevertheless, our data for $0.5\lesssim\Omega\lesssim0.75$ shows a clear trend for $dV/d\Omega$, which we will take to hold for $\Omega_\mathrm{min}\leq\Omega\leq\Omega_\mathrm{max}$. Let us assume that $dV/d\Omega$ is described by a power law in the latter range, and write
\begin{equation}\label{eq:powerlaw}
    \frac{dV}{d\Omega}:=-C\,\Omega^{-\alpha}.
\end{equation}
The combined set of 705 secondaries from the analyses in Secs \ref{subsec:10x} and \ref{subsec:100x} (the colored points in the inset of Fig.\,\ref{fig:100x}) is distilled to a set of 251 secondaries with overlaps $>0.5$; the distribution of this smaller set is fit well by Eq.\,\eqref{eq:powerlaw} with $C\approx26$ and $\alpha\approx7$ (see Fig.\,\ref{fig:powerlaw}). As it turns out, the results from the confusion study suggest that this power law is a reasonable approximation over the range $\Omega_\mathrm{min}\leq\Omega\leq\Omega_\mathrm{max}$. Assuming the root-mean-square pairwise overlap of $\sim10^{-3}$ from that study is indicative of the injection overlap for the majority of points in the full space, we might expect that $V(10^{-3})$ should be of a similar order of magnitude to $V_0\sim10^{18}$. This is indeed the case: for Eq.\,\eqref{eq:powerlaw} with $C=26$ and $\alpha=7$, we find $V(10^{-3})\approx4\times10^{18}$. Conversely, we also require $\Omega_\mathrm{min}\sim10^{-3}$ such that the probability density \eqref{eq:overlappdf} is properly normalized on the interval $[\Omega_\mathrm{min},\Omega_\mathrm{max}]$ (its integral depends negligibly on $\Omega_\mathrm{max}$).

Now we define $\beta:=\Omega_1+r\Omega_2$, where $r:=\rho_2/\rho_1\leq1$; this random variable takes values in the interval $[0,2]$, and $\beta\geq1$ corresponds to the case of interest $\rho_{\boldsymbol{\theta}}\geq\rho_1$. Since $\Omega_1$ and $\Omega_2$ are i.i.d.\! random variables, we have
\begin{align}
p(\beta)=&\int_{\Omega_\mathrm{min}}^{\Omega_\mathrm{max}}\int_{\Omega_\mathrm{min}}^{\Omega_\mathrm{max}}d\Omega_1\,d\Omega_2\,\frac{C^2}{V_0^2}\Omega_1^{-\alpha}\Omega_2^{-\alpha}\nonumber\\
&\times\delta(\beta-\Omega_1-r\Omega_2)\nonumber\\
=&\;\frac{C^2}{V_0^2}r^{\alpha-1}\int_{\Omega_\mathrm{min}}^{\Omega_\mathrm{max}}d\Omega\,(\beta-\Omega)^{-\alpha}\Omega^{-\alpha}\nonumber\\
&\times\vartheta(\beta-\Omega-r\Omega_\mathrm{min})\,\vartheta(r\Omega_\mathrm{max}+\Omega-\beta),\label{eq:betapdf}
\end{align}
where $\delta$ and $\vartheta$ denote the Dirac-delta and Heaviside-theta functions respectively. The most conservative (largest) estimate of the probability $\mathrm{P}(\beta\geq1)$ occurs for the equal-SNR case $r=1$; with $V_0=10^{18}$, $C=26$, $\alpha=7$, $\Omega_\mathrm{min}=10^{-3}$ and $\Omega_\mathrm{max}=0.8$, we find $\mathrm{P}(\beta\geq1)\sim10^{-30}$. 

\begin{figure}[!tbp]
\centering
\includegraphics[width=\columnwidth]{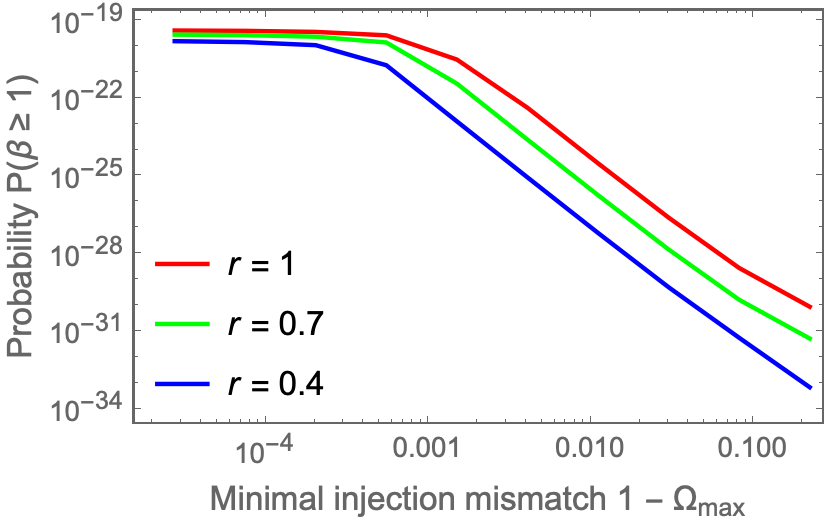}
\caption{Probability that the sum of two injections yields a non-local template with a higher detection SNR than the optimal SNR of either injection (i.e., the integral of Eq.\,\eqref{eq:betapdf} over the interval $[1,2]$). This is plotted as a function of the minimal injection--template mismatch $1-\Omega_\mathrm{max}$ across the signal space, for several values of the injection-SNR ratio $r$.}
\label{fig:omegamax}
\end{figure}

To what extent does the above estimate for the probability $\mathrm{P}(\beta\geq1)$ depend on both the maximal injection overlap $\Omega_\mathrm{max}$ and the SNR ratio $r$? In Fig.\,\ref{fig:omegamax}, $\mathrm{P}(\beta\geq1)$ is plotted as a function of the (minimal) mismatch $1-\Omega_\mathrm{max}$ for $0.8\leq\Omega_\mathrm{max}<1$ and $r\in\{1,0.7,0.4\}$. When $\Omega_\mathrm{max}<1-10^{-3}$, the probability scales approximately as $r^6$ (in agreement with Eq.\,\eqref{eq:betapdf}) and $(1-\Omega_\mathrm{max})^{-4.6}$. As $\Omega_\mathrm{max}\to1$, it asymptotes instead to a maximal value of $\approx4\times10^{-20}$ for $r=1$. In other words, even if we assume that the power law \eqref{eq:powerlaw} is valid up to $\Omega_\mathrm{max}=1$, the probability that $\beta\geq1$ is still completely negligible. From Eqs \eqref{eq:overlappdf} and \eqref{eq:powerlaw}, the maximal probability value corresponds approximately to $\mathrm{P}(\Omega\geq0.998)$ for a single injection--template overlap; essentially, if there exists a template closely resembling $h_1$ with a mismatch of $\lesssim10^{-3}$, virtually any $h_2$ can push $\beta$ over unity.

We started this analysis by considering in isolation the sum of two signal injections. For a set of $N$ injections, there are $N(N-1)/2$ such sums. The random variables $\beta_i$ associated with these sums are of course correlated, but it is conservative to treat them as i.i.d.\! in this context. A plausible middle-ground estimate for the number of resolvable EMRI signals that might be present in the LISA data stream is $N=200$, as considered in Sec.\,\ref{sec:confusion}. For such a set of signals, the probability that any $\beta_i\geq1$ is increased only by a factor of $\sim10^4$ (to $\sim10^{-26}$, if we take $r=1$ and $\Omega_\mathrm{max}=0.8$). Thus it is highly improbable for the sum of any two signals in the data to mimic a non-existent third signal (in the continuum of possible signals) so well that the best-fit template actually corresponds to the latter, rather than either of the two actual signals.

What about the possibility of three actual signals coincidentally summing to mimic a non-existent fourth signal? Intuition indicates that such an occurrence is even more improbable than the two-signal case, although an extension of the above analysis to three signals is somewhat non-trivial. Nevertheless, we may construct a back-of-the-envelope argument by restricting to specific sub-cases of the two scenarios. For a small set of $M$ injections $h_j$ with the same optimal SNR $\rho$, consider the probability that the overlap $\Omega_j$ of each injection with a given template $h(\boldsymbol{\theta})$ is $\geq1/M$, which is a sufficient condition for $\rho_{\boldsymbol{\theta}}=\rho\sum_j\Omega_j\geq\rho$. From Eqs \eqref{eq:overlappdf} and \eqref{eq:powerlaw}, we have
\begin{equation}\label{eq:specialcase}
    \mathrm{P}\left(\Omega\geq\frac{1}{M}\right)^M=\left(\frac{C(M^{\alpha-1}-1)}{V_0(\alpha-1)}\right)^M.
\end{equation}
With $V_0=10^{18}$, $C=26$ and $\alpha=7$ as before, we find a probability of $\sim10^{-31}$ for $M=2$; this is just below the estimate of $\mathrm{P}(\beta\geq1)\sim10^{-30}$ for the more general treatment with $r=1$ (as expected).

Now, for a set of $N=200$ actual signals in the data, we may estimate how likely it is that any $M\leq N$ signals fulfill the above sufficient condition, and then examine the ratio of probabilities between the two representative cases $M=3$ and $M=2$. This ratio evaluates to
\begin{equation}\label{eq:probratio}
    \frac{{N\choose3}\mathrm{P}\left(\Omega\geq\frac{1}{3}\right)^3}{{N\choose2}\mathrm{P}\left(\Omega\geq\frac{1}{2}\right)^2}\approx(3\times10^7)V_0^{-1}=3\times10^{-11},
\end{equation}
indicating that $\mathrm{P}(\rho_{\boldsymbol{\theta}}>\rho)$ in the case of three signals is infinitesimal relative to its value in the case of two signals (which we have already established as practically negligible). More generally, the ratio of probabilities between the case of any $M\leq N$ and that of $M=2$ scales as $\gamma V_0^{2-M}$, where $\lg{\gamma}$ is only slightly super-linear with $M$ and has a much smaller absolute gradient than $-18$ throughout (i.e., the coefficient of $M$ in the common logarithm of the other factor). Thus the logarithm of the ratio decreases monotonically with $M$, which convincingly rules out the likelihood of any number of actual signals coincidentally summing to mimic a non-existent signal. In summary, we conclude that the interaction of degeneracy with multiple signals is unlikely to pose any fundamental difficulties for the extraction and characterization of resolvable EMRI signals in LISA data.

\subsection{Search strategies}
\label{subsec:search}

The search for signals in GW data really involves two distinct tasks: detection, which establishes the presence of a signal; and what we shall refer to as ``identification'', which maps the detected signal to astrophysically relevant source parameters (approximately or otherwise). Detection technically includes the assessment of statistical significance for candidate signals, but may be treated in principle as simply the attainment of some threshold value for the detection SNR. Identification is the step that connects to inference; it poses less of a separate issue when the parameters have an easily invertible map to the waveform observables (as is the case for Galactic binaries \cite{Littenberg:2020bxy}), or when standard inference techniques can already cover the entire space of possible signals (as is the case for massive-black-hole mergers \cite{Katz:2020hku,Marsat:2020rtl}). For these sources and those of ground-based observing, the task of identification is either unnecessary or trivial, leading to the familiar dichotomy of detection and inference (the latter is traditionally known as ``parameter estimation'').

In the case of EMRIs, detection is relatively straightforward. The presence of a signal can be uncovered with crude waveform models, with completely phenomenological ones, or even without models at all (see Sec.\,\ref{subsec:emrispecific} for a short overview of such approaches). Parameter degeneracy may give rise to spurious candidate signals, but does not affect detection of any actual signal itself. On the other hand, identification becomes non-trivial since degeneracy endows the inverse problem with a highly disjoint set of near-solutions. It remains unclear whether identification will eventually be necessary as an independent step, since a sophisticated sampling algorithm that is well tuned to the EMRI problem might conceivably allow inference with a global prior; however, this has not been demonstrated before, much less in a reliable fashion. If it proves impractical to move directly from detection to inference, identification will be required to provide a good initial guess or to narrow down the prior region. Approximate waveform models might be used in rough searches to supply such information, but as shown in Sec.\,\ref{subsec:representativeness}, even a weakly physical model can still admit a multitude of near-solutions to its inversion. If the model is overly simplistic, it might succeed in smoothing out the search surface, but then may not provide sufficiently precise and/or accurate prior localization for inference.

The search for EMRIs is further complicated by the potential presence of more than one such signal in the data. Transdimensional methods \cite{green1995reversible}, as used in Galactic-binary searches to find sets of signals with unspecified cardinality, are less viable for EMRIs due to degeneracy and the greater expense of waveform models. They are also not as crucial in this context, since an astrophysically realistic set of EMRI signals is far less likely to ``self-confuse'' (as shown in Sec.\,\ref{sec:confusion}). A more direct approach for EMRIs would be to conduct a single-source search for the global best-fit template, corresponding to the strongest signal in the data. After performing precise inference with accurate models, this signal is subtracted from the data, which may then be searched for the second-strongest signal. (The same approach might be taken for massive-black-hole-merger signals, since these are expected to be more resolvable from one another in the time domain.) However, the search for the strongest signal is still hindered by the presence of a second strong signal, which would manifest in the search surface as another set of apparent secondaries (but see Sec.\,\ref{sec:suggestions}).

\subsubsection{General sampling algorithms}
\label{subsec:generalsampling}

We now touch broadly on the implications of our results for the two main classes of sampling algorithms that will be important in LISA signal searches. The sampling of source posterior distributions with these algorithms is of course a key component of inference as well, but we may view any non-local aspect of inference as being part of the search (identification) stage. Most forms of modeled search rely on the global exploration of some search surface---not necessarily the posterior---that is defined in terms of cross-correlations (Eq.\,\eqref{eq:detsnr}) or residuals (Eq.\,\eqref{eq:mle}) between data and template. As is well known, stochastic sampling is required to explore such a surface for EMRIs, even without accounting for the presence of degeneracy. This is due to the information volume of the signal space, which rules out grid sampling with pre-computed template banks \cite{Gair:2004iv}. Let us focus on the case of a single-source search for the global probability maximum, since any difficulties encountered there are only going to be amplified for strategies that look for multiple candidates at the same time (e.g., transdimensional sampling, or the online clustering of posterior samples).

Parallel-tempering MCMC methods \cite{2005PThPS.157..317W} utilize multiple Markov chains of ``walkers'' to explore variants of the likelihood at different temperatures $T_i$ (essentially $L^{1/T_i}$), doing so in parallel and with the exchange of information. Such algorithms allow walkers to move more efficiently across the search region than in traditional MCMC, and thus aid sampling convergence for complex distributions. They are well suited to coping with the ``noisy'' tails of GW likelihoods \cite{Littenberg:2009bm,2016MNRAS.455.1919V}, and should also be useful for posterior-based EMRI search. In practice, the strength and prevalence of EMRI degeneracy, coupled with the miniscule extent of the posterior bulk relative to the global prior, is likely to necessitate more walkers and/or temperatures than usual. As a rough indicator of the sort of numbers required for the EMRI problem, we report that for the prior region $\mathcal{R}_2$ in Sec.\,\ref{subsec:100x}, the posterior bulk cannot be found within $10^6$ iterations using a standard geometric temperature ladder \cite{2005PCCP....7.3910E} (with $T_\mathrm{max}=T_{10}=\rho_\mathrm{inj}^2=400$), 10 walkers per temperature, and an assortment of jump proposals \cite{justin_ellis_2017_1037579}. A fine-tuning of the ladder might be informed by exploratory studies to determine the pertinent temperature scales in the EMRI likelihood, and to do so for a representative range of injections across the full parameter space.

Nested sampling \cite{skilling2006nested} is a more recent paradigm for Bayesian posterior sampling and evidence calculations. It involves the iterative replacement of a set of ``live'' points with new points of higher likelihood, which provides an exponential contraction of the prior to the posterior. In the context of EMRI search, a very large number of live points is needed for sufficient resolution to chance upon the region containing the posterior bulk---which can easily be $<10^{-12}$ times the size of the prior region, even for a modest degree of prior localization. This might be computationally prohibitive, and furthermore it is not clear how to incorporate problem-specific tuning in the proposal of new live points (the real engine under the hood of nested sampling, which only assumes that such points can be generated). Again, to indicate the required sampling resolution for the prior region $\mathcal{R}_2$ in Sec.\,\ref{subsec:100x}, a slice-nested-sampling implementation \cite{2015MNRAS.453.4384H} with $10^4$ live points fails to converge on the posterior bulk for the (considerably smoother) likelihood $L_\mathrm{match}$ in Eq.\,\eqref{eq:matchloglike}.

\subsubsection{EMRI-specific methods}
\label{subsec:emrispecific}

A small number of modeled-search methods for EMRIs have been proposed in the literature. All of these were developed using the earliest kludge model \cite{Barack:2003fp}, which supplies both the injection and the templates in each study. Nevertheless, the basic ideas that underpin such methods will apply to the search for realistic signals as well, provided that models with sufficiently accurate frequency evolution are used in the analysis. Both tasks of detection and identification are addressed simultaneously in modeled search, since each candidate signal is associated with some high-detection-SNR template from the model, which in turn has ``known'' parameters of interest (that must be specified in order to generate the template).

Semi-coherent search is a broadly applicable strategy that is similar in effect to annealing-type sampling methods. As sketched out for EMRIs in \cite{Gair:2004iv}, it is a modeled search with ``flexible'' templates that are phase-matched against $N$ segments of the data with duration $T/N$, rather than the full data stream over its duration $T$. Note however that there are several other (mis)conceptions of semi-coherent EMRI search---none of which have been described properly in the literature, if at all. The general idea of semi-coherent filtering originates from continuous-wave searches in ground-based observing \cite{Brady:1998nj,Cutler:2005pn}. In the context of a pre-computed template bank, it increases the ``reach'' of each template so that the density of the bank can be reduced. When used in stochastic searches (as required for EMRIs), it works essentially by smoothing out the search surface, making high-probability regions more extensive and easier to find. In terms of its interaction with degeneracy, the semi-coherent approach will likely congeal secondaries and thus be useful for prior localization; the question is whether this localization is sufficient to transition directly into inference, and if not, then how many iterations of semi-coherent search (with gradually increased segment durations) might be required.

Another class of method that is more specific to EMRIs exploits the fact that any point $\boldsymbol{\theta}_\mathrm{sec}$ with high injection overlap generally satisfies Eq.\,\eqref{eq:system} for some $\{j,j',k\}$, which can thus be used to inform MCMC jump proposals. This is a central idea in \cite{Cornish:2008zd}, where the inversion of Eq.\,\eqref{eq:system} with $\boldsymbol{\theta}_\mathrm{inj}\to\boldsymbol{\theta}_\mathrm{sec}$ (more precisely, the minimization of the associated objective function, as in Eq.\,\eqref{eq:objective}) is proposed for combinations of the azimuthal frequency $\omega_\phi$ and the Lense--Thirring frequency $\omega_\phi-\omega_\theta$, along with their time derivatives up to second order. Only $j\neq j'$ is considered in the original proposal, which effectively puts the focus on jumping between templates where the dominant mode in one is aligned with a sideband of the dominant mode in the other; it is straightfoward to extend this to include the $j=j'$ case as well, in light of our results that highlight its importance. Finding all minima in the objective function is no less intractable than finding all posterior maxima, of course, and so only a single ``root'' is identified per jump---however, this may be of limited benefit if secondaries are extremely numerous. It is shown in \cite{Cornish:2008zd} that such jumps help with localization to a broad region in parameter space, but it remains unclear whether they are sufficient on their own to cope with the denser set of secondaries due to intrinsic degeneracy, when homing in on the posterior bulk.

In \cite{Babak:2009ua}, similarly constrained MCMC jump proposals are integrated with ideas from the semi-coherent approach in a search pipeline. Shorter half-year segments of the data are analyzed with either simplified or fully phenomenological templates in a number of separate analysis stages, but a key feature of the method is the inclusion of jumps that are informed by fixing the three fundamental frequencies (and not their derivatives), i.e., the inversion of Eq.\,\eqref{eq:system} with $j\in\{r,\theta,\phi\}$ and $j=j'$.
Segment information is combined by tracing the evolution of candidate signals in different segments to the same reference time $t_0$, and the overall list of candidates is then used to seed a localized search with full-duration templates. The entire method is stated to involve some degree of manual intervention and tuning. As in the case of \cite{Cornish:2008zd}, it is demonstrated on data sets with high-SNR injections as part of the Mock LISA Data Challenges \cite{MockLISADataChallengeTaskForce:2009wir}; the final parameters reported in \cite{Babak:2009ua} do not seem to agree with most of the injection parameters within their expected uncertainties (this is only an estimation on our part, since no inference results are presented). Nevertheless, the method appears to provide solid localization for follow-up analysis.

Apart from approaches that use astrophysically parametrized EMRI models, there is also the option of performing minimally modeled or unmodeled searches. One such proposal \cite{Wang:2012xh} features a phenomenological model that is parametrized directly by the observable quantities in the signal: the dominant mode amplitudes (treated as constant), the three initial fundamental frequencies, and their initial time derivatives to third order. This approach could be useful for search since the model is extremely efficient, and can overfit physical signals such that detection sensitivity is not impaired. Its main difficulty lies in mapping the phenomenological quantities to the source parameters for identification, which ultimately still requires the inversion of a physical model. Another proposed strategy is to directly search for EMRI-like frequency tracks in spectrograms of the data, using edge-detection algorithms \cite{Gair:2008ec}. Such excess-power searches are more limited for EMRIs, due to their low instantaneous SNR and their place in the SNR hierarchy of the global LISA catalog. They could be feasible for the detection of high-SNR sources, but again would have to rely on physical models for identification, and thus remain susceptible to the issue of degeneracy as well.

\subsection{Inference (and modeling) strategies}
\label{subsec:inference}

By partitioning the EMRI data-analysis problem into non-local search and local inference, all that is required in the latter stage for each candidate signal is to map out the vicinity of its source parameters in a localized prior region. Standard posterior sampling, as well as the multitude of techniques proposed for its acceleration, should be adequate for this purpose (but may still be less efficient if the prior region is too large). A high detection SNR for the candidate is insufficient to provide confidence that the examined parameter point corresponds to the parameters of some physical signal, but the nature of EMRI degeneracy actually works to our benefit here. One of the key conclusions from Sec.\,\ref{sec:degeneracy} is that secondary posterior peaks are expected to feature pathologies that are absent in the posterior bulk, which could be a simple way of vetoing such solutions (see Sec.\,\ref{subsec:vetoes}).

However, we have up till now assumed a fitting factor \cite{apostolatos1995search} of unity between the injection and the template manifold, i.e., $h_\mathrm{inj}\in\mathcal{S}$ such that $F:=\max_{\boldsymbol{\theta}}\Omega(h_\mathrm{inj},h(\boldsymbol{\theta}))=1$. This ensures that the posterior is well behaved near the best-fit point $\mathrm{argmax}_{\boldsymbol{\theta}}\,\Omega(h_\mathrm{inj},h(\boldsymbol{\theta}))$, but it may not be achievable in practice even with future post-adiabatic waveforms. A low fitting factor can impact on search through the reduction of expected detection SNRs, and might pose a problem for inference even in the case where $F\gtrsim0.5$. The latter is evident from the posterior around secondary nodes with injection overlaps as high as $\approx0.8$ (indicative of putative models where those nodes are the best-fit points)---and also from the results in \cite{Katz:2021yft}, where the parameters of an injected signal with fully relativistic mode amplitudes are inferred from a template model with identical phasing but semi-relativistic amplitudes (leading to a best-fit overlap of $\approx0.5$ and a multi-modal posterior). To complicate matters, $F$ alone contains no information about the template manifold away from the best-fit point, and so is insufficient as a determinant of whether and how much the posterior is deformed. Assuming the source parameters are still correctly and sufficiently localized by the search stage, $F<1$ does not significantly challenge posterior estimation during inference, but may of course lead to an unacceptable degree of bias for precision applications of EMRI observations.

The well-understood need for low or manageable bias in high-precision inference \cite{Cutler:2007mi}, as well as the strong dependence of EMRI-specific search strategies on models with the correct frequency evolution, will impose constraints on modeling accuracy that have yet to be determined. Degeneracy does not impact directly on accuracy requirements, but is still relevant through its influence on data-analysis approaches. Studies on modeling accuracy are required at a level beyond simple dephasing arguments or fitting-factor calculations, and are presently being undertaken as part of the LISA Science Group's work-package activities. In terms of modeling strategies, degeneracy would appear to severely handicap approaches such as the construction of reduced-order-modeling surrogates \cite{Field:2011mf,Field:2013cfa,Rifat:2019ltp} in the time or frequency domain (at least in the global sense). These waveform-level fits are already unlikely to be practical for EMRIs due to the length and complexity of signals, and the added severity of non-local correlations seems to rule out their viability beyond highly localized regions in parameter space. The data compression provided by reduced-order modeling itself is still useful in EMRI modeling, however; as demonstrated in \cite{Chua:2020stf}, it only needs to be applied to the set of instantaneous mode amplitudes over the space of Kerr geodesics.

\section{Suggestions for data analysis}
\label{sec:suggestions}

\subsection{Simple post-hoc vetoes}
\label{subsec:vetoes}

Notwithstanding the efficacy of search strategies, the main concern for EMRI search is that secondaries of actual signals might be falsely identified as candidate signals. There are two specific sub-scenarios to consider: i) the actual signals themselves are found, and ii) they are not. The former implies the existence of a set of $\geq2$ candidate signals; a simple check would then be to compute the pairwise overlaps among all of them (effectively, to cluster the candidates using the direct overlap as a measure of connectivity, as mentioned in Sec.\,\ref{subsec:clustering}). From the results of our confusion study in Sec.\,\ref{sec:confusion}, we do not expect $\gtrsim1\%$ overlap between any two candidates. If this occurs, the pair should be flagged as a possible manifestation of parameter degeneracy for follow-up analysis.

In the latter sub-scenario (which could arise after the above veto, or for a single incorrectly identified candidate), there exists a set of $\geq1$ candidate signals that are virtually uncorrelated with one another. If any of these candidates is a secondary for some actual signal that was not found due to search error, this should immediately be evident when attempting to estimate the posterior in the vicinity of its associated parameters. The obtained posterior is expected to be highly non-Gaussian and difficult to localize; further checks could involve comparison to either a Fisher-matrix analysis, or to a simulated posterior with the spurious candidate as an injection.

\subsection{Veto ``likelihood'' function}
\label{subsec:modlike}

Annealing-type sampling methods such as parallel-tempering MCMC involve exploring the likelihood at different temperatures ($L^{1/T_i}$) to efficiently converge on the posterior bulk. However, the relative log-likelihood gradients near the injection parameters and near some secondary node are unaltered by annealing, i.e., the ratio between $\ln{L}(\boldsymbol{\theta}_\mathrm{inj})-\tau(\boldsymbol{\theta}_\mathrm{inj})$ and $\ln{L}(\boldsymbol{\theta}_\mathrm{sec})-\tau(\boldsymbol{\theta}_\mathrm{sec})$ for local tail values $\tau(\boldsymbol{\theta}):=-\rho_{\boldsymbol{\theta}}^2$ (see Eq.\,\eqref{eq:apploglike2} and related discussion) is preserved for all $T_i$. This means that secondary likelihood peaks remain present at all temperatures, and a large number of high-overlap ones can still cause problems in practice. Here we suggest a modified ``likelihood'' that suppresses secondary peaks by accounting for the expected spread of individual-mode detection SNRs, in order to provide improved identification in the search stage. Specifically, we define a sampling density $L'(\boldsymbol{\theta})$ such that for a typical secondary node $\boldsymbol{\theta}_\mathrm{sec}$, the difference $\ln{L'}(\boldsymbol{\theta}_\mathrm{sec})-\tau(\boldsymbol{\theta}_\mathrm{sec})$ is much reduced from its original counterpart $\ln{L}(\boldsymbol{\theta}_\mathrm{sec})-\tau(\boldsymbol{\theta}_\mathrm{sec})$. We also demand that $L'(\boldsymbol{\theta}_\mathrm{inj})\approx L(\boldsymbol{\theta}_\mathrm{inj})$ in the vicinity of $\boldsymbol{\theta}_\mathrm{inj}$, so as to facilitate a smooth transition into inference. More precisely, the volume of the corresponding ``posterior bulk'' should neither be much larger (to give good prior localization), nor much smaller (such that there is no need to compensate with increased sampling resolution).

We begin with a generic angular and frequency-based decomposition of any EMRI waveform into a small number $M(\boldsymbol{\theta})$ of strong harmonic modes:
\begin{equation}\label{eq:decomp}
    h(\boldsymbol{\theta})\approx\sum_{m=1}^{M(\boldsymbol{\theta})}h_m(\boldsymbol{\theta}),
\end{equation}
where $M(\boldsymbol{\theta})$ varies slowly over parameter space and really only depends strongly on the orbital eccentricity. For any given analysis, we may fix $M(\boldsymbol{\theta})$ to some maximal value $M$ in practice. Throughout most of this work, we have used the AAK model with $m\equiv j$ and $M=4$ (see Eq.\,\eqref{eq:aakdecomp}), but the decomposition \eqref{eq:decomp} is compatible with all other standard approaches in EMRI modeling (see Sec.\,\ref{subsec:models}). Note however that a decomposition into angular harmonics alone (as done for comparable-mass-binary waveforms) is insufficiently discriminative for our proposal, which relies on there being a distinctive spread in individual-mode optimal SNRs for each source.

Our desired likelihood $L'$ must naturally make use of the discrepancy between the mode structure of a putative signal at each $\boldsymbol{\theta}$, i.e., the vector of mode optimal SNRs
\begin{equation}\label{eq:modeoptsnr}
    [v_\mathrm{opt}(\boldsymbol{\theta})]_m:=\rho_\mathrm{opt}( h_m(\boldsymbol{\theta}))
\end{equation}
with $\rho_\mathrm{opt}$ as defined in Eq.\,\eqref{eq:optsnr}, and the mode information that is actually recovered from the data $x$ at that same point, i.e., the vector of mode detection SNRs
\begin{equation}\label{eq:modedetsnr}
    [v_\mathrm{det}(\boldsymbol{\theta})]_m:=\rho_\mathrm{det}(h_m(\boldsymbol{\theta}))
\end{equation}
with $\rho_\mathrm{det}$ as defined in Eq.\,\eqref{eq:detsnr}. This allows the point $\boldsymbol{\theta}_\mathrm{inj}$ (where $\mathrm{E}[v_\mathrm{det}]=v_\mathrm{opt}$) to be differentiated from any typical $\boldsymbol{\theta}_\mathrm{sec}$ (where only a single component of $v$ is matched). In fact, the discrepancy between $v_\mathrm{opt}$ and $v_\mathrm{det}$ is already partly accounted for in the standard log-likelihood \eqref{eq:fullloglike}, which may be written more evocatively as
\begin{align}\label{eq:expandloglike}
    \ln{L}=&\;\langle x|h\rangle-\frac{1}{2}\langle h|h\rangle-\frac{1}{2}\langle x|x\rangle\nonumber\\
    \approx&\;v_\mathrm{opt}\cdot v_\mathrm{det}-\frac{1}{2}\langle h|h\rangle-\frac{1}{2}\langle x|x\rangle,
\end{align}
where all dependence on $\boldsymbol{\theta}$ is implicit. The approximation in Eq.\,\eqref{eq:expandloglike} is due only to the truncation of modes in Eq.\,\eqref{eq:decomp}, whereas $\langle h|h\rangle\approx|v_\mathrm{opt}|^2$ if we further assume $\langle h_m|h_{m'}\rangle\approx0$ for all $m\neq m'$ (which is generally valid).

Computation of the constant final term in Eq.\,\eqref{eq:expandloglike} is not needed in practice except for evidence calculations, and the penultimate term varies much more slowly than the first with respect to most parameters that affect phasing. Thus the first term is the main determinant of the likelihood profile over parameter space---its value is closer to $\rho_\mathrm{inj}^2$ at $\boldsymbol{\theta}_\mathrm{sec}$, and closer to zero at $\boldsymbol{\theta}_\mathrm{tail}$. A direct strategy is then to suppress the first term to zero at $\boldsymbol{\theta}_\mathrm{sec}$, using a similarity-like quantity $Q:\Theta\to[0,1]$ that depends on some measure of discrepancy between $v_\mathrm{opt}$ and $v_\mathrm{det}$:
\begin{align}\label{eq:modloglike}
    \ln{L'}:=&\;Q\langle x|h\rangle-\frac{1}{2}\langle h|h\rangle-\frac{1}{2}\langle x|x\rangle\nonumber\\
    =&\;\ln{L}-(1-Q)\langle x|h\rangle.
\end{align}
The form of Eq.\,\eqref{eq:modloglike} seems to preclude statistically motivated definitions where $L'$ is derived from natural probabilistic statements; nevertheless, such a modification should facilitate EMRI search, at the cost of invalidating the interpretation of $L'$ as a proper Bayesian likelihood.

\begin{figure}[!tbp]
\centering
\includegraphics[width=\columnwidth]{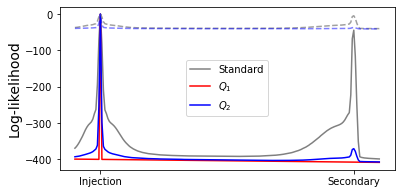}
\caption{Standard and veto log-likelihood values along an extended connecting line through the injection parameters and a strong secondary node from Sec.\,\ref{subsec:10x}---specifically, the set-I node of pair 2 (red point in middle panel of Fig.\,\ref{fig:pairs}). The approximate inner product \eqref{eq:appinner} is used for clarity, although the general concept holds for the full inner product \eqref{eq:fullinner}. Dashed curves are annealed versions of the respectively colored standard and veto ($Q_2$) log-likelihoods, with a temperature of 10.}
\label{fig:modlike}
\end{figure}

Our first suggestion for $Q$ is
\begin{equation}\label{eq:chisquaredveto}
    Q_1:=1-F_{\chi^2}\left(|v_\mathrm{opt}-v_\mathrm{det}|^2\right),
\end{equation}
where $F_{\chi^2}$ denotes the cumulative distribution function of a chi-squared random variable with $M$ degrees of freedom. The veto likelihood $L'$ with $Q_1$ works well to ``flatten out'' secondaries (see Fig.\,\ref{fig:modlike}), while the cross-sectional profile of the density bulk around $\boldsymbol{\theta}_\mathrm{inj}$ is only marginally narrower at the Gaussian-analogous 3-$\sigma$ value $\ln{L}=-9/2$ (and well beyond). Another option for $Q$ is
\begin{equation}\label{eq:cosineveto}
    Q_2:=\left(\hat{v}_\mathrm{opt}\cdot\hat{v}_\mathrm{det}\right)^{2q},
\end{equation}
where an overhat denotes normalization with respect to the Euclidean inner product on $\mathbb{R}^M$ ($\hat{v}\cdot\hat{v}=1$). The quantity $q$ is empirically determined; it might be defined to depend on $\boldsymbol{\theta}$ via the properties of $v_\mathrm{opt}$, but here we treat it as a tunable hyperparameter. This version of $L'$ is also shown in Fig.\,\ref{fig:modlike}, for $q=20$. There are of course many other viable choices for the functional form of $Q$, which only needs to be $\approx1$ near $\boldsymbol{\theta}_\mathrm{inj}$ and $\approx0$ everywhere else (or at least near $\boldsymbol{\theta}_\mathrm{sec}$). Such veto likelihoods might be combined with other likelihood-smoothing approaches such as annealing or semi-coherent filtering (which essentially broadens all likelihood peaks), in order to further aid EMRI search. We leave optimization of the basic concept in Eqs \eqref{eq:modloglike}--\eqref{eq:cosineveto} for future follow-up studies.

As decompositions of the form \eqref{eq:decomp} are central to EMRI waveform modeling, the set of modes $\{h_m\}$ for the veto likelihood can first be pre-determined (in terms of membership), then obtained as a byproduct of template generation during data analysis. Similarly, the inner-product operations $\langle h_m|h_m\rangle$ in $v_\mathrm{opt}$ (and $v_\mathrm{det}$) may be bypassed with a pre-computed fit. The veto likelihood is however penalized with added online cost through its reliance on $v_\mathrm{det}$, which entails $M$ evaluations of $\langle x|h_m\rangle$. All other operations have negligible cost relative to the inner product at full sampling resolution, and thus the final cost of the veto likelihood is approximately $M$ times that of the standard likelihood. This may seem like a significant penalty even for modest $M$, but the hope is that it will be more than offset by the increased efficiency of search. Sampling algorithms will no longer have to explore a large number of probability peaks en route to the global peak; this in turn will increase the reliability of candidate detections, and accelerate the transition to performing inference on individual signals.

\section{Conclusion}
\label{sec:conclusion}

Self-confusion and degeneracy for EMRIs, as we have defined them in Sec.\,\ref{subsec:keyconcepts}, are both manifestations of the fact that sources with very different parameters can have signals that strongly resemble one another. The former term is used in the context of a finite set of putative EMRI signals that might plausibly be present in LISA data, while the latter refers instead to the continuum of possible EMRI signals described by a physical waveform model. We have shown in Sec.\,\ref{sec:confusion} that self-confusion is unlikely to arise in reality---for a typical set of $\approx200$ detectable sources distributed according to a representative astrophysical model, the root mean square of pairwise overlaps among their two-year signals is $\sim10^{-3}$. On the other hand, degeneracy is one of the main hindrances to the search for EMRI signals in LISA data, and we have devoted the bulk of this work to its characterization.

Several new analysis tools have been introduced in Sec.\,\ref{sec:tools} for our study of degeneracy, where existing methods are inadequate. These include an approximate noise-weighted inner product that acts on highly downsampled waveform amplitude/phase trajectories, as well as a bespoke clustering algorithm that uses overlaps along connecting lines in parameter space to define the degree of connectivity between pairs of signal templates. In Sec.\,\ref{subsec:mapping}, we have conducted an extensive mapping analysis of the posterior surface over parameter space for a representative signal injection, which has yielded a variety of qualitative results. The most notable of these are: i) secondary posterior peaks are both strong (injection--template overlaps of $\lesssim0.8$) and numerous ($\sim10^3$ of them) in an encompassing region that is $\gtrsim10^{12}$ times larger than the posterior bulk; ii) they come in different shapes and sizes; and iii) they are inter-connected and thus cannot be localized. Furthermore, the conditions that give rise to secondaries are very natural, as discussed in Sec.\,\ref{subsec:corollaries}, and thus they are expected to be a generic feature of the EMRI signal space.

Our results hold several implications for EMRI data analysis, as we have discussed at length in Sec.\,\ref{sec:implications}. The fundamental interaction of degeneracy with detector noise or with multiple actual signals is unlikely to cause false positives or negatives in the search for candidate signals, but this may still arise in practice due to the technical limitations of stochastic search methods. Regardless, previously proposed strategies for EMRI search are a promising first step in addressing degeneracy; we have provided a few complementary suggestions of our own in Sec.\,\ref{sec:suggestions}, to rule out false signals and to better identify the source parameters of actual ones. We have however mostly limited the scope of the current study on degeneracy to illuminate the problem itself, rather than to develop possible practical solutions. The new tools and perspectives we have introduced here should help to inform and direct follow-up work on the latter.

There is also utility in an improved characterization of the degeneracy problem, although this is arguably unneeded before the advent of efficient and extensive next-generation waveform models. For example, the global distribution of secondaries over the full-dimensional space of intrinsic and extrinsic parameters is still unknown for a single injection---much less a representative set of injections with different source parameters and analysis durations. Scaling up the mapping analysis in this way will require computational enhancements such as fast LISA-response models, alternative algorithms for obtaining a large set of high-overlap points, and GPU acceleration of the sampling and clustering steps. Once fast and fully generic Kerr models with adiabatic evolution schemes and mode content become available, it might be worthwhile to perform comprehensive surveys of secondaries, with the focus shifting to the quantitative characterization of degeneracy for these specific models. Such studies could inform the tuning of sampling algorithms and strategies for EMRI search. Proxy post-adiabatic models that incorporate approximate or phenomenological resonant jumps might also be explored using the tools we have introduced, in order to investigate how degeneracy changes in the presence of transient resonances.

\begin{acknowledgments}
AJKC broadly thanks members of the LISA and self-force communities, for any relevant discussions that might have taken place over the past four years. Specific gratitude goes out to Michele Vallisneri and Yanbei Chen for their financial and moral support. Both AJKC and CJC acknowledge support from the NASA LISA Preparatory Science grants 18-LPS18-0027 and 20-LPS20-0005, from the NSF grant PHY-2011968, and from the Jet Propulsion Laboratory (JPL) Research and Technology Development program. Parts of this work were carried out at JPL, California Institute of Technology, under a contract with the National Aeronautics and Space Administration (80NM0018D0004).
\end{acknowledgments}

\bibliographystyle{unsrt}
\bibliography{references}

\end{document}